\documentclass[
superscriptaddress,
amsmath,amssymb,aps,prb,twocolumn,
floatfix
]{revtex4-2}

\usepackage{graphicx}% Include figure files
\usepackage{dcolumn}% Align table columns on decimal point
\usepackage{bm}% bold math
\usepackage{hyperref}% add hypertext capabilities
\usepackage{float}
\usepackage{xcolor}
\usepackage{comment}
\usepackage[normalem]{ulem}
\usepackage[USenglish,american]{babel}
\graphicspath{ {./Images/} }
\begin{document}

\title{Extended Bose-Hubbard Model on Small Grids:
\\Exact Diagonalization and Monte Carlo Studies}

\author{Gabriele Costa}
\email{gabriele.costa@studenti.unime.it}
\affiliation{Dipartimento di Scienze Matematiche e Informatiche, Scienze Fisiche e Scienze della Terra, Università degli Studi di Messina, Viale F. Stagno d’Alcontres 31, 98166 Messina, Italy}
\affiliation{Institute for Chemical-Physical Processes, National Research Council of Italy (IPCF-CNR),
98158 Messina (Italy)}
                
\author{Matteo Ciardi}
\email{matteo.ciardi@tuwien.ac.at}
\affiliation{Institute for Theoretical Physics, TU Wien, Wiedner Hauptstraße 8-10/136, 1040 Vienna, Austria}

\author{Fabio Cinti}
\email{fabio.cinti@unifi.it}
\affiliation{Dipartimento di Fisica e Astronomia, Universit\`a di Firenze, I-50019, Sesto Fiorentino (FI), Italy}
\affiliation{INFN, Sezione di Firenze, I-50019, Sesto Fiorentino (FI), Italy}

\author{Santi Prestipino}
\email{sprestipino@unime.it}
\affiliation{Dipartimento di Scienze Matematiche e Informatiche, Scienze Fisiche e Scienze della Terra, Università degli Studi di Messina, Viale F. Stagno d’Alcontres 31, 98166 Messina, Italy}
\affiliation{Institute for Chemical-Physical Processes, National Research Council of Italy (IPCF-CNR),
98158 Messina (Italy)}

\begin{abstract}
The superfluid-insulator transition in systems of lattice bosons is usually analyzed in the framework of the Bose-Hubbard model, and has been extensively studied by theory and simulations.
Less attention has been paid to the remnants of the transition in truncated lattices, with or without periodic boundary conditions.
Here we consider the hard-core limit of the extended Bose-Hubbard model on small square and triangular grids --- i.e., sections of the square and triangular lattices containing up to 13 sites.
By mapping out the zero-temperature phase diagram through exact diagonalization, we find ground-state characteristics that are markedly different from those emerging in the thermodynamic limit, together with similarities.
The dichotomy between superfluid-like and insulating-like behavior is then investigated in two-dimensional systems of a few interacting bosons in the continuum, subject to confining and optical-lattice potentials mimicking the $3\times 3$ square grid.
Using path-integral Monte Carlo simulations, we compute kinetic and potential energies, as well as superfluidity and exchange-cycle statistics, finding hints of Bose-Hubbard behavior even in systems of just five particles.
\end{abstract}
\date{\today}
\maketitle

\section{Introduction}

A better knowledge of the mechanisms underlying the onset of emergent behavior in complex many-particle systems will make it easier to design materials with prescribed properties.
Yet, a hallmark of complexity is that macro-scale properties are often hard to predict from micro-scale features.
Besides, the challenge of characterizing phase equilibria becomes increasingly more difficult when moving from classical to quantum many-body systems, for which analytic methods are limited and numerical simulations are considerably more expensive.
From a theoretical point of view, it would be appropriate to work at zero temperature ($T=0$), where the relevant quantum phases are already visible and their features stand out more clearly, since free from thermal noise.
At $T=0$, phase transitions may still occur in the ground state of a many-body system as functions of non-thermal parameters like pressure, magnetic-field strength or particle mass.
Such ``quantum phase transitions''~\cite{sachdev1999,vojta2003} typically arise from the competition between various terms in the Hamiltonian, each prompting a different kind of system self-organization.

A paradigmatic case is the Bose-Hubbard (BH) model~\cite{fisher1989}, probably the simplest model of interacting bosonic particles on a lattice with a non-trivial phase diagram.
Here, each particle experiences two competing forces:
One pushes it to jump to neighboring sites, while the other pushes it away from neighboring particles.
The interplay between these opposing tendencies is further regulated by the chemical potential, which decides how many particles the lattice is to contain. 
The ground state of this system is either insulating or superfluid, depending on the ratio between the hopping amplitude and the strength of repulsion.
A large repulsion favors insulating (solid-like) behavior, characterized by integer boson densities and a gapped spectrum.
Instead, highly mobile particles develop overall phase coherence and a non-zero superfluid response.

The BH model describes the behavior of an ultracold gas of bosonic atoms in an optical lattice~\cite{jaksch1998,greiner2002,trefzger2011}.
At very low temperature, a gas of bosons is condensed.
Turning on the lattice potential, the system remains in the condensed phase as long as the potential wells are shallow.
In this regime, the kinetic energy is dominant and a delocalized wave function minimizes the total energy.
In the opposite limit where the lattice potential is strong, the interatomic repulsion overcomes kinetic energy and the total energy is minimized by filling each lattice site with the same number of atoms.

The phase diagram of the BH model at $T=0$ has been worked out numerically in both three~\cite{capogrosso-sansone2007} and two dimensions~\cite{capogrosso-sansone2008}, and is well reproduced by a number of theories~\cite{rokhsar1991,krauth1992,sheshadri1993,vanoosten2001,kuebler2019}.
Other phases appear when the range of interparticle repulsion is extended to further neighbors.
In this case, supersolid phases (combining a non-uniform density with superfluidity) can also become stable~\cite{ng2010,iskin2011,kimura2012,ohgoe2012,kurdestany2012,zhang2022}.
Extended BH models provide a more realistic approximation than the original model to condensates of dipolar bosonic atoms in optical lattices~\cite{iskin2009,pollet2010,baier2016,su2023,lagoin2022a}.
Other generalizations of the BH model are possible~\cite{dutta2015,suthar2020,chanda2025}, leading to a wealth of spatially modulated phases.

However, genuine phase transitions only occur when the lattice is infinite in every direction.
In large samples, sharp crossover behaviors occur in place of proper, non-analytic transitions.
In turn, such crossovers do not arise from nowhere, but rather would be the continuation of trends that are present even in much smaller systems.
In this paper, we put this expectation to the test on the simplest extended BH model (that is, the original BH model with the addition of first-neighbor repulsion), which will be examined on two-dimensional grids (i.e., truncated lattices) small enough that diagonalization of the model can be done exactly.
To accomplish this task, we must assume an infinite on-site repulsion, i.e., single site occupancy --- despite this simplification, the gross features of the model stay unaltered.
We shall see that BH behavior is quite different on small grids, whereby the structure of insulating ``phases'' deviates from the infinite-size limit, but the superfluid-insulator interplay nevertheless remains.
Exact diagonalization studies have been carried out for the Bose-Hubbard model by various authors, see e.g. Refs.~\cite{zhang2010,sowinski2012,szabados2012,raventos2017}.

The rivalry between BH phases on a grid can also be analyzed in simulation, using the ``language'' of bosonic particles in continuous space.
In this case we must apply an external field having the same symmetry of the grid and employ a box-shaped potential to prevent particles from exiting the grid region.
At low but non-zero temperature, this system will be studied by path-integral Monte Carlo (PIMC) simulations, which provide reliable indications on the structure and energy of the sample, as well as on its superfluid response.

The paper is organized as follows:
After presenting the model and the method in Section II, we discuss exact-diagonalization results in Section III, where the model is examined on various grids, with and without periodic boundary conditions.
Then, in Section IV we comment the results of boson simulations, and compare them with the ascertained behavior of the discrete model.
Section V is devoted to conclusions.

\section{Model and method}

On a regular lattice, the (grand) Hamiltonian of the extended BH model reads
\begin{eqnarray}
H&=&-t\sum_{\langle i,j\rangle}\left(a_i^\dagger a_j+a_j^\dagger a_i\right)+\frac{U}{2}\sum_in_i(n_i-1)
\nonumber \\
&+&V\sum_{\langle i,j\rangle}n_in_j-\mu\sum_in_i\,,
\label{eq1}
\end{eqnarray}
where the first and third sums run over all pairs (each counted once) of nearest-neighbor (NN) sites, $a_i$ and $a_i^\dagger$ are bosonic field operators, and $n_i=a_i^\dagger a_i$.
Furthermore, $t>0$ is the hopping parameter, $U>0$ is the interaction energy between two particles on the same site, $V>0$ is the interaction energy between two NN particles, and $\mu$ is the chemical potential.

As discussed in the Introduction, in order to make the dimension of state space finite, we must truncate the lattice {\em and} apply an upper cutoff $n_{\rm cut}$ on site occupancy.
A finite $n_{\rm cut}$ will substantially affect the phase diagram only at low $U/V$ and high $\mu/V$ (see Section III).
The simplest choice is taking $U\rightarrow\infty$ in Eq.~(\ref{eq1}), corresponding to $n_{\rm cut}=1$ (hard-core limit).
This is the model version we consider from now onward.

Focusing on the two-dimensional case, the phase diagram of hard-core bosons at $T=0$ is known on the square~\cite{capogrosso-sansone2010,yamamoto2012} and the triangular lattice~\cite{wessel2005,zhang2011} (in the latter case there is also a supersolid region).
On the square lattice, the only non-trivial insulating phase (i.e., other than the empty and fully occupied lattices) has checkerboard structure.
This phase is stable in a lobe-shaped region of the $t$-$\mu$ plane.
On the triangular lattice, two insulating phases (and lobes) instead exist, where either one or two triangular sublattices out of three are filled and the other(s) are empty.
On a small-sized grid, homogeneity is lost and we expect (beside the removal of every singularity and sharp distinction between the phases) some more changes in the ``insulating'' sector, which will be especially evident at $t=0$~\cite{prestipino2020,prestipino2021,degregorio2021}.
Other differences will result from the application of periodic boundary conditions (PBC).

\begin{figure}[t]
\includegraphics[width=\linewidth]{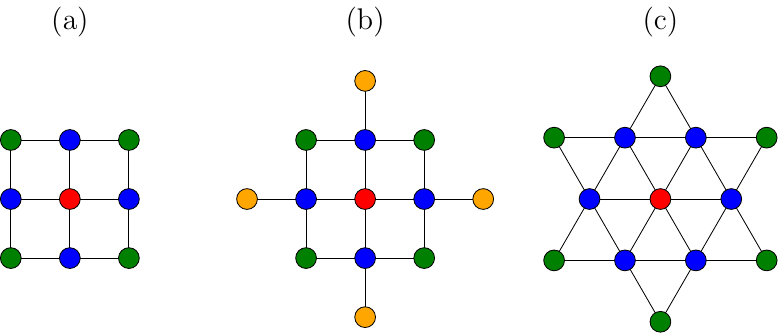}
\caption{\label{fig1}
The three grids on which we map out the phase diagram of the model by exact diagonalization: (a) Square. (b) Diamond. (c) Star. We use different colors to mark first, second, and third neighbors of the central site. All these grids can be periodically replicated without overlaps or gaps.}
\end{figure}

In Fig.~\ref{fig1} we show the grids considered in our study:
Two different sections of the square lattice (``Square'' and ``Diamond''), with 9 and 13 sites respectively, and a small fragment of the triangular lattice (``Star'') containing 13 sites.
All such grids show discrete rotational symmetry around the central site.
We have thus used different colors in Fig.~\ref{fig1} to identify shells of neighbors at same distance from the central site.
However, we cannot exclude that rotational symmetry is spontaneously broken in the actual ground state of the system, which will then be multiple (i.e., associated with a degenerate eigenvalue), making it impossible to describe, e.g., the local density just in terms of shell values (we shall give examples in Section III).
We point out that the infinite lattice can be covered without overlaps or gaps with each tile in Fig.~\ref{fig1} (see Appendix A), meaning that for all such tiles we can contrast the effects of periodic and open boundary conditions on the system behavior.

For our model, the dimension of the Hilbert space on a grid of $M$ sites is $2^M$ (with $M$ equal to 9 or 13), i.e., small enough that we can compute a few exact energy eigenvalues and eigenstates in reasonable time.
To this aim, we represent the Hamiltonian on the Fock basis $\{\vert x_1,\ldots,x_M\rangle\}$ (with $x_i=0$ or 1) and diagonalize the ensuing matrix numerically.
In particular, the ground state $\vert g\rangle$ and its eigenvalue, the zero-temperature grand potential $\Omega$ of the entire grid, are obtained as functions of $t$ and $\mu$ (in particular, the chemical potential is typically increased in steps of $0.01V$).
Once $\vert g\rangle$ has been determined, we calculate the average occupancies of all nodes and the condensate fraction $\rho_{\rm c}$~\cite{vanoosten2001,yamamoto2009}:
\begin{equation}
\rho_{\rm c}=\frac{1}{M}\langle g\vert\widetilde a_{\bf 0}^\dagger\widetilde a_{\bf 0}\vert g\rangle\,,
\label{eq2}
\end{equation}
where $\widetilde a_{\bf 0}=(1/\sqrt{M})\sum_{i=1}^Ma_i$ is the zero-momentum field operator.
In the superfluid phase of an infinite two-dimensional lattice system $\rho_{\rm c}>0$, since a non-zero superfluid response (as measured, in a simulation with PBC, through the winding-number fluctuations~\cite{ceperley1995}) also entails a non-zero condensate density, and vice versa.
Since the Hamiltonian commutes with the total number of particles, $\sum_in_i$, the $(t,\mu)$ plane is divided in sectors where the number of particles takes a constant value $N$, from 0 to $M$.
Clearly, $N=0$ is the empty grid and $N=M$ is the fully occupied grid.
In each $N$-sector, the only non-zero Fourier coefficients of $\vert g\rangle$ are those relative to basis states with $\sum_ix_i=N$.

A distinguishing feature of insulating phases is the non-zero energy gap, in contrast to the zero gap of a superfluid or a supersolid phase (see, e.g., Ref.~\cite{bloch2008b}).
A finite gap signals that low-energy excitations are rather costly, implying a substantial robustness of the ground state to thermal fluctuations.
In order to see to what extent this is recovered in our finite system, in addition to the lowest eigenvalue $\Omega$, we also compute the second smallest eigenvalue ($\Omega_2$), which finally yields the gap as $\Delta=\Omega_2-\Omega$.

\section{Zero-temperature results}

In this section, we present results for the hard-core extended BH model on various two-dimensional grids.
We fix $T=0$ and thus consider only properties of the ground state as functions of $t/V$ and $\mu/V$.
When we apply open boundary conditions the system is intrinsically inhomogeneous, due to boundary sites having a smaller number of first neighbors.
Since the NN interaction is repulsive ($V>0$), we expect a general tendency of the dense system to gather particles at the boundary of the grid.
Homogeneity is recovered when PBC are applied, but again peculiarities are expected on small grids.

\subsection{Square}

\begin{figure*}[t]
    \centering
        \begin{minipage}[t]{0.24\linewidth} 
        \centering
        (a)\par\medskip
        \includegraphics[width=0.42\linewidth]{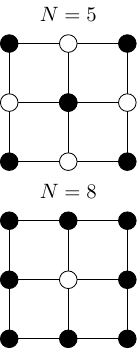}
    \end{minipage}
    \begin{minipage}[t]{0.24\linewidth} 
        \centering
        (b)\par\medskip
        \includegraphics[width=1.0\linewidth]{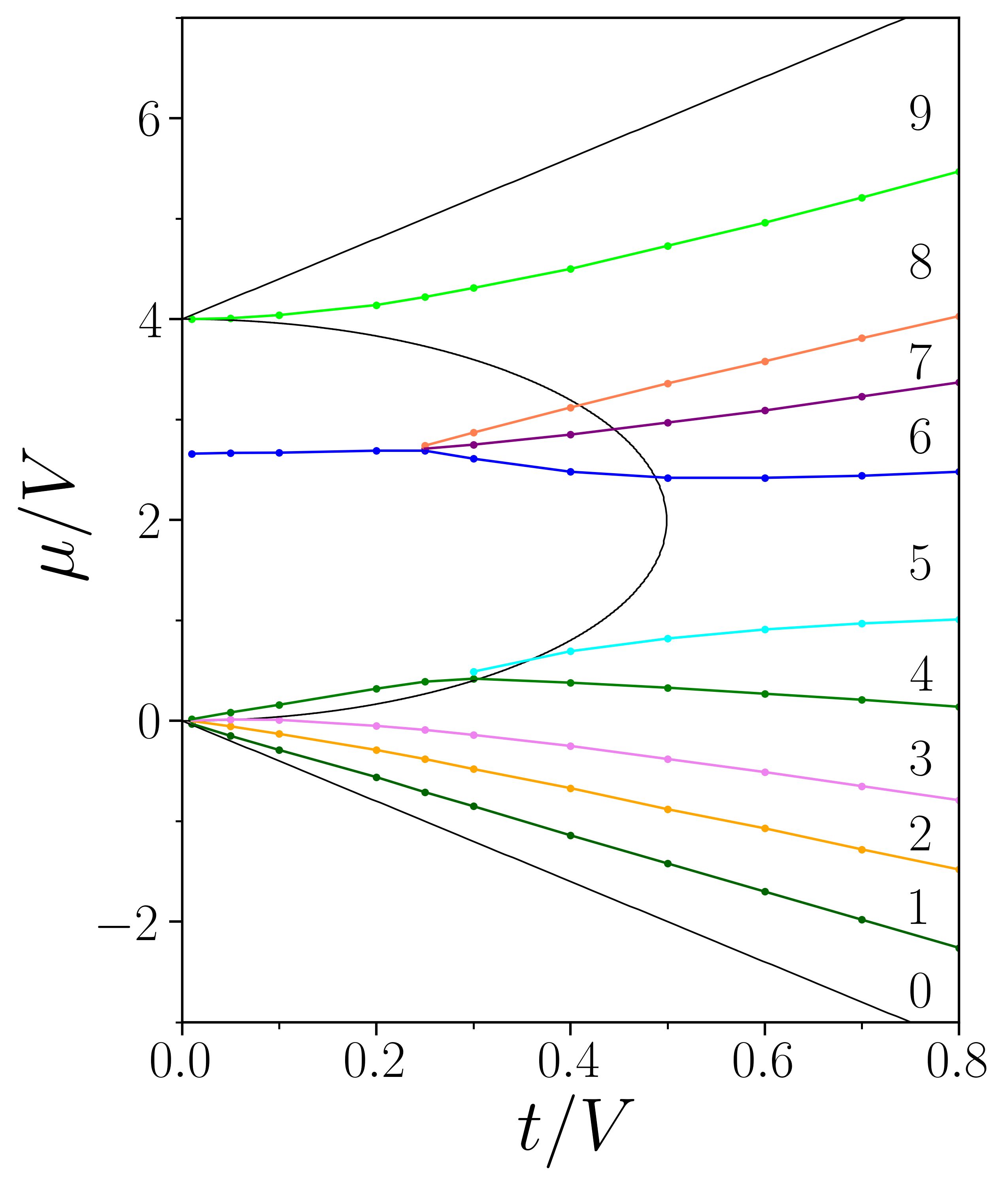}
    \end{minipage}
    \begin{minipage}[t]{0.24\linewidth}
        \centering
        (c)\par\medskip
        \includegraphics[width=1.0\linewidth]{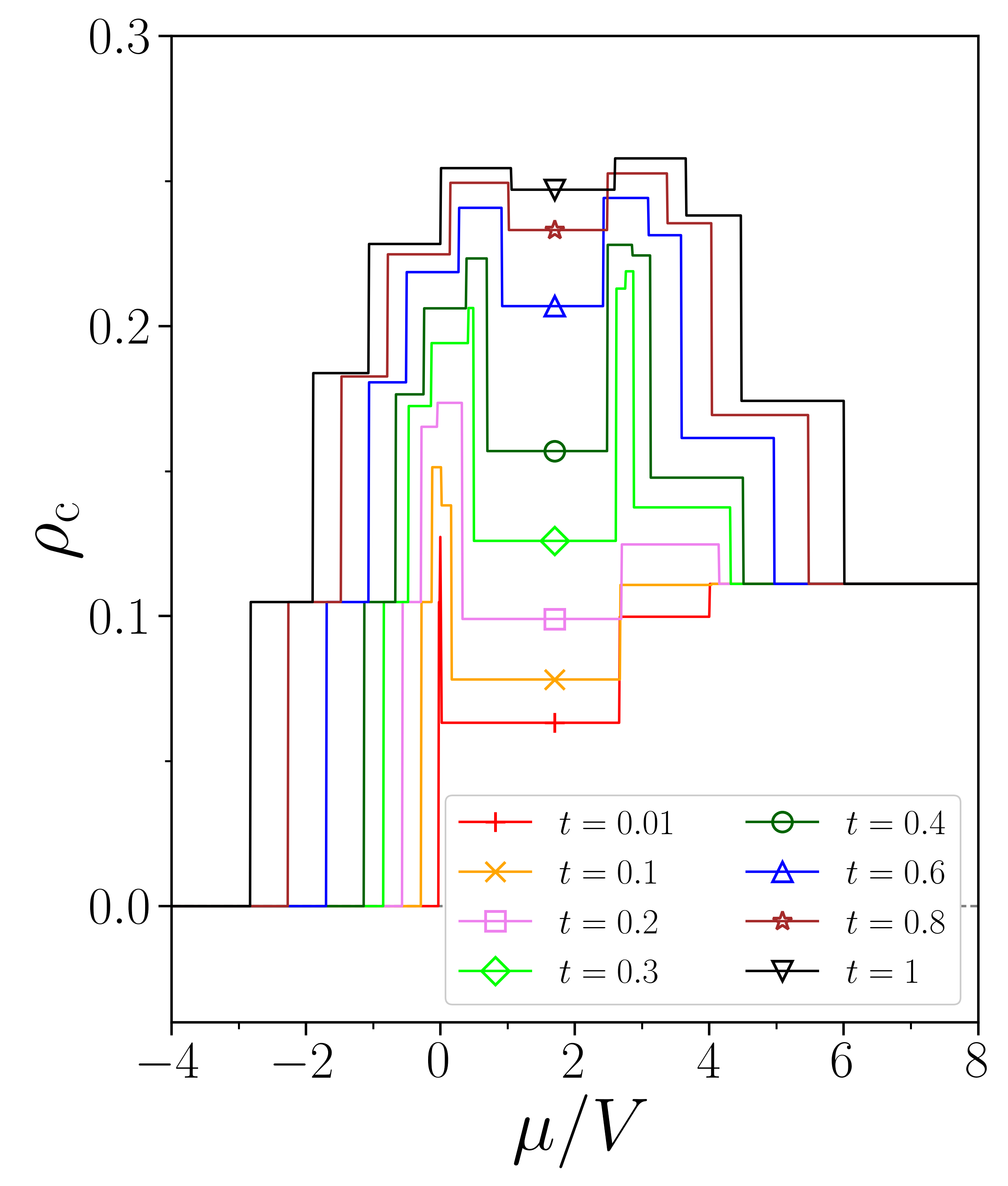}
    \end{minipage}
    \begin{minipage}[t]{0.24\linewidth} 
        \centering
        (d)\par\medskip
        \includegraphics[width=1.0\linewidth]{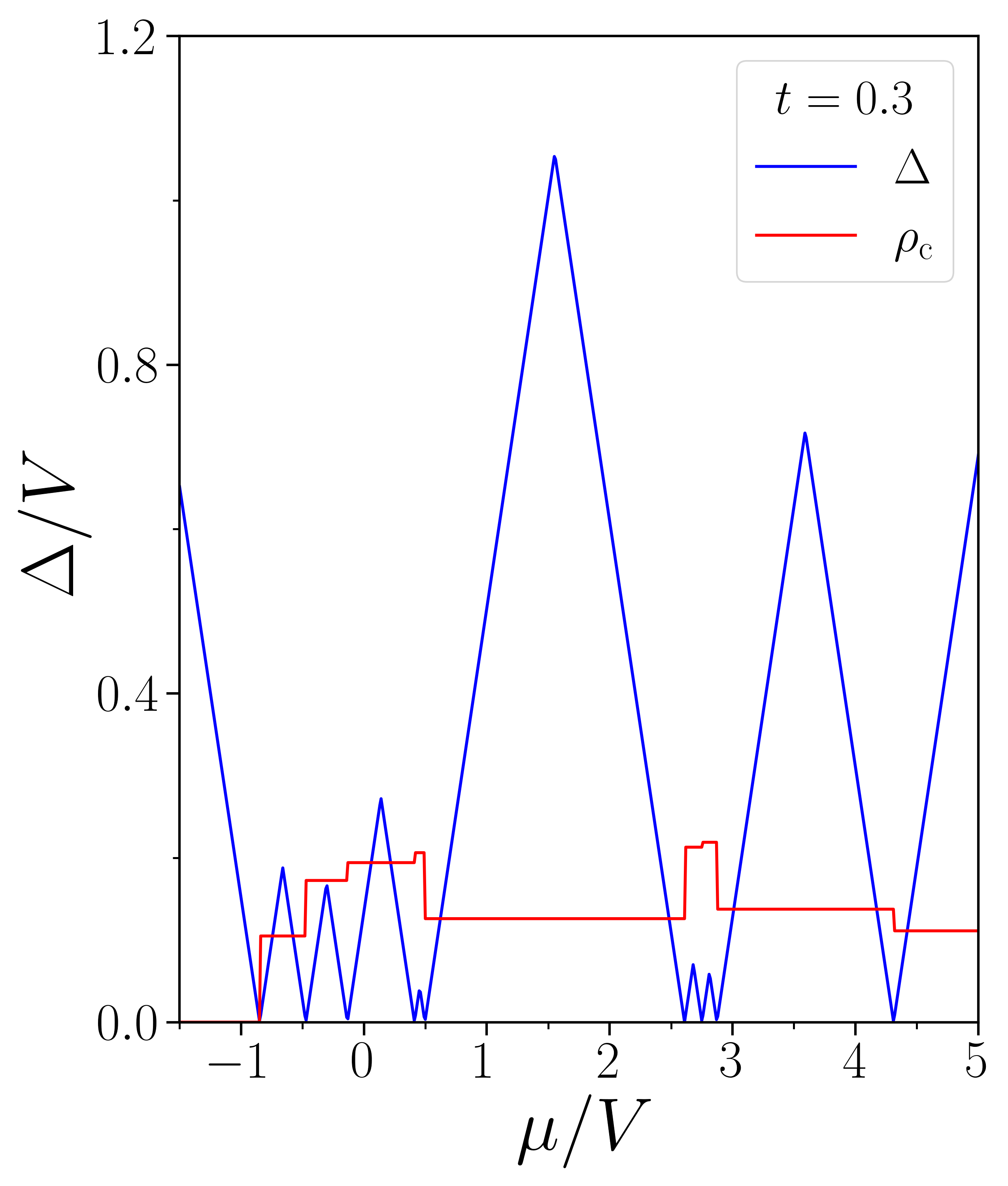}
    \end{minipage}
    
    \caption{\label{fig2}
    Square. (a) Ground states of the system for $t=0$ and $0<\mu<4V$. (b) Exact phase diagram at $T=0$. The broken lines through the points are boundary lines of $N$-sectors. The black thin lines are mean-field phase boundaries~\cite{gheeraert2015}. (c) Condensate fraction $\rho_{\rm c}(\mu)$ for various $t$. As $t$ grows, the condensate fraction grows as well and the central dip becomes more shallow. (d) Energy gap $\Delta(\mu)$ for $t=0.3$ (blue line). Also shown is the condensate fraction for the same $t$ (red line).}
\end{figure*}

The $3\times 3$ square grid in Fig.~\ref{fig1}a consists of a central site, together with its first- and second-neighbor sites.
At $t=0$, the system ground state corresponds to the occupancies yielding the lowest value of
\begin{equation}
h(x_1,\ldots,x_M)=V\sum_{\langle i,j\rangle}x_ix_j-\mu\sum_ix_i\,,
\label{eq3}
\end{equation}
for $x_i=0,1$.
Clearly, the minimum-$h$ state depends on $\mu$:
Aside from the trivial cases of the empty grid ($\mu<0$) and the fully occupied grid ($\mu>4V$), the only other possibilities are depicted in Fig.~\ref{fig2}a.
While for $0<\mu<(8/3)V$ the ground state (``chessboard'') mirrors the infinite-size structure of minimum energy on a smaller scale (though with an average occupation of 5/9, instead of 1/2), in the ground state for $(8/3)V<\mu<4V$ (``ring'') the central site is empty and the boundary sites are all filled.
The stability of the ring is a finite-size effect:
The ring provides a smaller $h$ than the chessboard when
\begin{equation}
8V-8\mu<-5\mu\,\,\,\Rightarrow\,\,\,\mu>\frac{8}{3}V\,.
\label{eq4}
\end{equation}

The full ``phase diagram'' at $T=0$, obtained by exact diagonalization, is plotted in Fig.~\ref{fig2}b.
In stark contrast with the mean-field phase boundaries~\cite{iskin2011,gheeraert2015} (reported in the same figure as black thin lines), there are no sharp phases in Fig.~\ref{fig2}b;
rather, the $t$-$\mu$ plane is divided in $N$-sectors (see Section II) which, with the relevant exception of the line separating the 5-sector from the 8-sector, are vaguely indicative of the location of iso-density lines (constant-density profiles) in the thermodynamic limit (see, e.g., Ref.~\onlinecite{alet2004}).

A quantity which more clearly discriminates between superfluid-like and insulating-like behavior is $\rho_{\rm c}$, plotted in Fig.~\ref{fig2}c as a function of $\mu$ along a few iso-$t$ lines.
Each jump discontinuity of $\rho_{\rm c}$ occurs in coincidence with one of $N$.
$\rho_{\rm c}$ develops a peak when we cross the ``superfluid'' region, while it takes a smaller value (in relative terms) when crossing an ``insulator'' region.
Eventually, when $N=M=9$ and $\vert g\rangle=\vert 1,1,\ldots,1\rangle$, it is easy to obtain from Eq.~\ref{eq2} that $\rho_{\rm c}$ becomes $1/M$.
As $t$ grows, the condensate fraction gradually increases and flattens out.
A feature of the superfluid phase of hard-core bosons is that the maximum possible value of $\rho_{\rm c}$ is definitely smaller than 1.
For example, we have verified numerically for $t=20$ that the maximum $\rho_{\rm c}$ is approximately 0.3 (see Fig.~\ref{fig11}b below).
The same is found in mean-field theory, where in models similar to the present one the asymptotic, $t\rightarrow\infty$ value of $\rho_{\rm c}$ is 0.25~\cite{prestipino2020,prestipino2021}.
Alternatively, we may consider the tight-binding model for hard-core bosons, where the condensate fraction equals 0.285 on Square, regardless of $\mu$. 

All in all, a clear correlation exists between $\rho_{\rm c}$ on Square and the condensate fraction on the square lattice, demonstrating that signatures of condensation are present even in a rather small boson system.

Then, we plot in Fig.~\ref{fig2}d the gap $\Delta$ as a function of $\mu$ for $t=0.3$.
$\Delta$ is larger in the insulating ``phases'' than in the superfluid region;
as a rule, the gap is wider the larger the distance in chemical potential from the nearest adjacent $N$-sector.
The non-monotonic behavior of $\Delta$ with $\mu$ has a simple explanation:
While the less-costly excitation is hole-like on the low-$\mu$ side of a sector, it is particle-like on the high-$\mu$ side. 
Looking more closely at the data, we indeed realize that $\partial\Delta/\partial\mu=\pm 1$, meaning that the first excited state, which is generally non-degenerate, is a linear combination of basis states with one particle more or less than those composing the ground state.

\begin{figure*}[t]
\centering
    \begin{minipage}[t]{0.24\linewidth} 
        \centering
        (a)\par\medskip
        \includegraphics[width=0.42\linewidth]{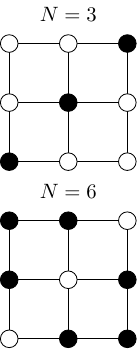}
    \end{minipage}
    \begin{minipage}[t]{0.24\linewidth} 
        \centering
        (b)\par\medskip
        \includegraphics[width=1.0\linewidth]{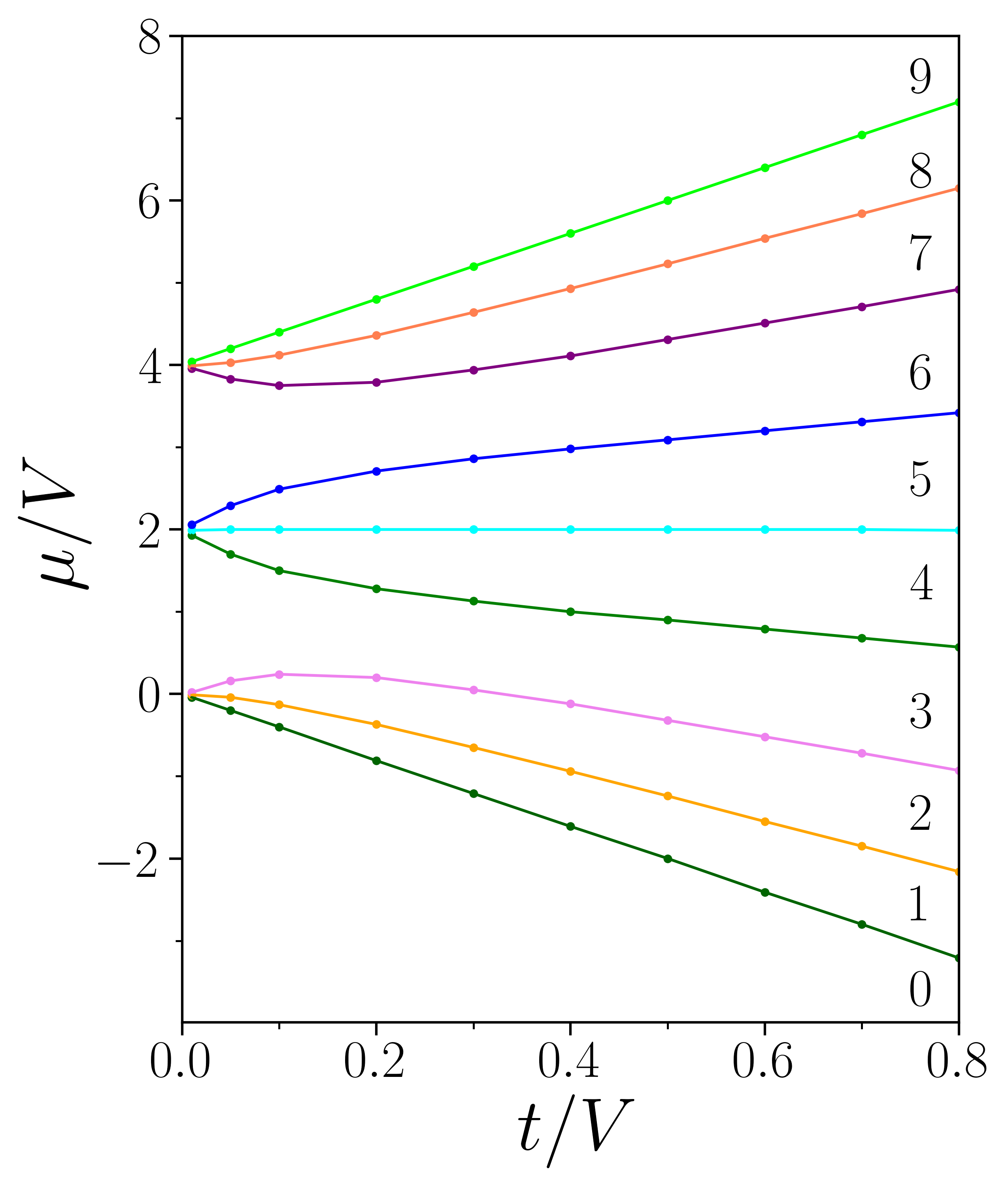}
    \end{minipage}
    \begin{minipage}[t]{0.24\linewidth}
        \centering
        (c)\par\medskip
        \includegraphics[width=1.0\linewidth]{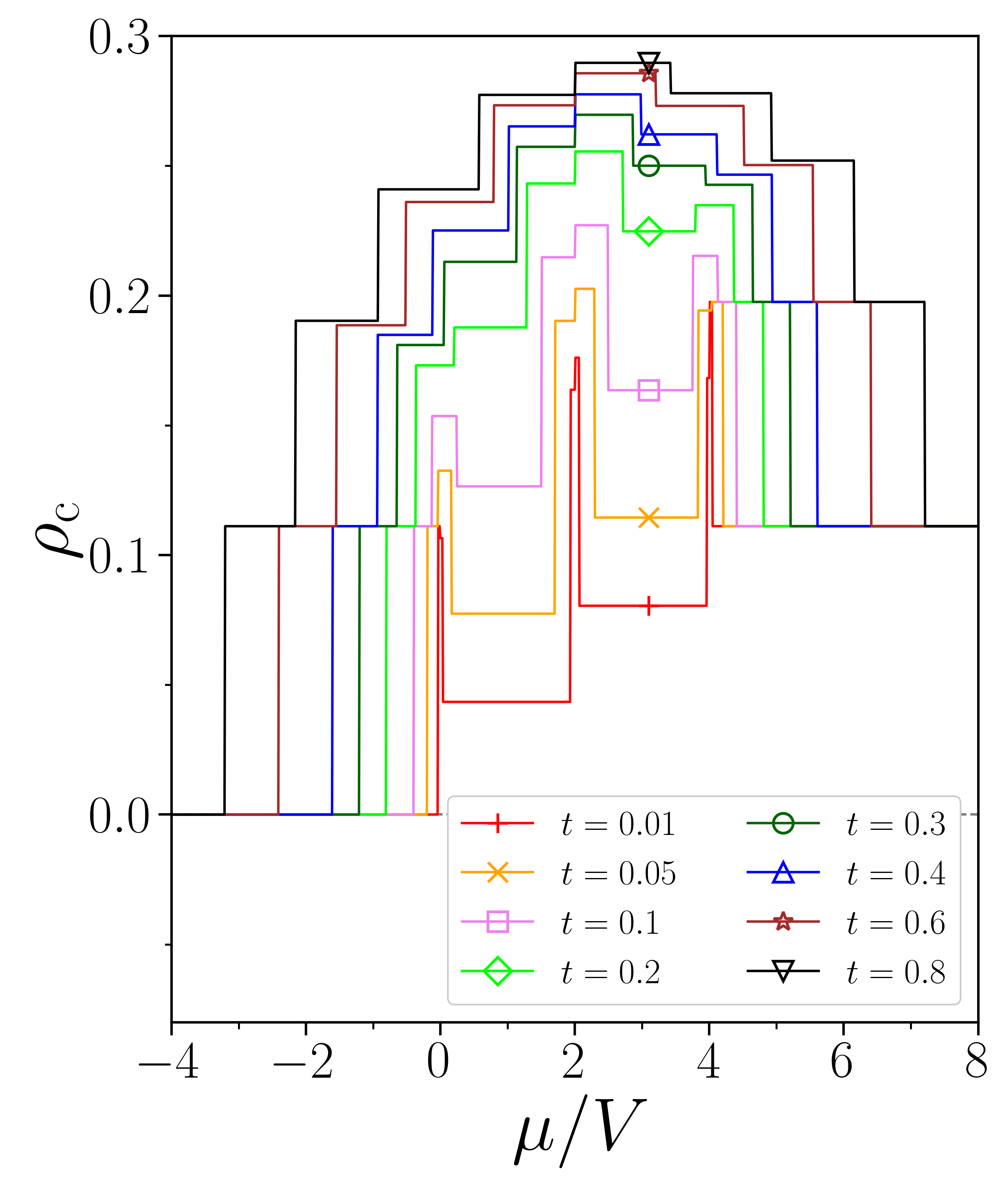}
    \end{minipage}
    \begin{minipage}[t]{0.24\linewidth} 
        \centering
        (d)\par\medskip
        \includegraphics[width=1.0\linewidth]{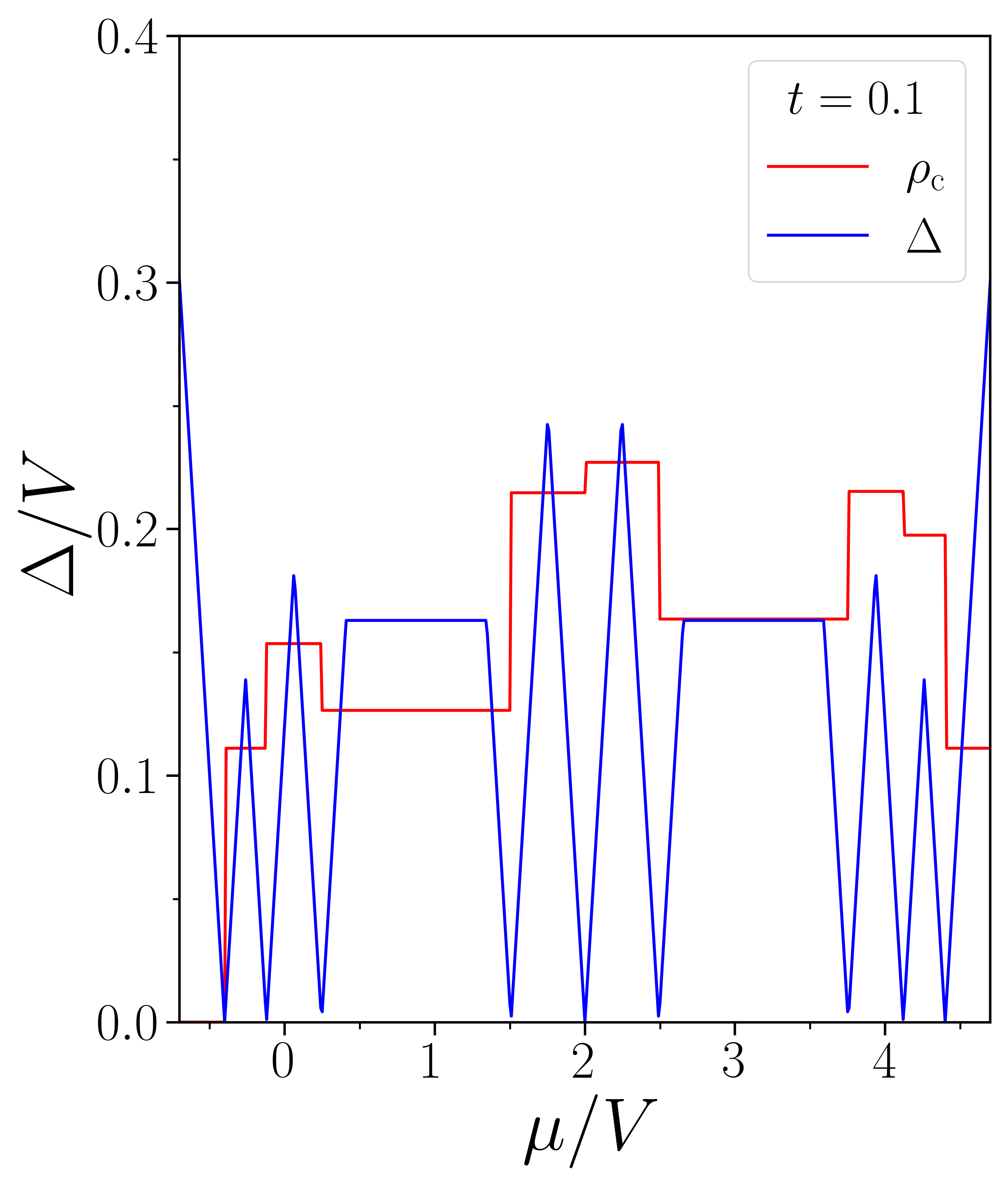}
    \end{minipage}
\caption{\label{fig3}
Square with PBC. (a) Ground states of the system for $t=0$ and $0<\mu<4V$. (b) Exact phase diagram at $T=0$. The broken lines through the points are boundary lines of $N$-sectors. (c) Condensate fraction $\rho_{\rm c}(\mu)$ for various $t$. As $t$ grows, the condensate fraction grows as well. (d) Energy gap $\Delta(\mu)$ for $t=0.1$ (blue line). Also shown is the condensate fraction for the same $t$ (red line).} 
\end{figure*}

Finally, we have looked at the same grid, but with periodic conditions applied at the boundary.
New boundary conditions bring about a redefinition of neighbors for boundary sites, resulting in a new phase portrait.
In Fig.~\ref{fig3}a we show the minimum-$h$ structures for $t=0$;
both these structures, which are negative images of one another, break rotational symmetry, implying degeneracy (here, equal to 6 for both).
%By drawing any of these ground states with a few periodic images, we realize that both structures in Fig.~\ref{fig3}a form stripes running along the diagonal direction.
However, the lowest energy eigenvalue for $t>0$ is non-degenerate and all symmetry-breaking Fock states of same species appear in the ground state with identical weight.
The overall phase diagram is illustrated in Fig.~\ref{fig3}b.
Compared to Fig.~\ref{fig2}b, the ``insulating'' regions are seemingly less extended in the $t$-$\mu$ plane, suggesting that (unless $t$ is very small) superfluid-like behavior is present in the whole range $0<N<9$.
This is also apparent from the condensate fraction (Fig.~\ref{fig3}c), which already for $t=0.3$ shows the typical bell shape of a superfluid.
We note that $\rho_{\rm c}=1/M=1/9$ in the 1-sector, since the ground state for $t>0$ is an equal-weight linear combination of Fock states with $\sum_ix_i=1$.
Lastly, we look at the gap as a function of $\mu$.
In Fig.~\ref{fig3}d (which refers to $t=0.1$) we see a flat region in both $N=3$ and $N=6$ sectors, which just indicates that for those values of $t$ and $\mu$ the first excited state is, as the ground state, a linear combination of basis states having $\sum_ix_i=3$ or 6.

The more widespread stability of the superfluid phase under PBC seems to contradict the common belief that, when the system is periodically replicated, finite-size effects are reduced and the behavior is more similar to the infinite system.
In fact, this cannot happen for Square because the number of sites on its edges is odd, and checkerboard ordering is thus excluded.
A small grid where the chessboard ground state is not frustrated under PBC is the $4\times 2$ rectangle (see Appendix A).
As is seen in Fig.~\ref{fig4}a, the topology of the phase diagram is indeed similar to the MF phase diagram.
Another example is the $4\times 3$ rectangle, where however the periodic images of the grid in the $y$ direction are shifted horizontally by one site to avoid frustration (Fig.~\ref{figA1}b).
Also in this case, as illustrated in Fig.~\ref{fig4}b, the $T=0$ phase diagram closely matches the MF one.
Interestingly, for both rectangles the boundaries of the 0- and $M$-sectors predicted by exact diagonalization perfectly coincide with those obtained from MF theory.

\begin{figure}[t]
\centering
\centering
    \begin{minipage}[t]{0.5\linewidth} 
        \centering
        (a)\par\medskip
        \includegraphics[width=1.0\linewidth]{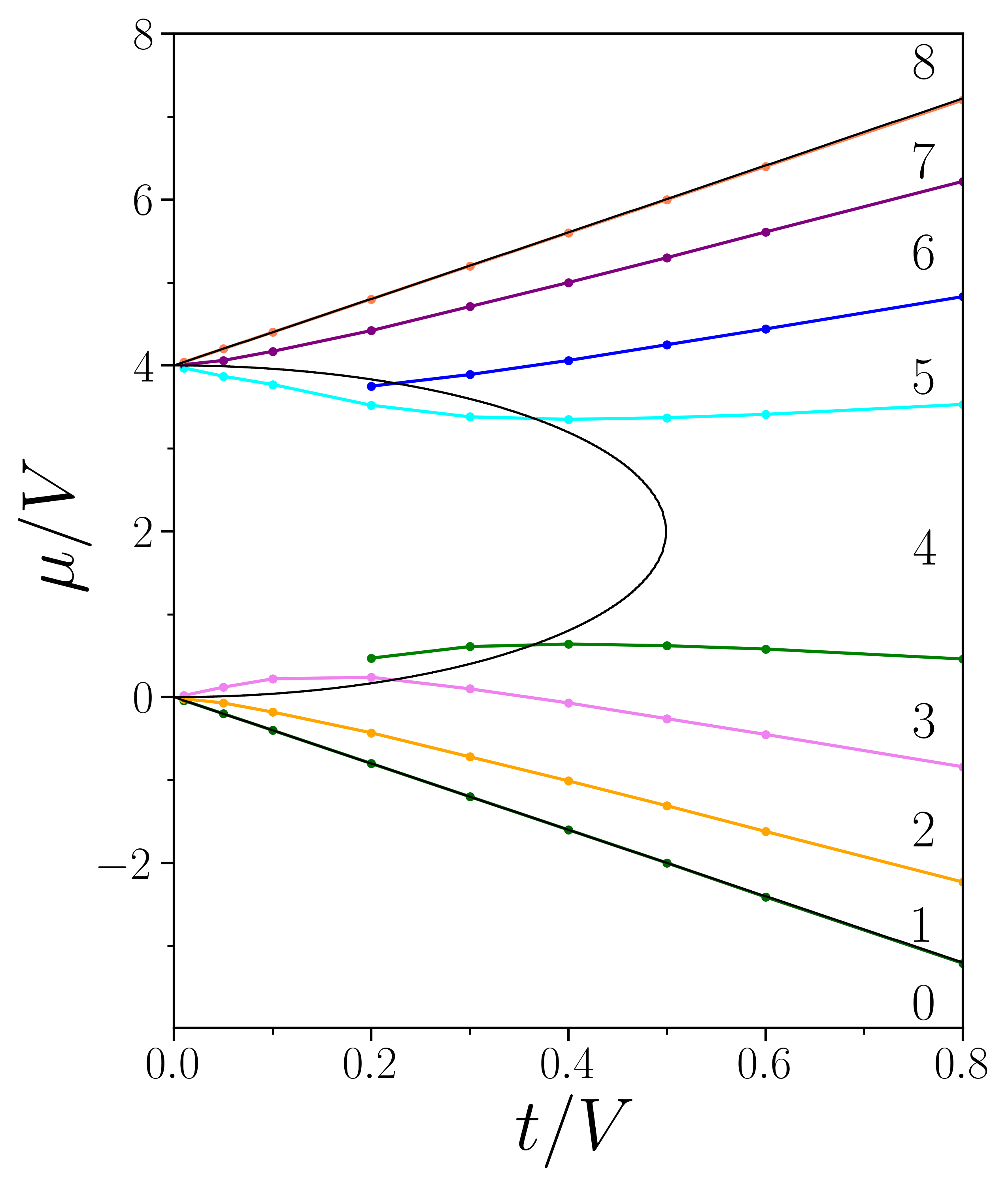}
    \end{minipage}\hfill
    \begin{minipage}[t]{0.5\linewidth} 
        \centering
        (b)\par\medskip
        \includegraphics[width=1.0\linewidth]{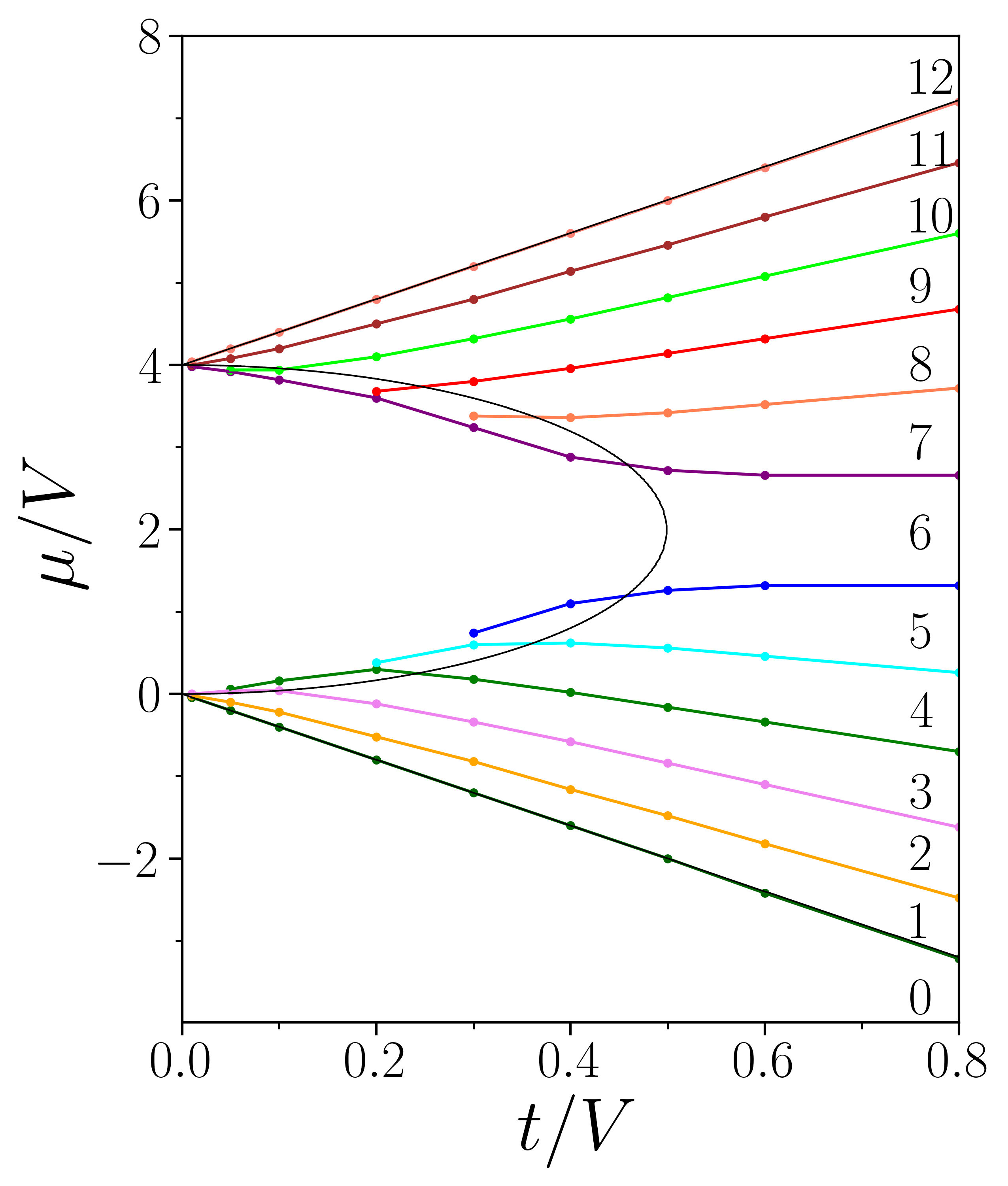}
    \end{minipage}
\caption{\label{fig4}
Exact phase diagram at $T=0$ of the $4\times 2$ (a) and the $4\times 3$ (b) rectangles with PBC. The broken lines through the points are boundary lines of $N$-sectors. The black thin lines are mean-field phase boundaries~\cite{gheeraert2015}.}
\end{figure}

\subsection{Diamond}

\begin{figure*}[t]
\centering
    \begin{minipage}[t]{0.24\linewidth} 
        \centering
        (a)\par\medskip
        \includegraphics[width=0.6\linewidth]{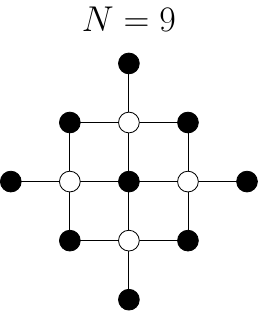}
    \end{minipage}
    \begin{minipage}[t]{0.24\linewidth} 
        \centering
        (b)\par\medskip
        \includegraphics[width=1.0\linewidth]{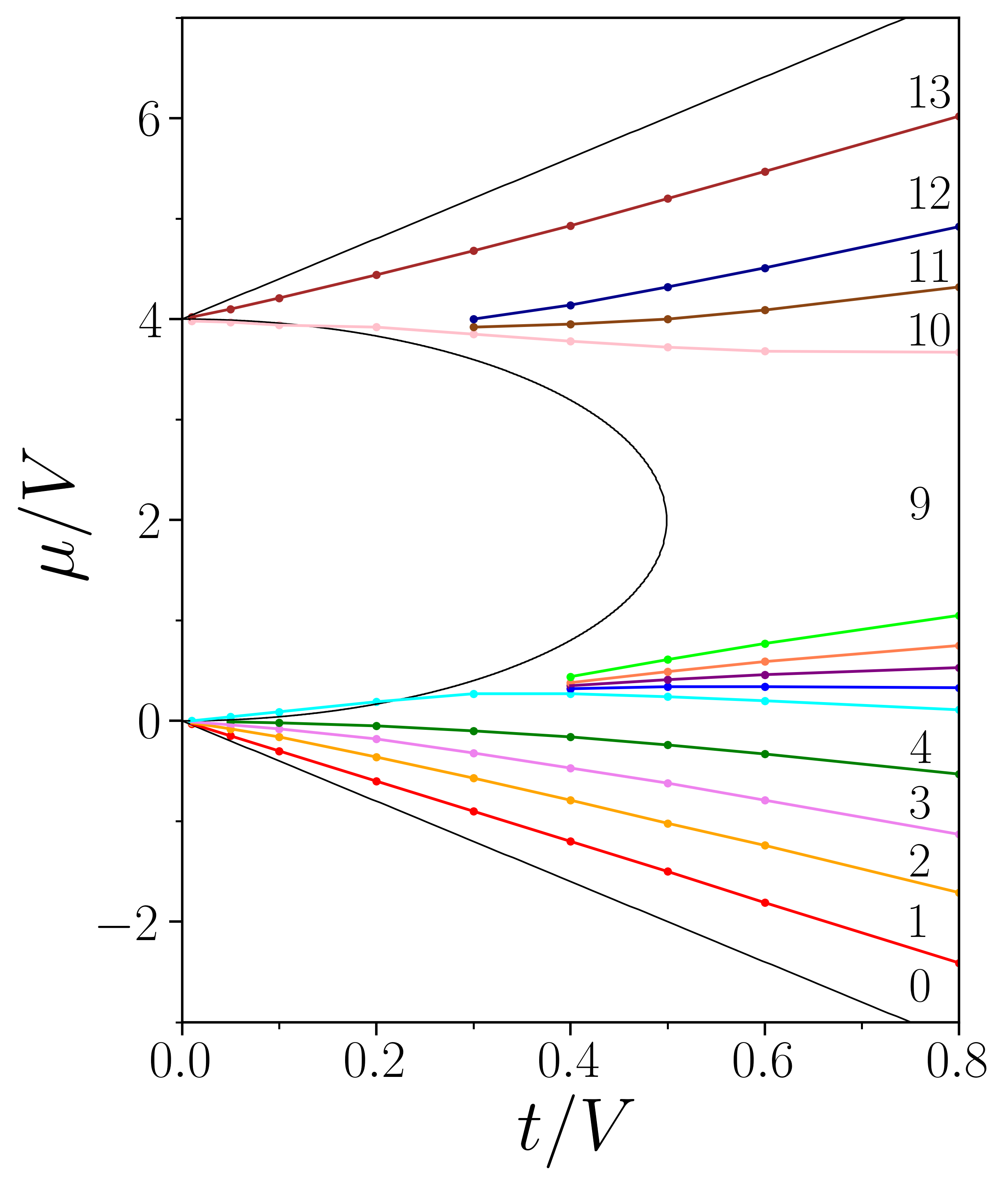}
    \end{minipage}    
    \begin{minipage}[t]{0.24\linewidth} 
        \centering
        (c)\par\medskip
        \includegraphics[width=1.0\linewidth]{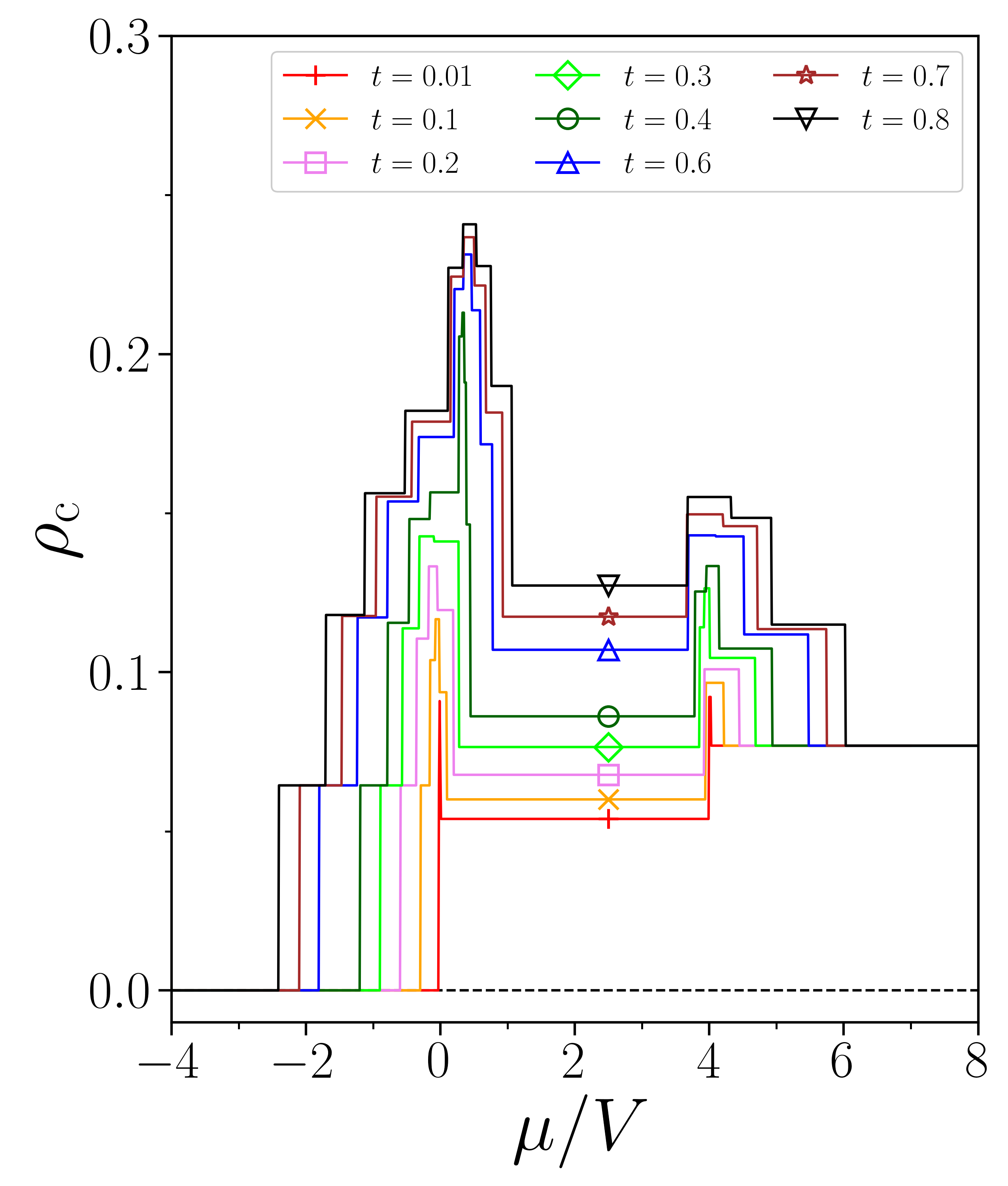}
    \end{minipage}
    \begin{minipage}[t]{0.24\linewidth}
        \centering
        (d)\par\medskip
        \includegraphics[width=1.0\linewidth]{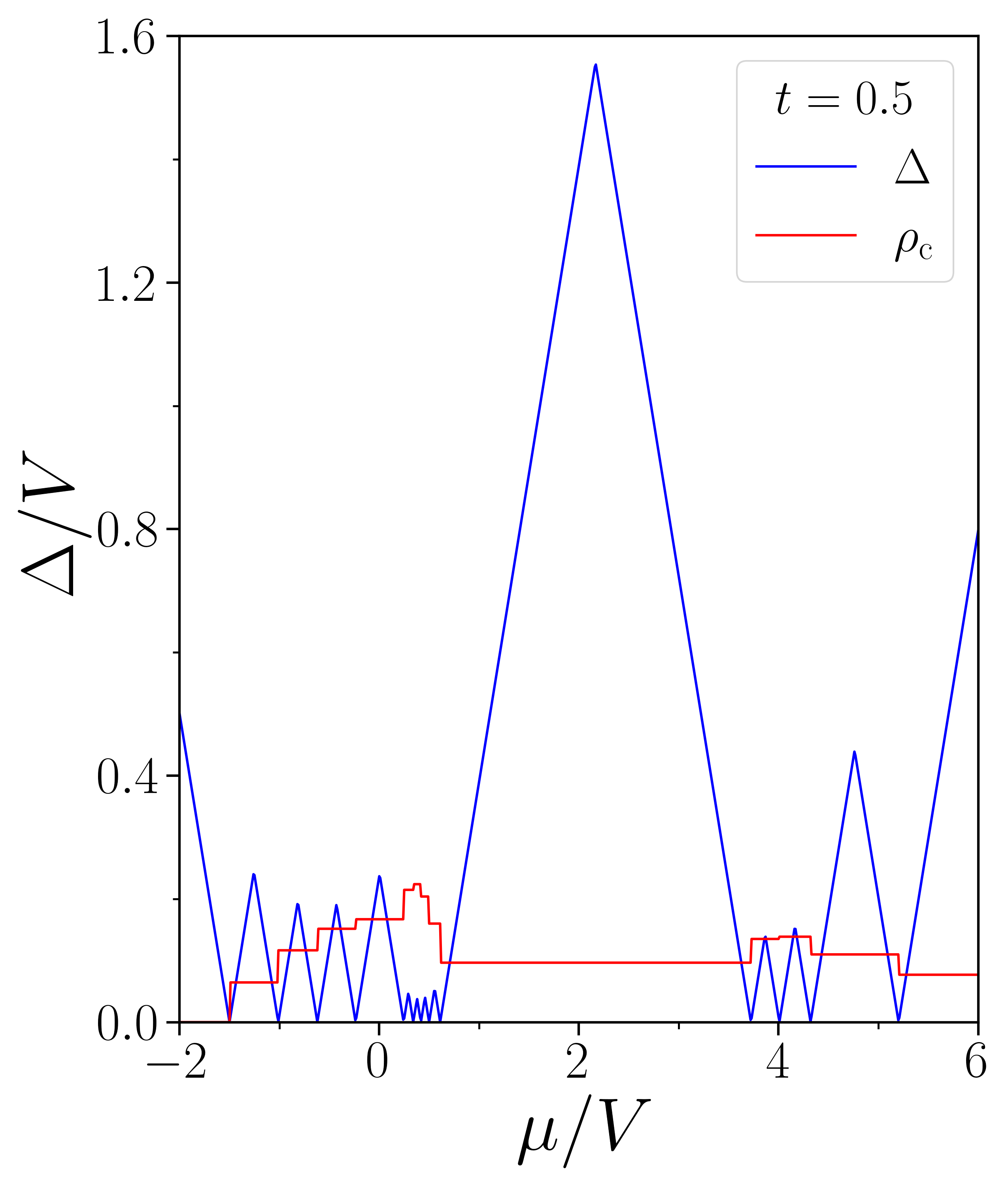}
    \end{minipage}
\caption{\label{fig5}
Diamond.
(a) Ground state of the system for $t=0$ and $0<\mu<4V$. (b) Exact phase diagram at $T=0$. The broken lines through the points are boundary lines of $N$-sectors. The black thin lines are mean-field phase boundaries~\cite{gheeraert2015}. (c) Condensate fraction $\rho_{\rm c}(\mu)$ for various $t$. As $t$ grows, the condensate fraction grows as well. (d) Energy gap $\Delta(\mu)$ for $t=0.5$ (blue line). Also shown is the condensate fraction for the same $t$ (red line).}
\end{figure*}

We obtain Diamond from Square by also including the third neighbors of the central site (see Fig.~\ref{fig1}b).
With four sites more, the phase diagram of the hard-core extended BH model results in better agreement with the phase diagram on the square lattice:
At $t=0$ the only non-trivial ground state has the checkerboard structure illustrated in Fig.~\ref{fig5}a.
The $T=0$ phase diagram, reported in Fig.~\ref{fig5}b, has the same topology of the mean-field phase diagram~\cite{iskin2011,gheeraert2015}, in turn pretty similar to the actual one.
Actually, the stability of the insulating ($N=9$) ``phase'' on Diamond is greatly enhanced relative to the square lattice, where the superfluid phase takes over the insulating phase at $t\approx 0.5V$.
Also for Square the stability of the insulating-like phase for $N=5$ overcomes $t=0.5$, but this effect is more pronounced for Diamond.
This is confirmed by the condensate fraction as a function of $\mu$ (Fig.~\ref{fig5}c), where the hollow centered at $\mu=2V$ is still clearly visible for $t=1$.
Also the energy gap is particularly large in the middle of the $N=9$ region, even at $t$ as large as 0.5 (Fig.~\ref{fig5}d).

\begin{figure*}[t]
\centering   
    \begin{minipage}[t]{0.24\linewidth} 
        \centering
        (a)\par\medskip
        \includegraphics[width=0.9\linewidth]{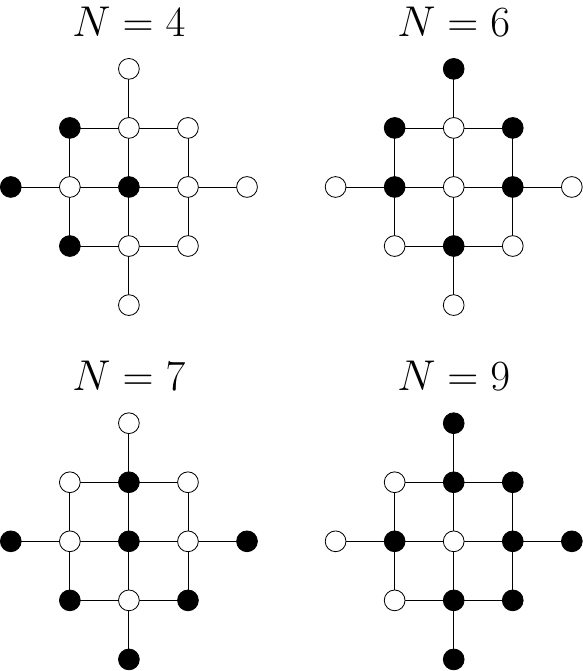}
    \end{minipage}
    \begin{minipage}[t]{0.24\linewidth} 
        \centering
        (b)\par\medskip
        \includegraphics[width=1.0\linewidth]{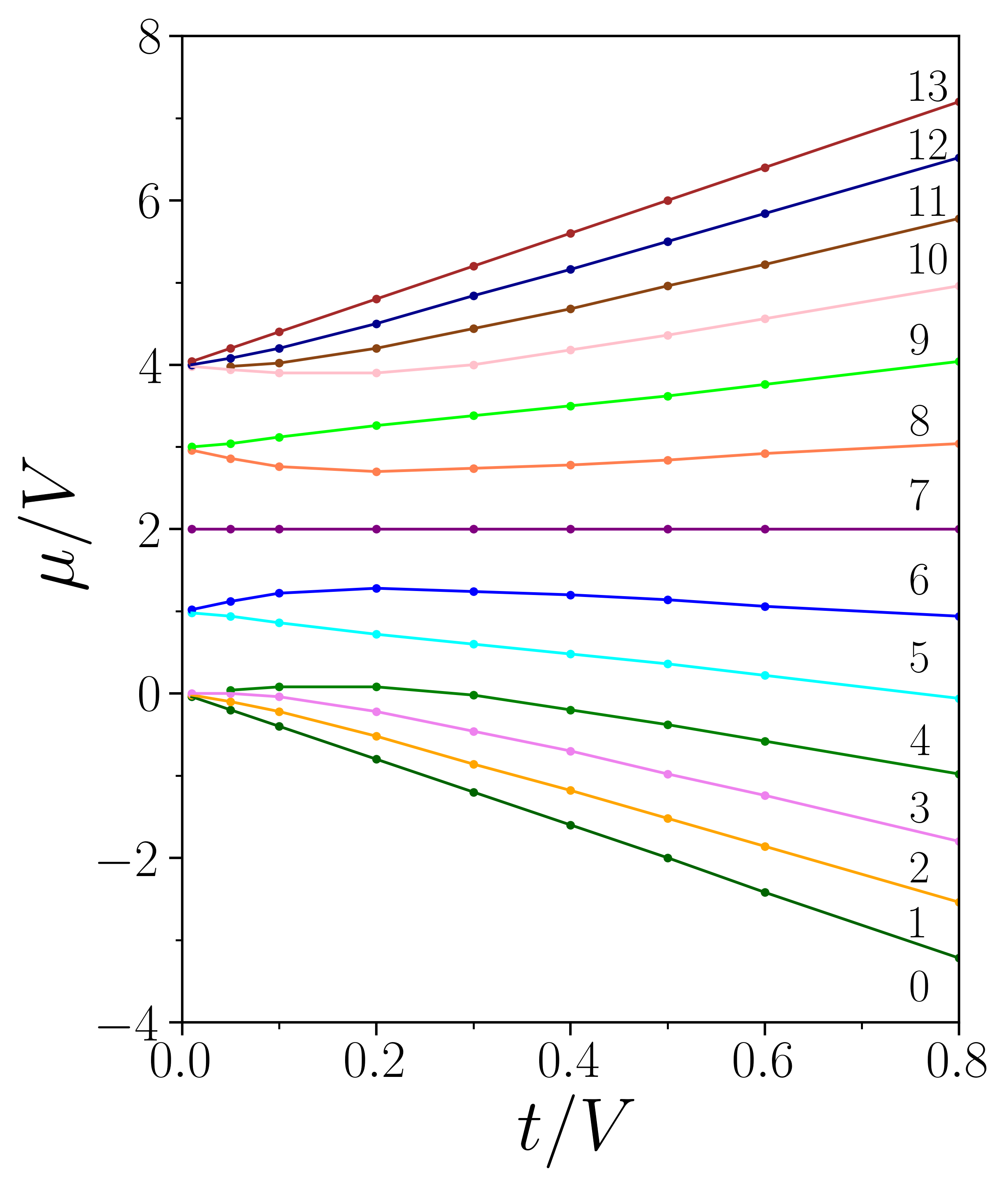}
    \end{minipage}
%        \vspace{0.5cm} 
    \begin{minipage}[t]{0.24\linewidth} 
        \centering
        (c)\par\medskip
        \includegraphics[width=1.0\linewidth]{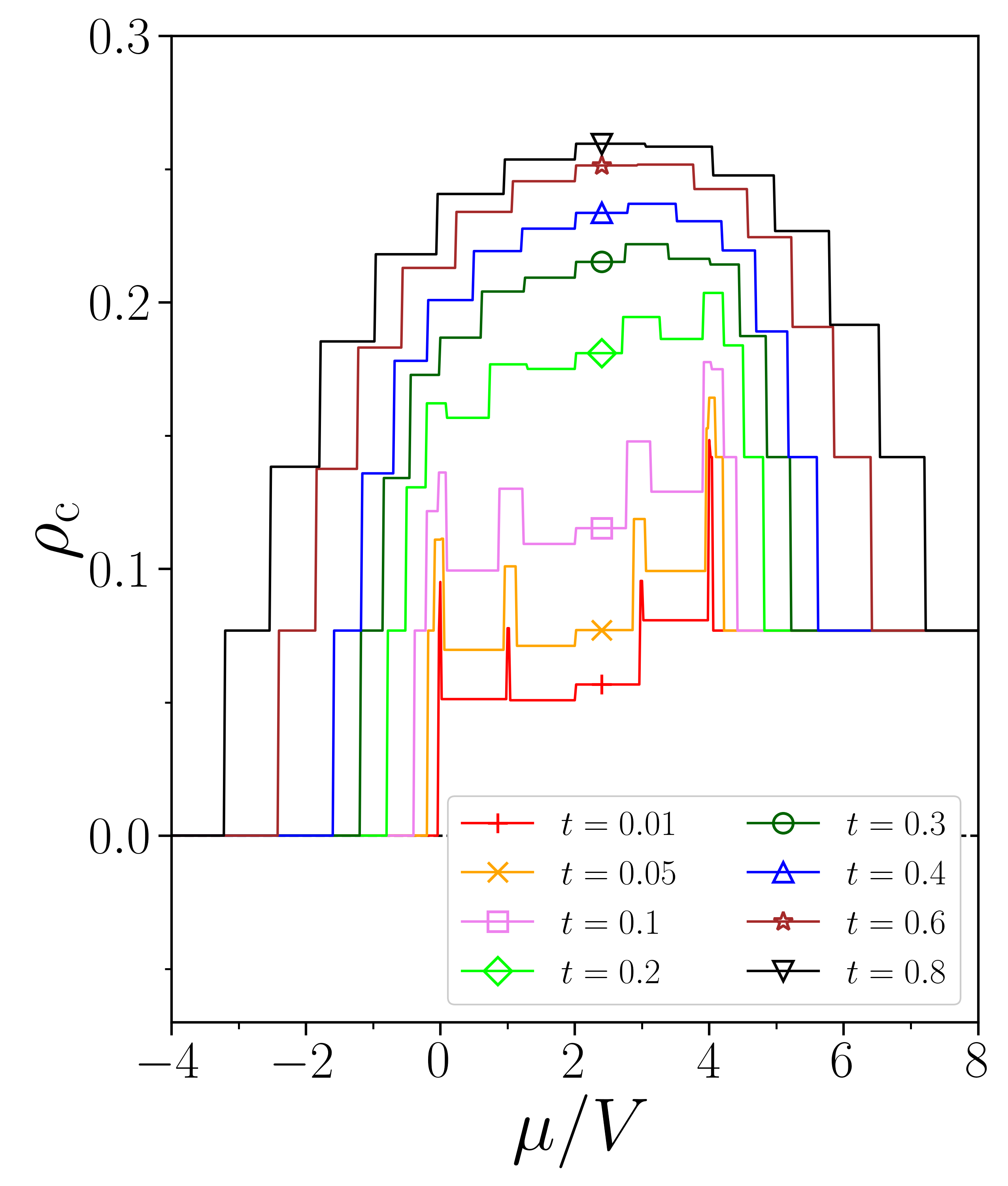}
    \end{minipage}
    \begin{minipage}[t]{0.24\linewidth}
        \centering
        (d)\par\medskip
        \includegraphics[width=1.0\linewidth]{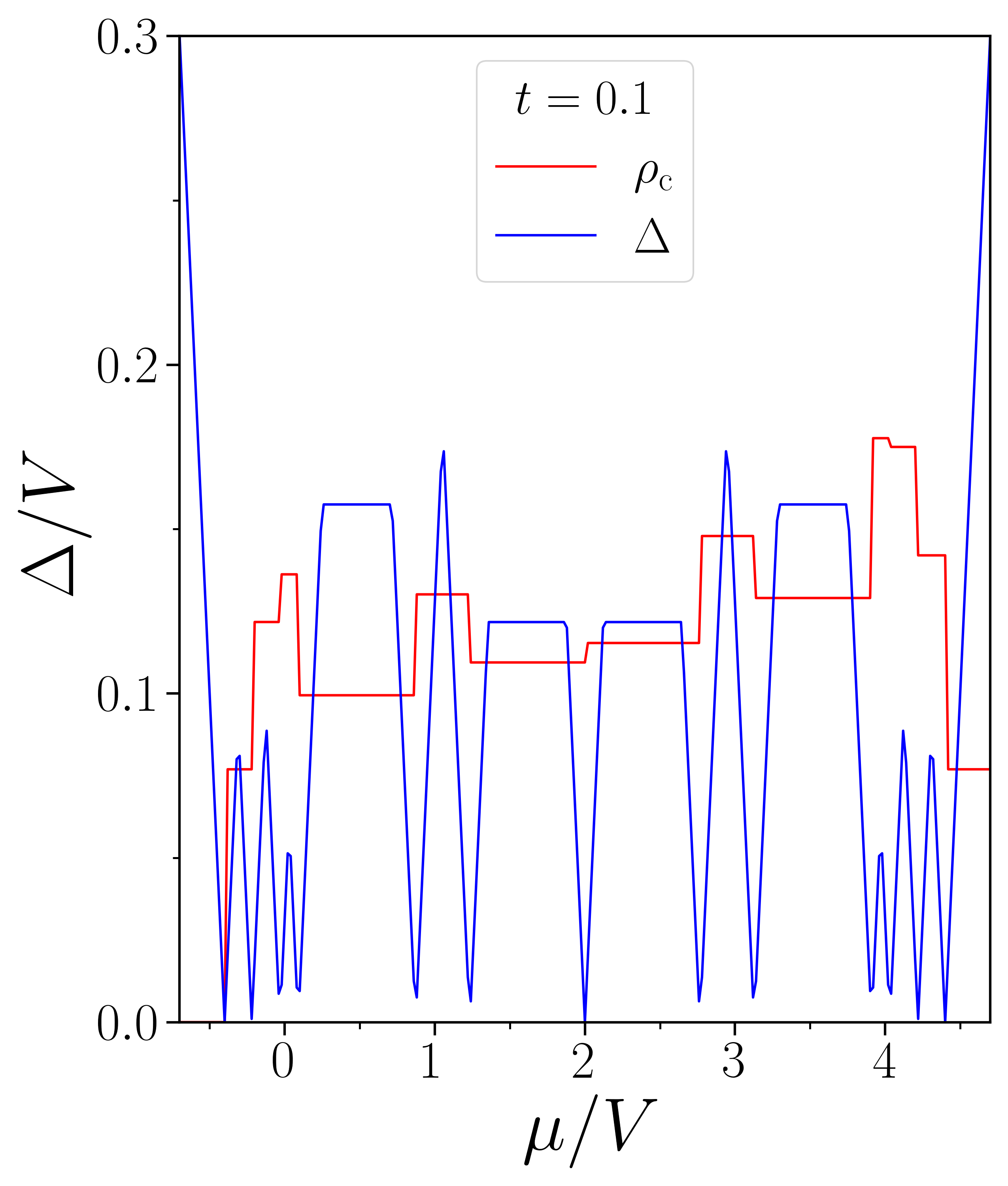}
    \end{minipage}
\caption{\label{fig6}
Diamond with PBC. (a) Ground states of the system for $t=0$ and $0<\mu<4V$. (b) Exact phase diagram at $T=0$. The broken lines through the points are boundary lines of $N$-sectors. (c) Condensate fraction $\rho_{\rm c}(\mu)$ for various $t$. As $t$ grows, the condensate fraction grows as well. (d) Energy gap $\Delta(\mu)$ for $t=0.1$ (blue line). Also shown is the condensate fraction for the same $t$ (red line).} 
\end{figure*}

When Diamond is replicated periodically (which can be done in either of two ways, see Fig.~\ref{figA2}) things change considerably.
At $t=0$ checkerboard order is ruled out and, in its place, the non-trivial ground states are four --- see Fig.~\ref{fig6}a, with two of them being negative images of the other two.
What is unusual in this case is the high degree of degeneracy of these ``phases'', namely 13 and 26 for the $N=4$ and the $N=6$ phase, respectively.
This multiplicity just reflects the freedom of translating (and, for $N=6$, also rotating) the filled sites rigidly on the grid without affecting the energy.
As already commented for Square with PBC, the boundary lines of $N$-sectors are distributed rather uniformly in the $0<N<13$ region, suggesting superfluid-like behavior for all not too small $t$.

\subsection{Star}

\begin{figure*}[t]
\centering
    \begin{minipage}[t]{0.24\linewidth} 
        \centering
        (a)\par\medskip
        \includegraphics[width=0.9\linewidth]{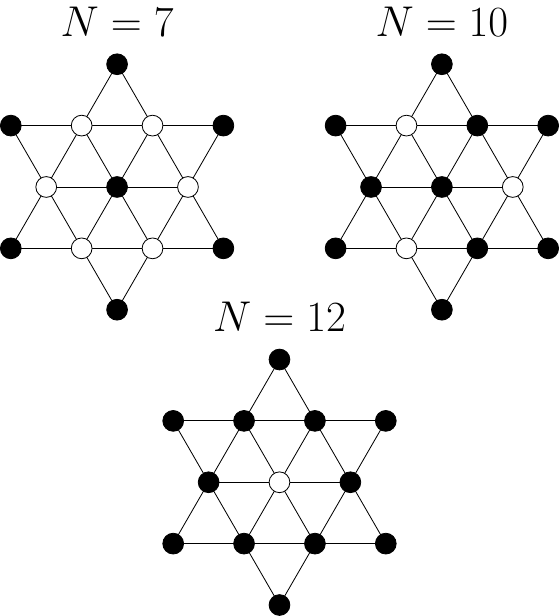}
    \end{minipage}
    \begin{minipage}[t]{0.24\linewidth} 
        \centering
        (b)\par\medskip
        \includegraphics[width=1.0\linewidth]{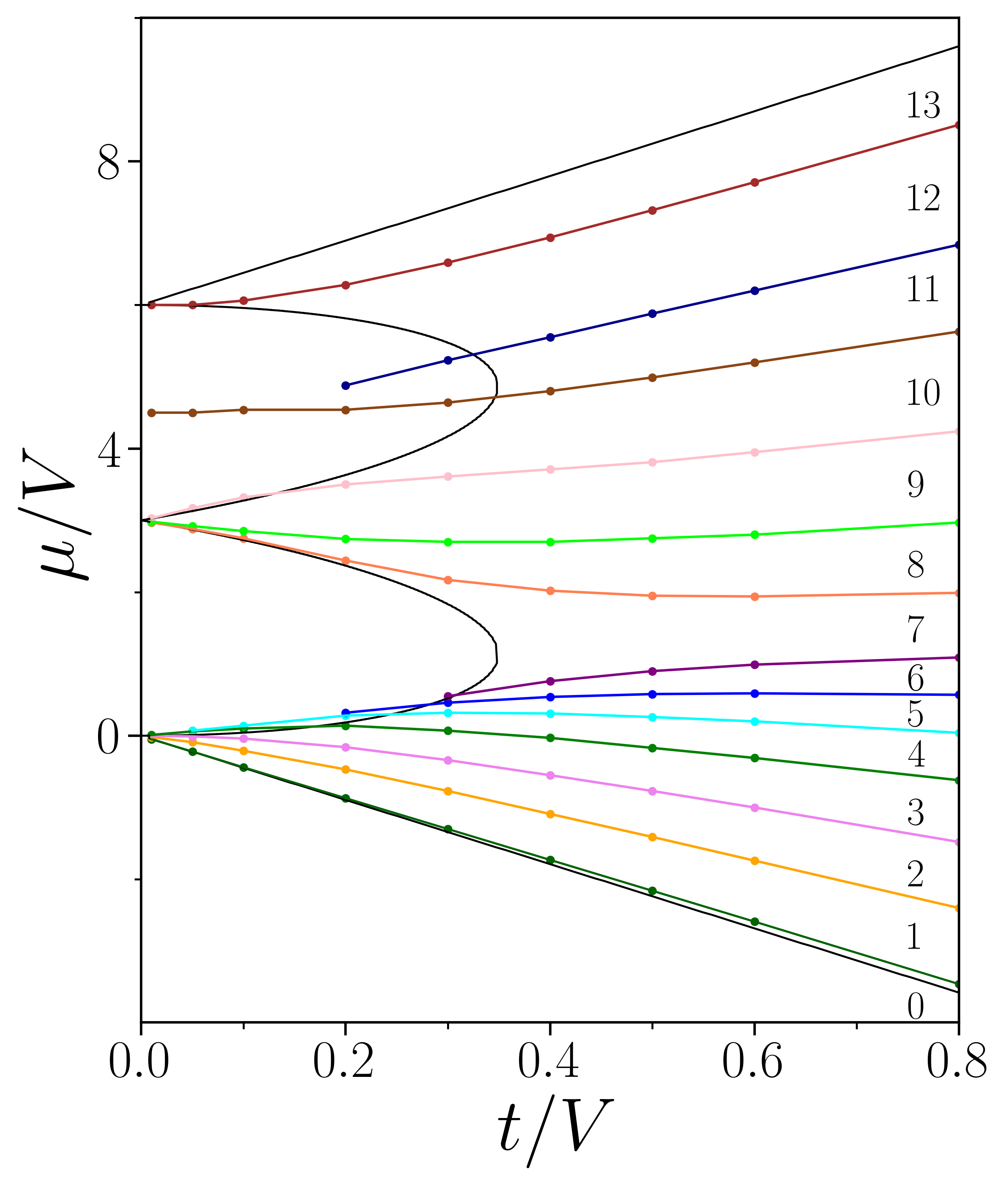}
    \end{minipage}
%    \vspace{0.5cm} 
    \begin{minipage}[t]{0.24\linewidth} 
        \centering
        (c)\par\medskip
        \includegraphics[width=1.0\linewidth]{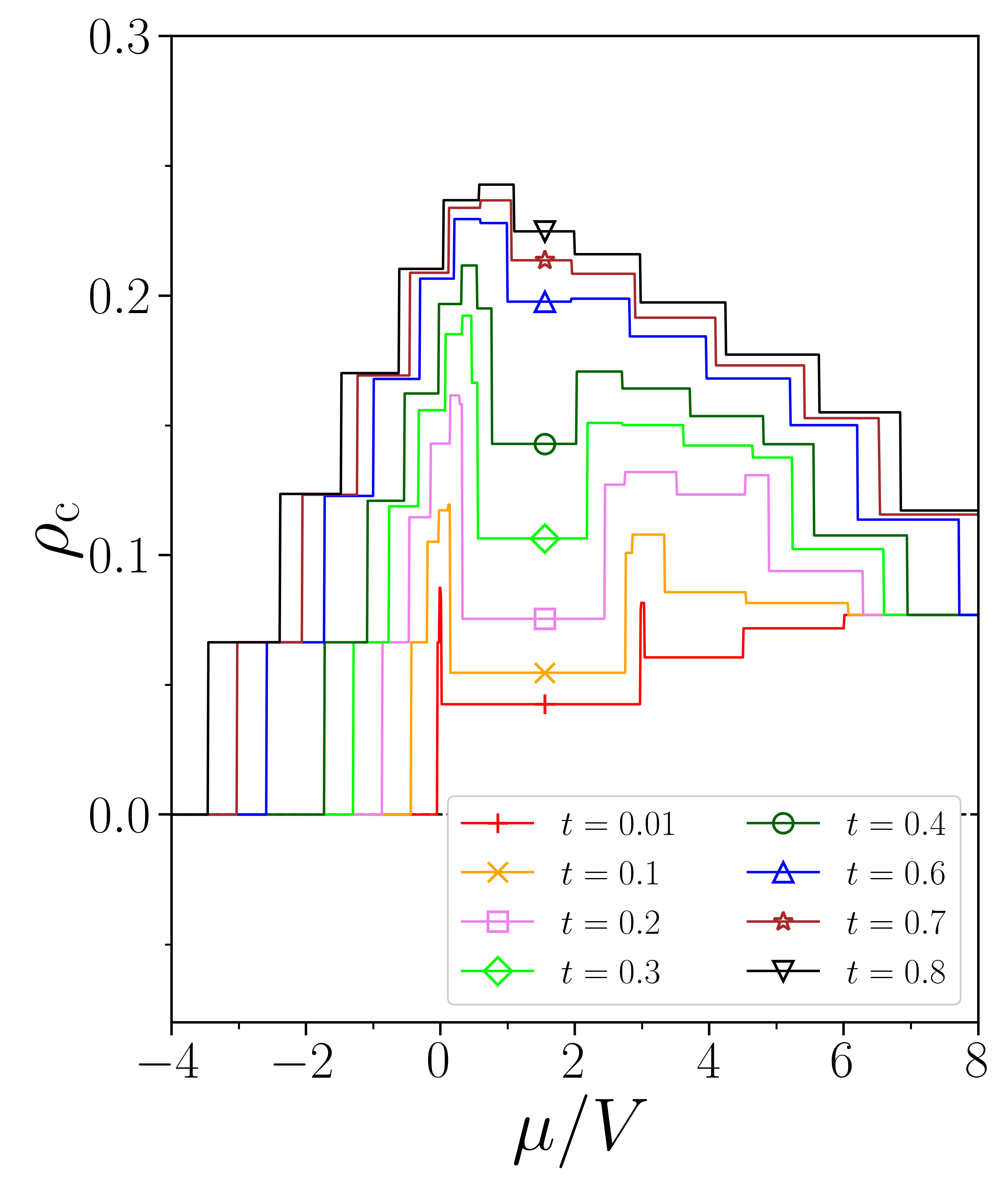}
    \end{minipage}
    \begin{minipage}[t]{0.24\linewidth}
        \centering
        (d)\par\medskip
        \includegraphics[width=1.0\linewidth]{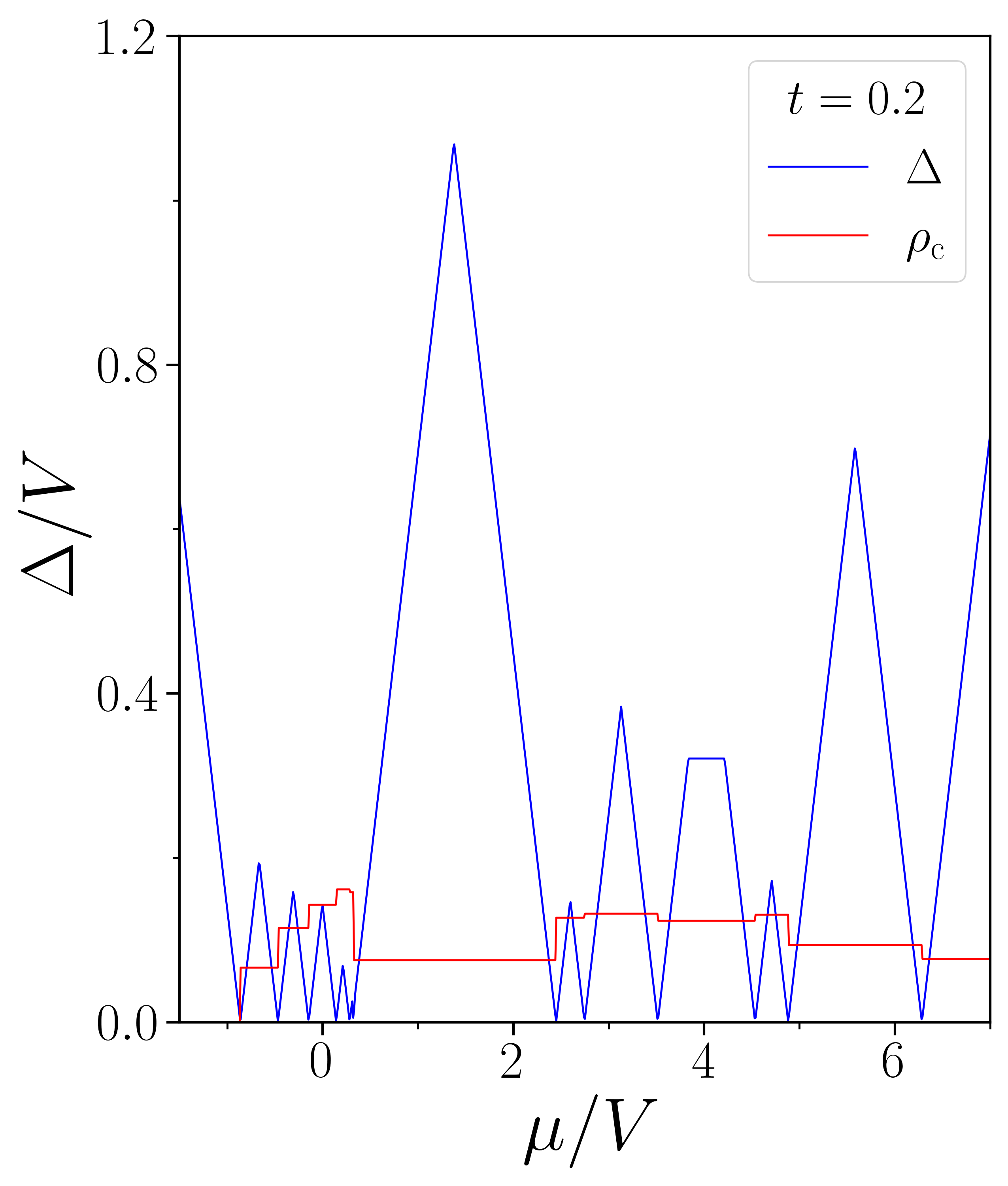}
    \end{minipage}
\caption{\label{fig7}
Star. (a) Ground states of the system for $t=0$ and $0<\mu<6V$ (see text). (b) Exact phase diagram at $T=0$. The broken lines through the points are boundary lines of $N$-sectors. The black thin lines are the transition boundaries according to mean-field theory (see Appendix B). (c) Condensate fraction $\rho_{\rm c}(\mu)$ for various $t$. As $t$ grows, the condensate fraction grows as well. (d) Energy gap $\Delta(\mu)$ for $t=0.2$ (blue line). Also shown is the condensate fraction for the same $t$ (red line).} 
\end{figure*}

The Star grid in Fig.~\ref{fig1}c has the same number of sites as Diamond, but the underlying symmetry is now triangular.
We have reported in Fig.~\ref{fig7}a the non-trivial ground states for $t=0$.
While the former two are reminiscent of the $1/3$ and $2/3$ ``solids'' of the triangular-lattice model~\cite{wessel2005,gheeraert2015}, the $N=12$ ``phase'' is totally new, and analogous to the ring of Fig.~\ref{fig2}a.
Moreover, the $N=7$ and $N=12$ ground states are unique, whereas the energy for $N=10$ is twofold degenerate.
For these three phases, the $h$ function reads:
\begin{eqnarray}
N=7: & \qquad h=-7\mu
\nonumber \\
N=10: & \qquad h=9V-10\mu
\nonumber \\
N=12: & \qquad h=18V-12\mu\,.
\label{eq5}
\end{eqnarray}
This implies that the transitions between them occur at $\mu=3V$ and $\mu=(9/2)V$.
The full $T=0$ phase diagram is depicted in Fig.~\ref{fig7}b.
Were it not for the ``spurious'' $N=12$ phase, this phase diagram would match the infinite-size one almost perfectly.
Like for Square and Diamond, also for Star $\rho_{\rm c}(\mu)$ behaves as expected for a system traversing insulating and superfluid regions (Fig.~\ref{fig7}c).
The same applies for the energy gap, which at $t=0.2$ takes larger values in the middle of ``insulating'' regions (Fig.~\ref{fig7}d).

\begin{figure*}[t]
\centering
    \begin{minipage}[t]{0.24\linewidth}
        \centering
        (a)\par\medskip
        \includegraphics[width=0.52\linewidth]{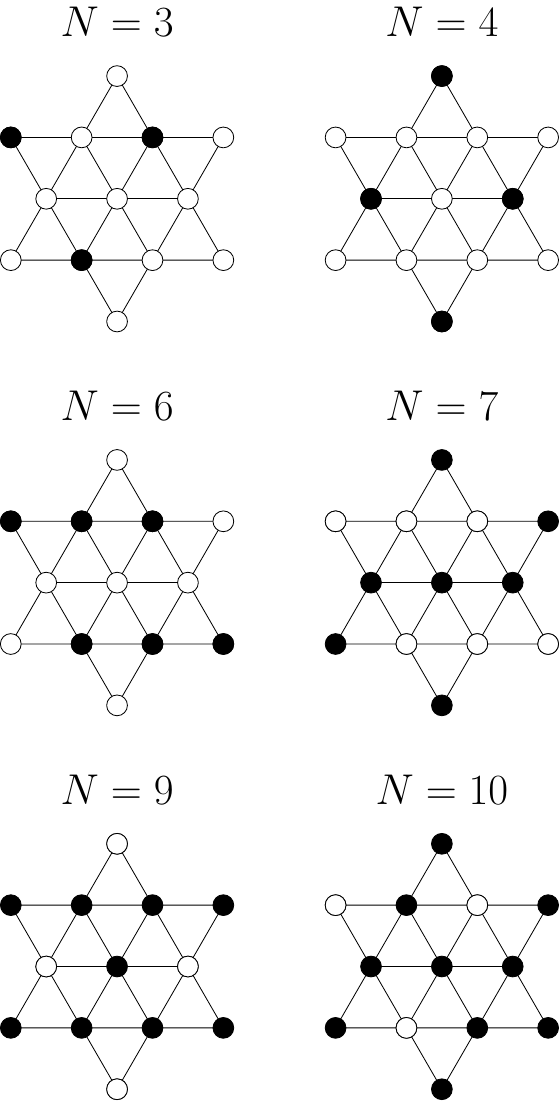}
    \end{minipage}
    \begin{minipage}[t]{0.24\linewidth} 
        \centering
        (b)\par\medskip
        \includegraphics[width=1.0\linewidth]{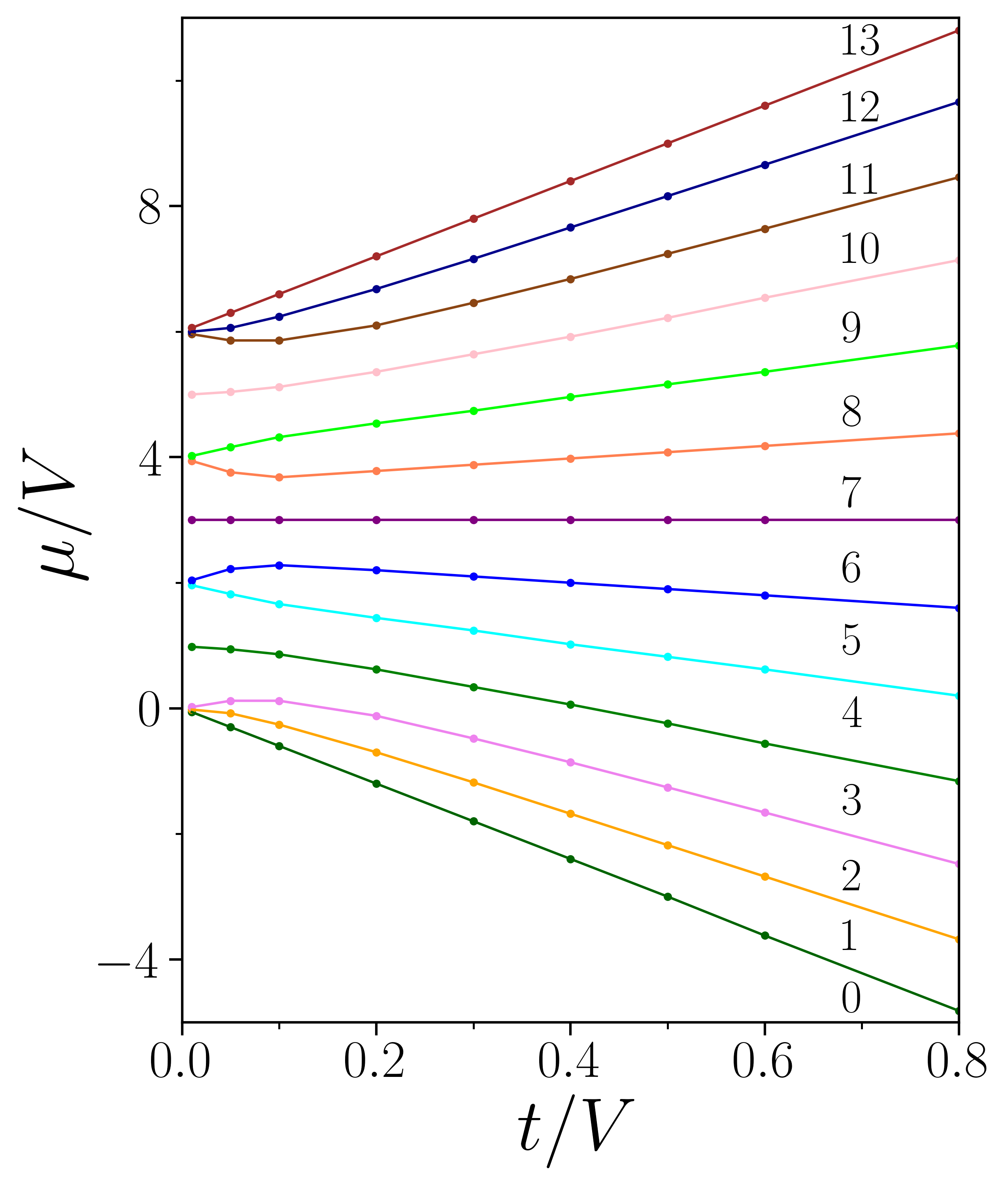}
    \end{minipage}
%    \vspace{0.5cm} 
    \begin{minipage}[t]{0.24\linewidth} 
        \centering
        (c)\par\medskip
        \includegraphics[width=1.0\linewidth]{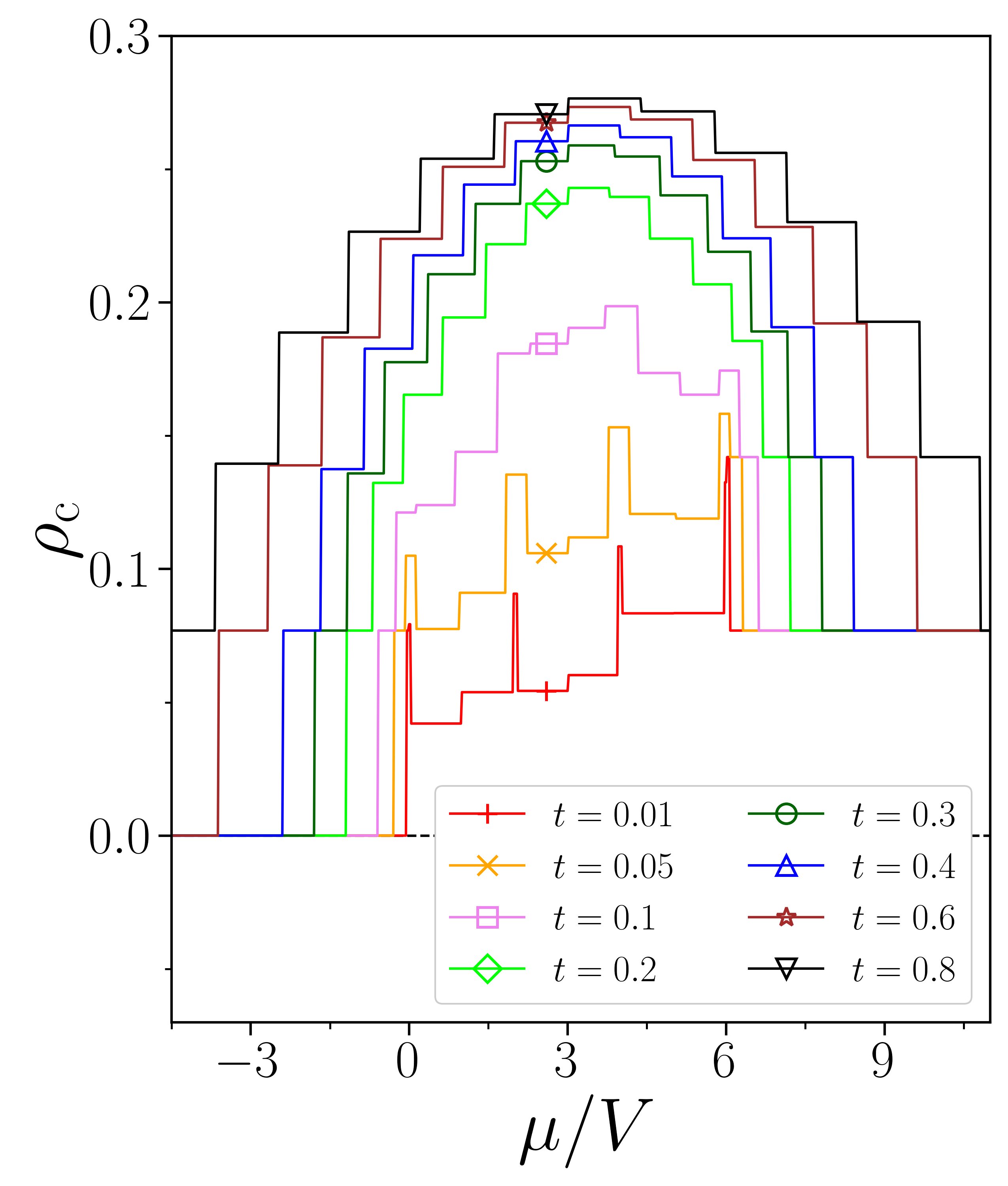}
    \end{minipage}
    \begin{minipage}[t]{0.24\linewidth}
        \centering
        (d)\par\medskip
        \includegraphics[width=1.0\linewidth]{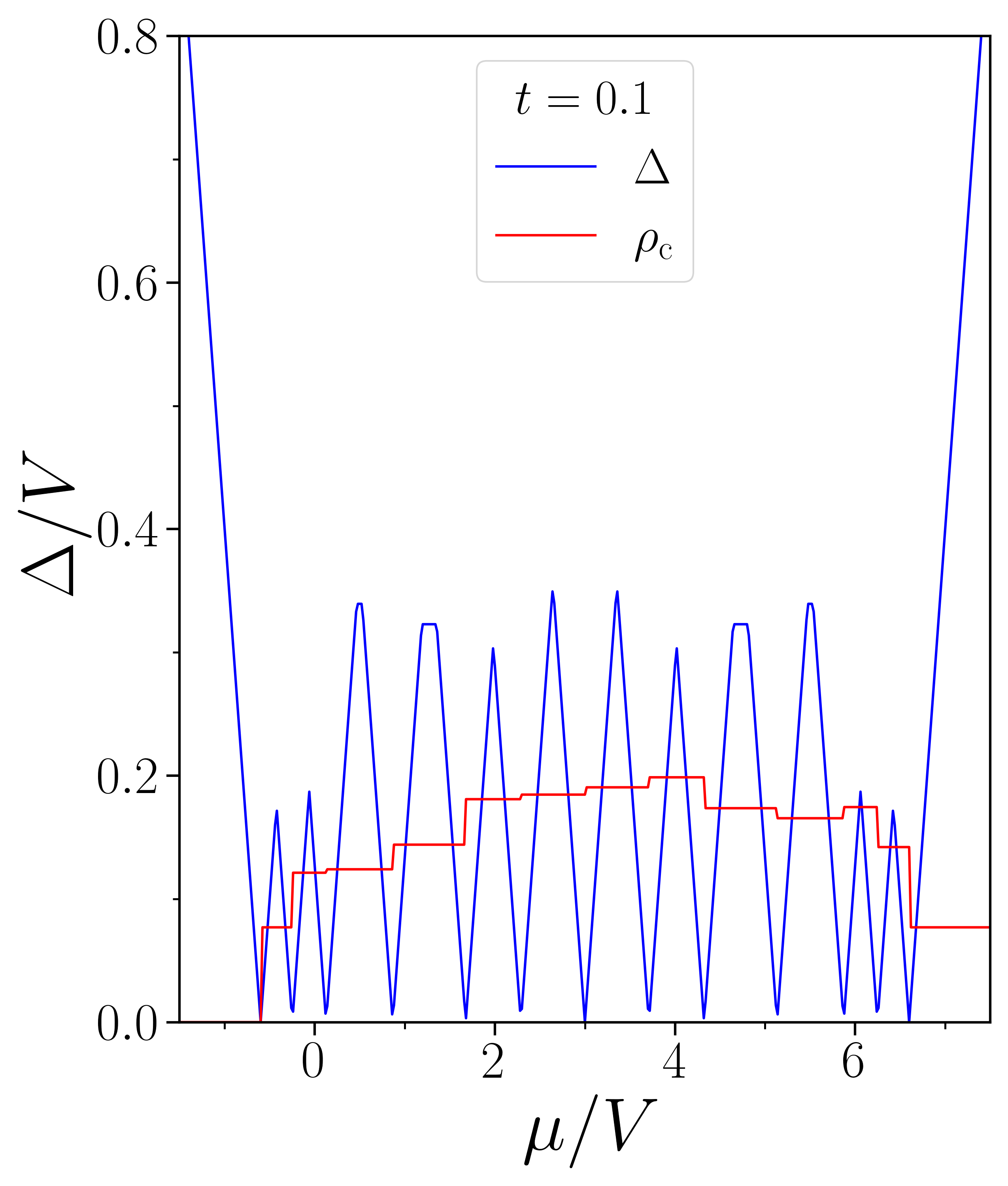}
    \end{minipage}
\caption{\label{fig8}
Star with PBC. (a) Ground states of the system for $t=0$ and $0<\mu<6V$. (b) Exact phase diagram at $T=0$. The broken lines through the points are boundary lines of $N$-sectors. (c) Condensate fraction $\rho_{\rm c}(\mu)$ for various $t$. As $t$ grows, the condensate fraction grows as well. (d) Energy gap $\Delta(\mu)$ for $t=0.1$ (blue line). Also shown is the condensate fraction for the same $t$ (red line).} 
\end{figure*}

Finally moving to Star with PBC (which, again, can be applied in two equivalent ways, see Fig.~\ref{figA3}), we plot in Fig.~\ref{fig8}a the non-trivial ground states for $t=0$.
The degree of degeneracy is 26 for the ground-state energy for $N=3$ and 39 for $N=4$ and $N=6$.
The other three phases are the negative images of the first three.
The phase diagram in Fig.~\ref{fig8}b is similar to the other two cases with PBC:
The ``insulating'' regions are rather limited in extent, and the behavior is thus predominantly superfluid-like.
Similar to the other grids, none of the ground states in Fig.~\ref{fig7}a survives periodic replication of the grid;
this greatly enhances the stability of the superfluid-like phase, while insulating-like behavior is confined to $t\lesssim 0.1$.

\begin{figure*}[t]
\centering
\includegraphics[width=1.0\linewidth]{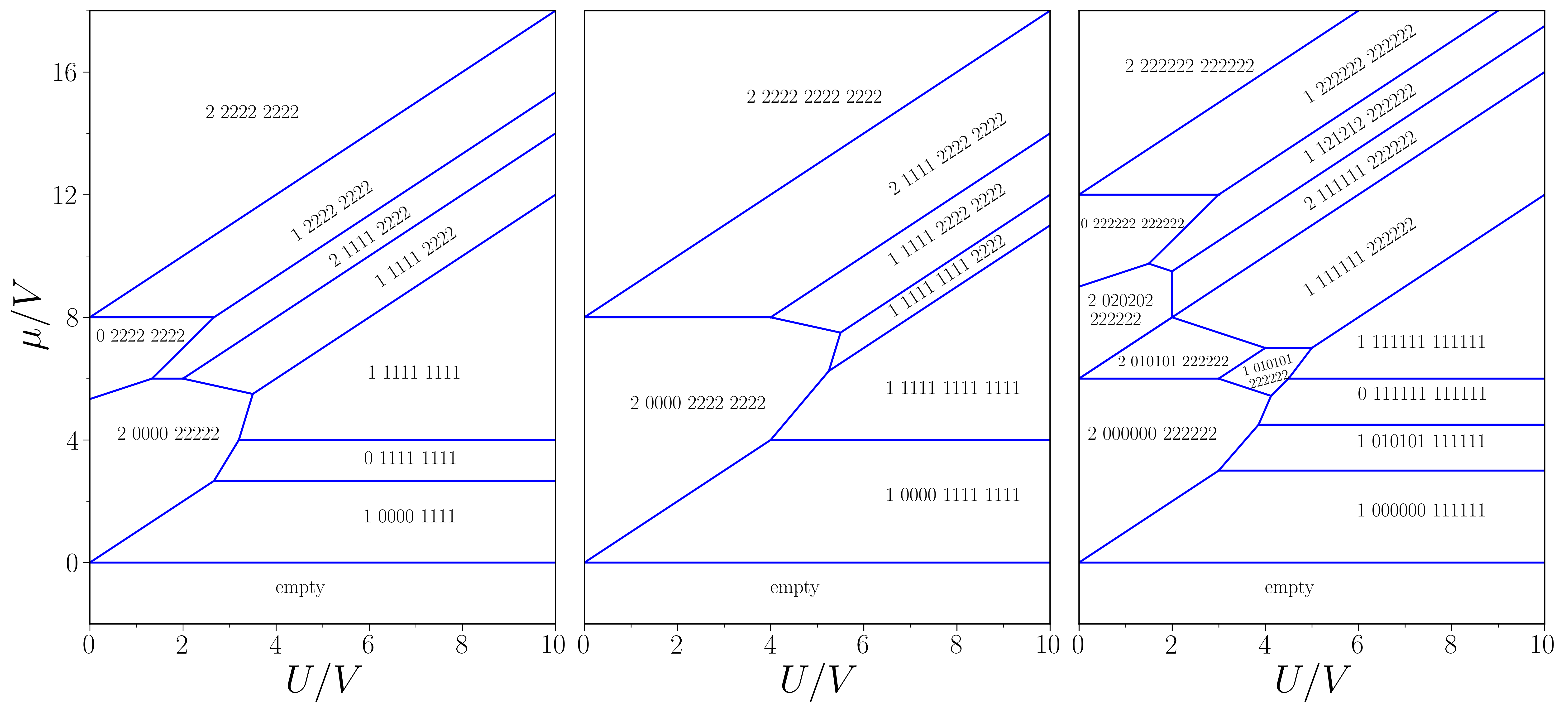}
\caption{\label{fig9}
Exact $t=0$ ground states for finite $U$ and $n_{\rm cut}=2$. Square (left).  Diamond (middle).  Star (right). Each ground state is labeled by a sequence of numbers, representing the occupancies of the central site, first-neighbor sites, second-neighbor sites, and (where relevant) third-neighbor sites.}
\end{figure*}

\vspace{3mm}
In closing this section, we discuss how the zero-temperature phase behavior of the extended BH model on a small grid changes when the $U$ parameter in Eq.~\ref{eq1} is finite.
For simplicity, only the case of $t=0$ {\em and} $n_{\rm cut}=2$ is considered below, which already gives a taste of the much higher complexity of the finite-$U$ case.
We report in Fig.~\ref{fig9} the exact ground states at $t=0$, as functions of $U/V$ and $\mu/V$, for the three grids in Fig.~\ref{fig1} (the hard-core limit is situated on the far right of each panel).
Clearly, the ``phases'' are now many more, with an intricate interplay between several of them in the star case.
Curious how, for very small $U/V$, singly-occupied sites disappear and only double occupancy becomes a possibility.
Moreover, the sequence of low-$U$ phases with increasing $\mu$ is almost identical to that for $U\rightarrow\infty$, but for the replacement of any $x_i=1$ with $x_i=2$.

\section{Quantum simulations}

\begin{figure*}
    \centering
    \begin{minipage}[t]{0.25\linewidth} 
        \centering
        \textbf{$V_0=0$}\par\medskip
        \includegraphics[width=0.99\linewidth]{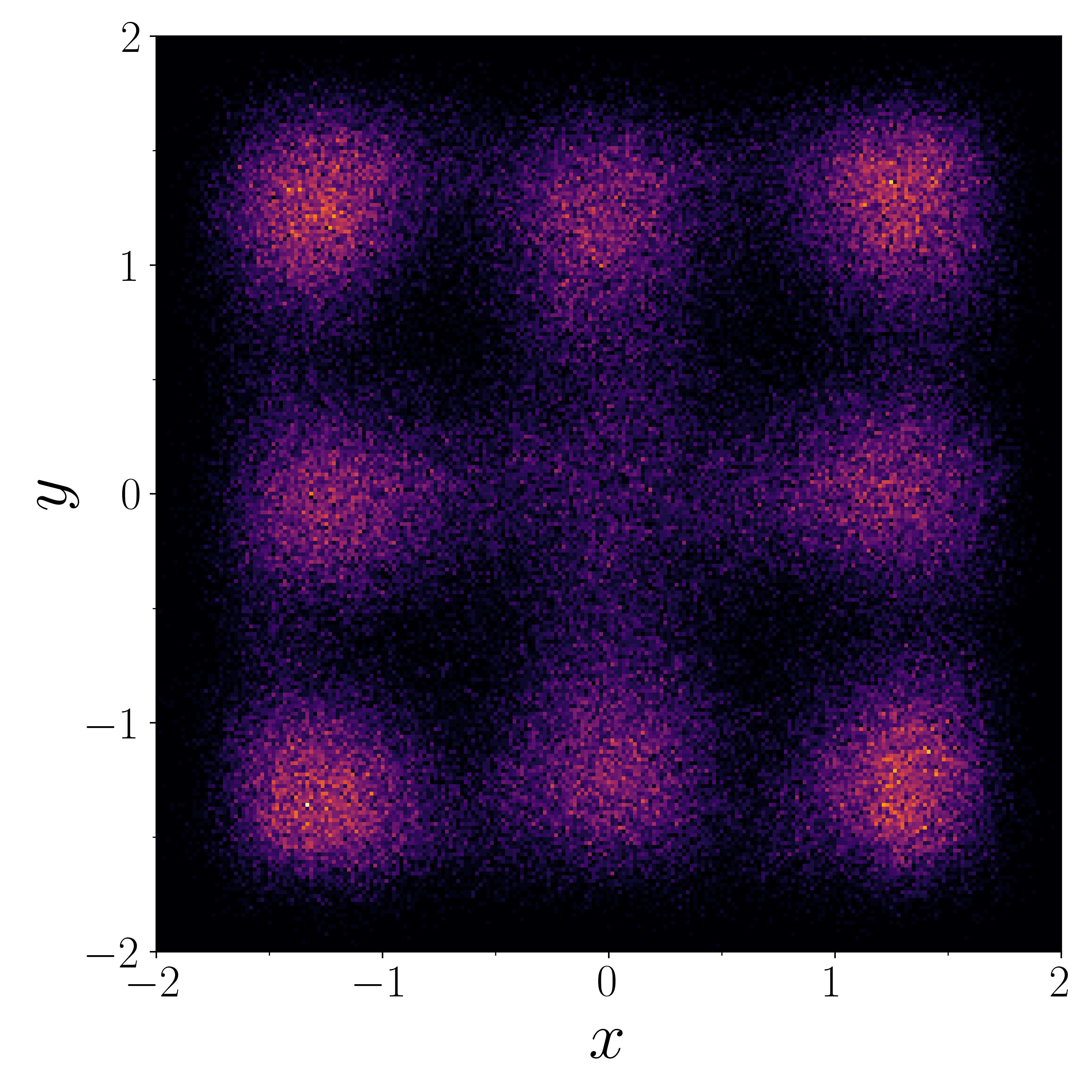}
    \end{minipage}\hfill
    \begin{minipage}[t]{0.25\linewidth} 
        \centering
        \textbf{$V_0=5$}\par\medskip
        \includegraphics[width=0.99\linewidth]{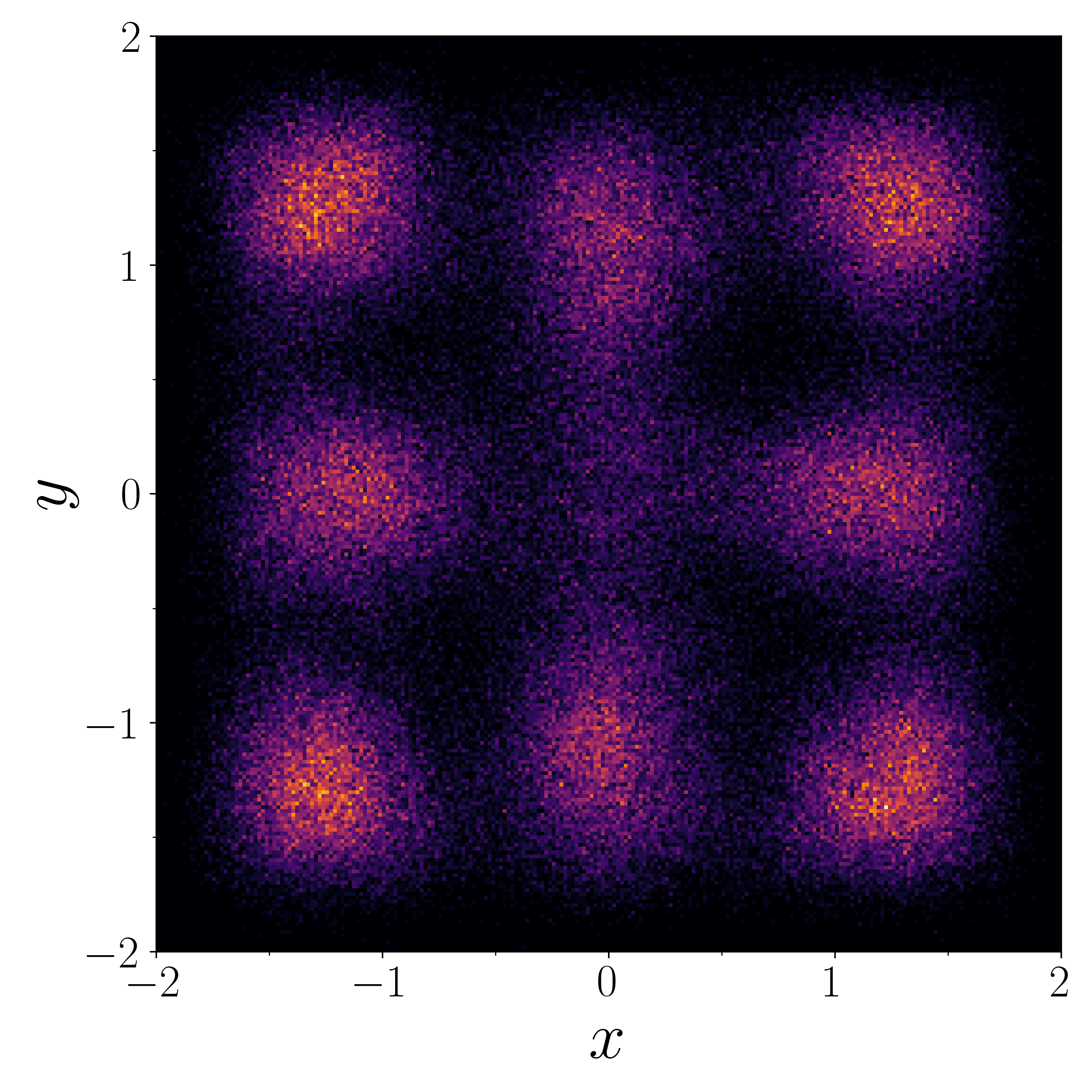}
    \end{minipage}\hfill
        \begin{minipage}[t]{0.25\linewidth}
        \centering
        \textbf{$V_0=10$}\par\medskip
        \includegraphics[width=0.99\linewidth]{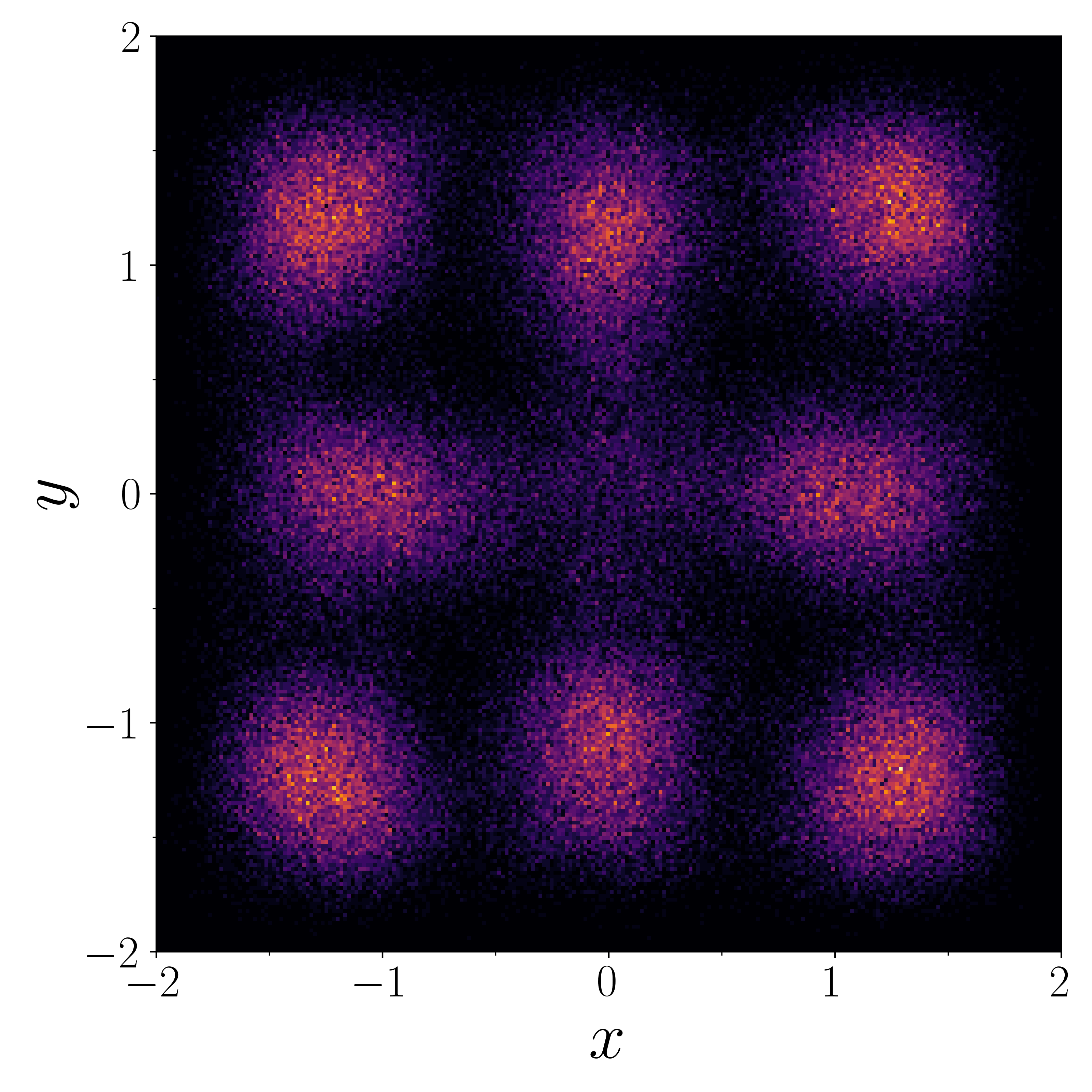}
    \end{minipage}\hfill
        \begin{minipage}[t]{0.25\linewidth} 
        \centering
        \textbf{$V_0=30$}\par\medskip
        \includegraphics[width=0.99\linewidth]{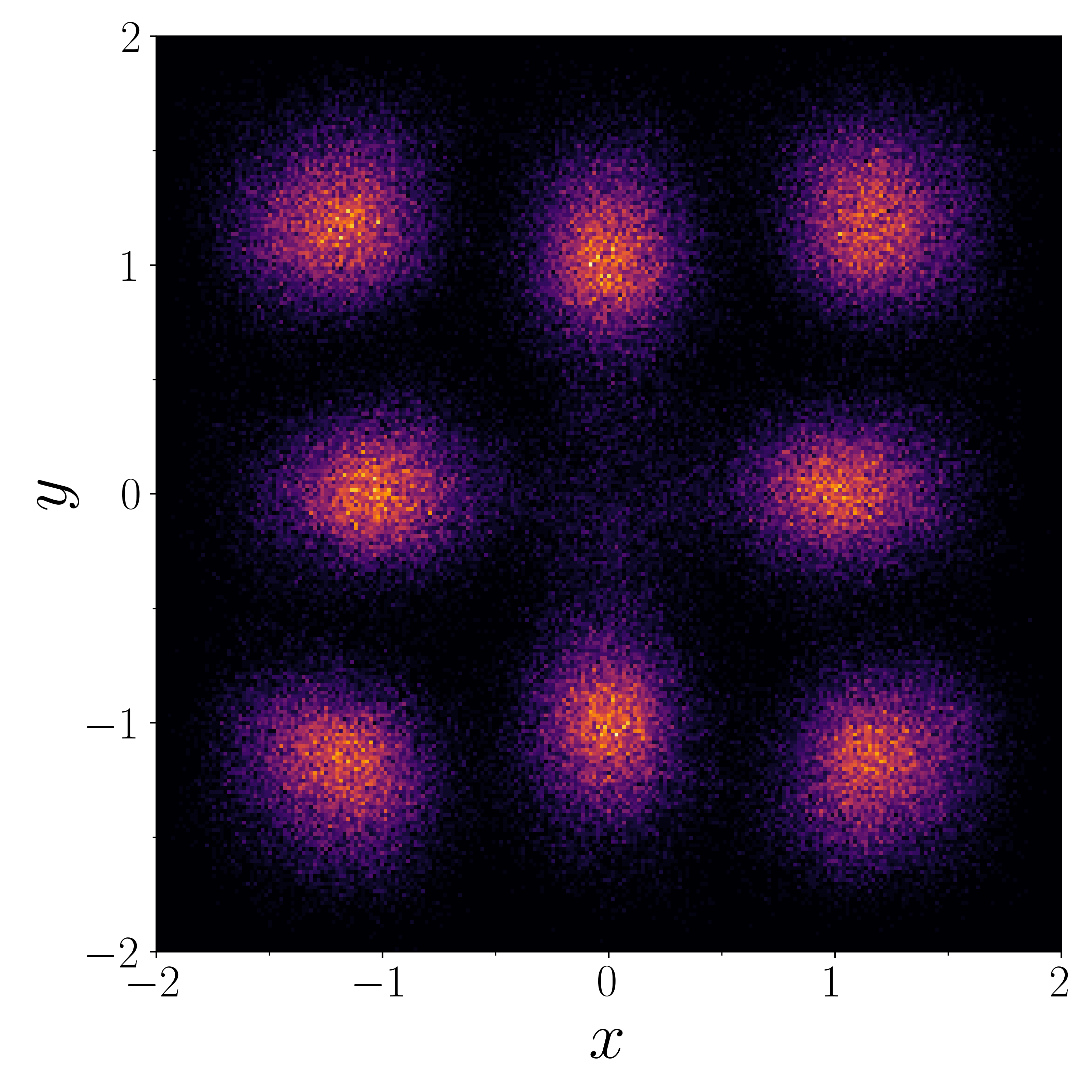}
    \end{minipage}\\
    \begin{minipage}[t]{0.25\linewidth} 
        \centering
        \includegraphics[width=0.99\linewidth]{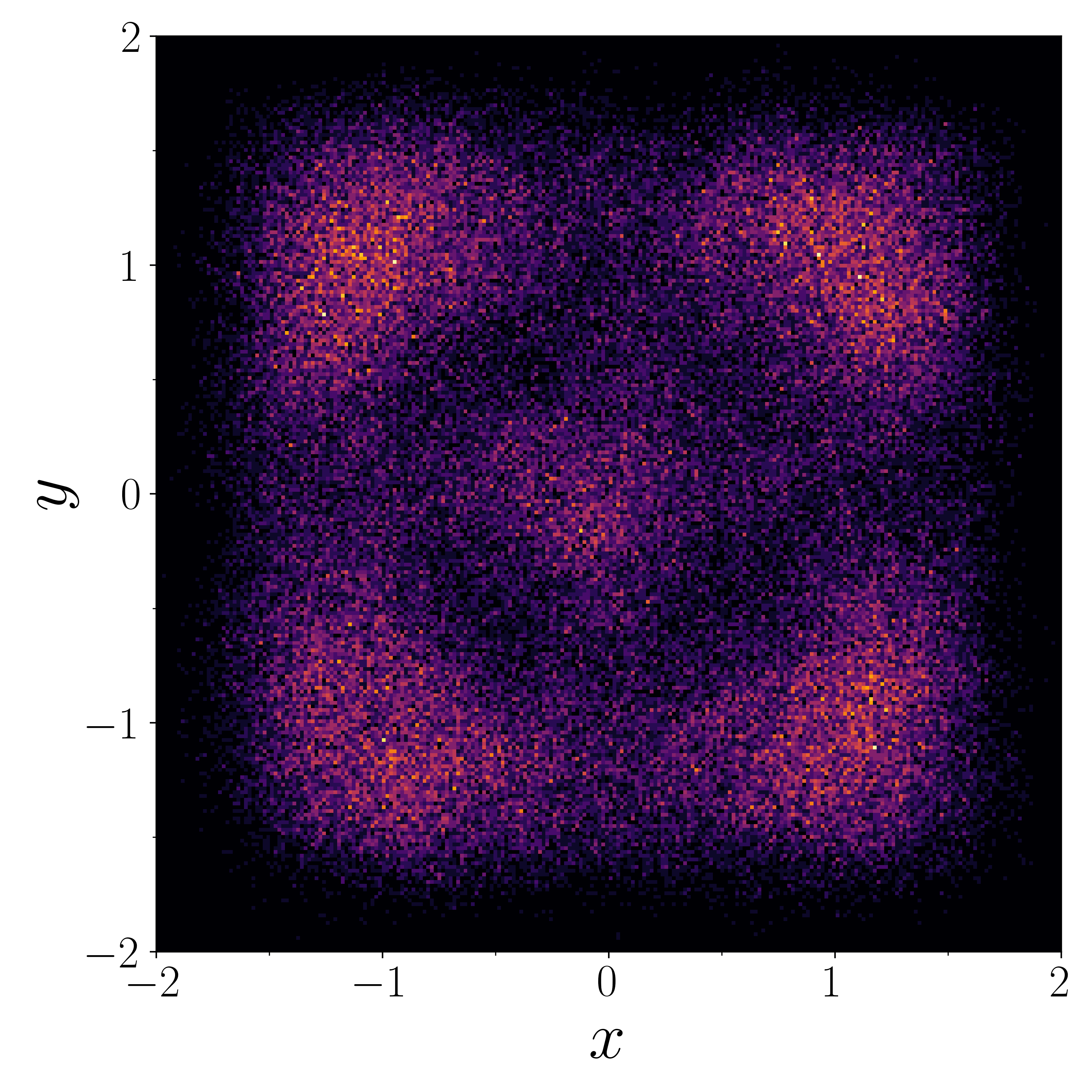}
    \end{minipage}\hfill
    \begin{minipage}[t]{0.25\linewidth} 
        \centering
        \includegraphics[width=0.99\linewidth]{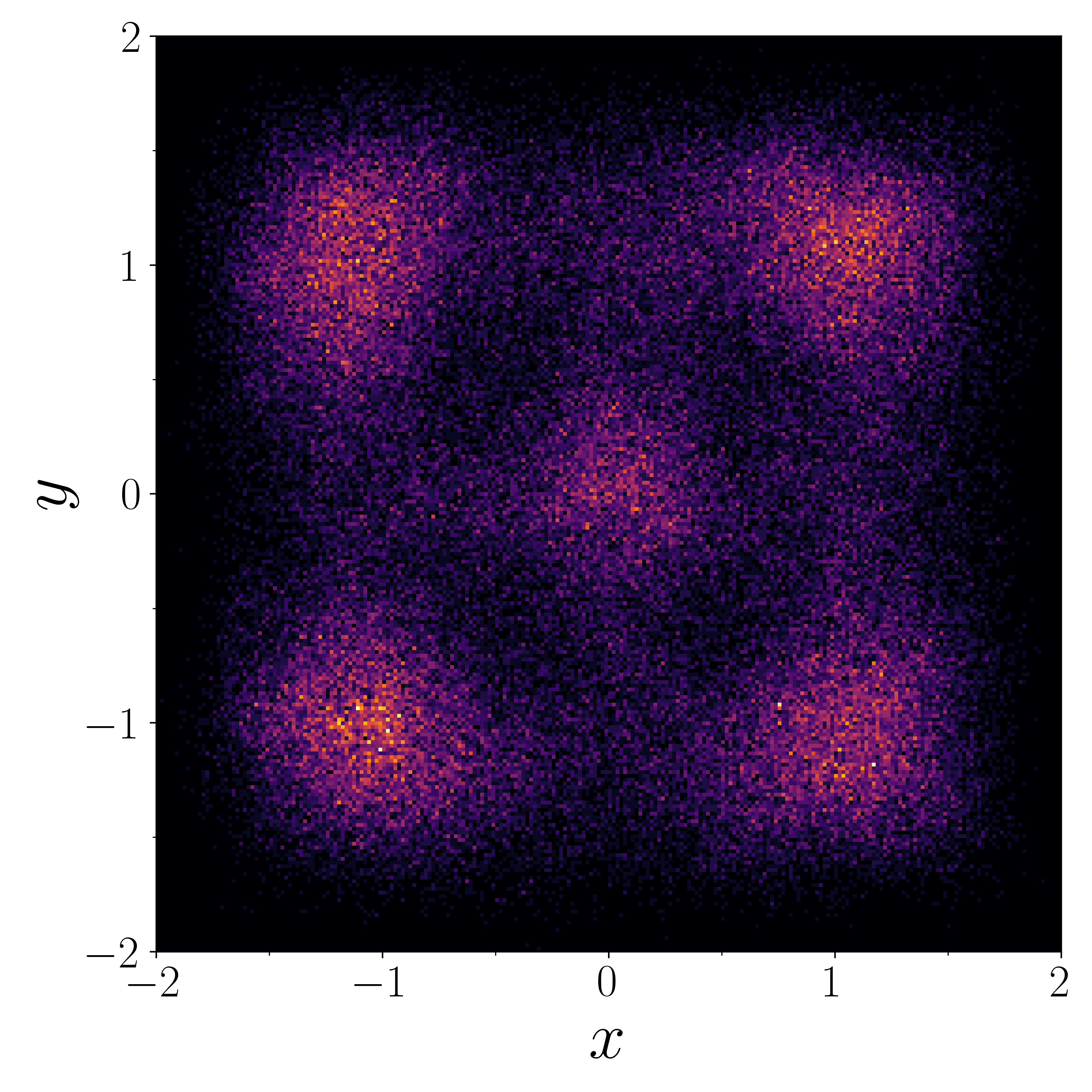}
    \end{minipage}\hfill
        \begin{minipage}[t]{0.25\linewidth}
        \centering
        \includegraphics[width=0.99\linewidth]{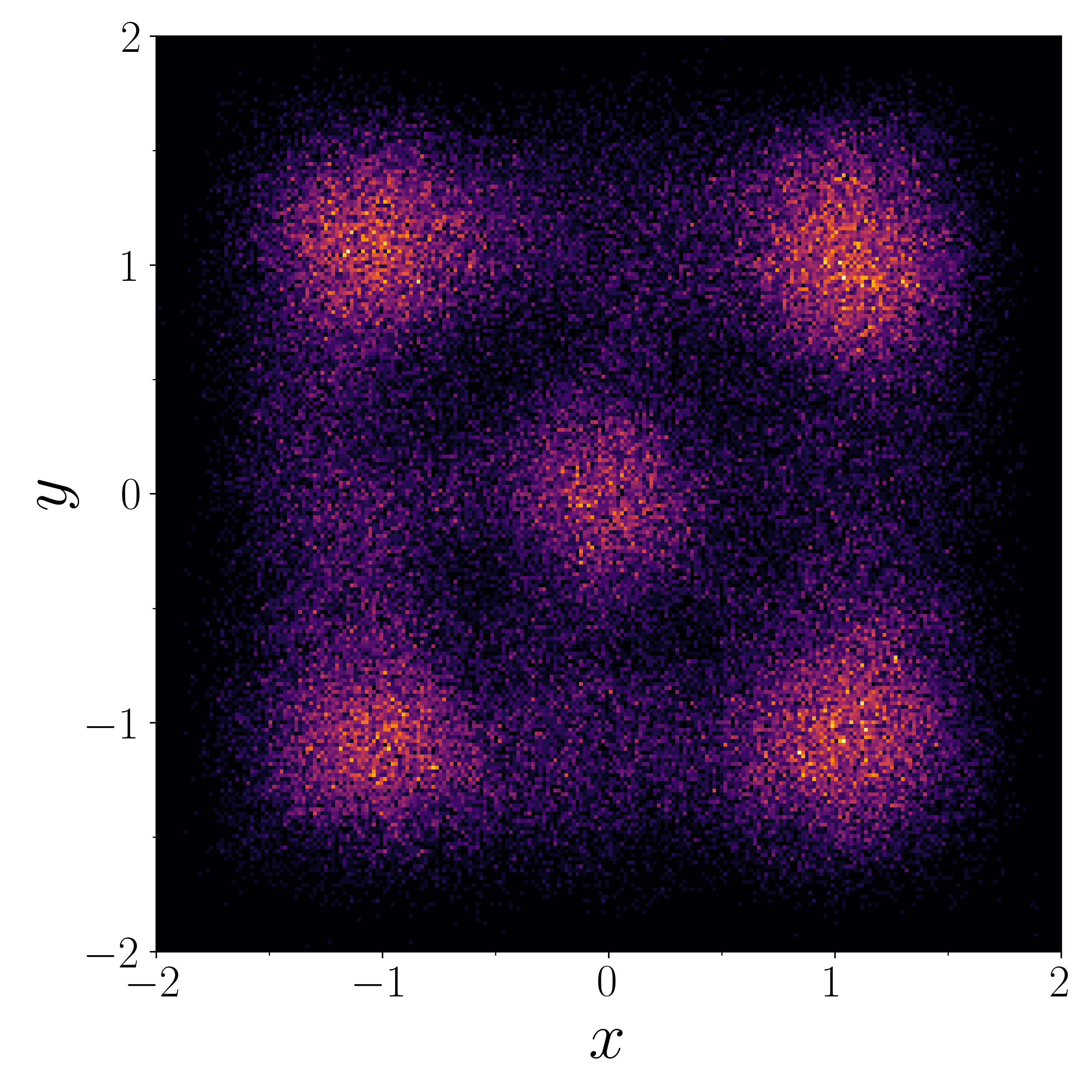}
    \end{minipage}\hfill
        \begin{minipage}[t]{0.25\linewidth} 
        \centering
        \includegraphics[width=0.99\linewidth]{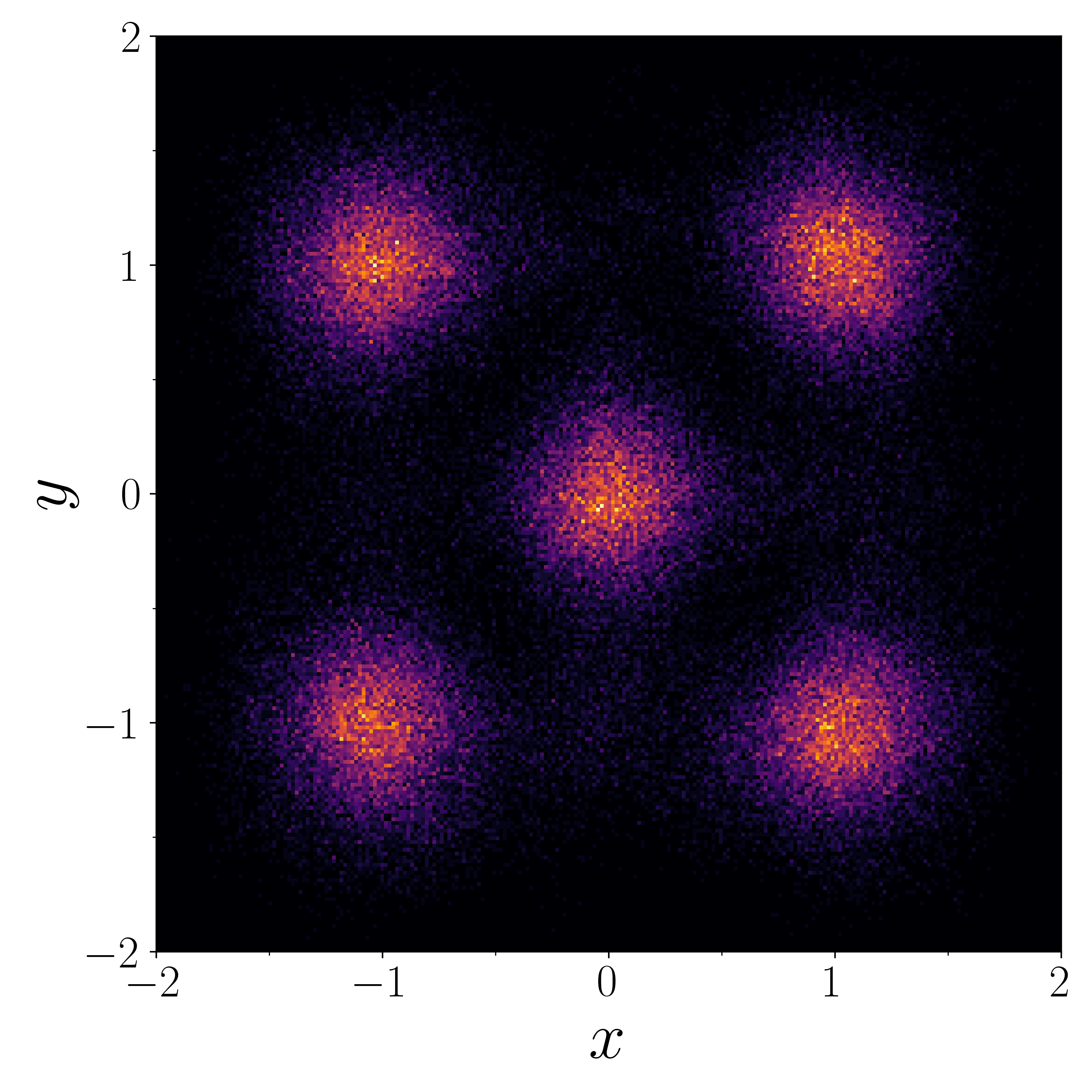}
    \end{minipage}
    \caption{\label{fig10}
    Density of beads for $N=8$ (top row) and $N=5$ (bottom row), for $\lambda=1$ and several values of $V_0$.}
\end{figure*}

In this section our aim is to show that the same oddities of the hard-core extended BH model on a small grid are seen in a system of bosonic particles in a convenient external field, designed to mimic the discrete system.
For this investigation, we employ the PIMC method~\cite{ceperley1995} in the canonical ensemble, which samples the equilibrium distribution at fixed $N$ and $T$ by leveraging on the isomorphism between quantum particles and imaginary-time paths (worldlines).
Furthermore, we use the worm algorithm~\cite{boninsegni2006b} for an efficient sampling of permutations.
This study demonstrates that, under appropriate conditions, even the behavior of a few-body system can be analyzed in terms of the interplay between superfluid-like and insulating-like properties.

We investigate a two-dimensional system of spinless bosons with an inverse-power, $\epsilon(\sigma/r)^{12}$ interaction (with $\epsilon=1$ and $\sigma=1$).
Classical particles interacting with this potential behave similarly to hard disks~\cite{kapfer2015}.
The $r^{-12}$ law serves a twofold purpose: 
First, a potential taking large positive values for distances below $\approx 1$ replicates the hard-core condition of the model studied in the previous section.
Second, the inverse-power potential features a unit strength at $r\approx 1$, followed by a rapid extinction at larger distances, thereby mimicking a ``contact'' repulsion between hard-core particles of unit diameter.

To enforce the preference of particle positions for discrete values we use an optical-lattice potential.
For square-lattice order, this is given by
\begin{equation}
u_{\rm opt}(x,y)=\frac{1}{4}V_0\left[2-\cos(2\pi x)-\cos(2\pi y)\right]
\label{eq6}
\end{equation}
with $V_0>0$ (notice that the minimum and maximum values of (\ref{eq6}) are 0 and $V_0$, respectively).
With this potential, $x$ and $y$ coordinates are pushed to take integer values.
Finally, in order to confine particles to e.g. a $3\times 3$ square box centered at the origin, encompassing a total of nine sites, we add the potential
\begin{equation}
u_{\rm box}(x,y)=\left(\frac{x}{1.5}\right)^{20}+\left(\frac{y}{1.5}\right)^{20}\,.
\label{eq7}
\end{equation}
Under the action of $u_{\rm box}$, the boson system ``acquires'' a definite area of 9 (units of $\sigma^2$).
The combination of (\ref{eq6}) and (\ref{eq7}) will force particle worldlines to stay close to the sites of a grid identical to Square.
Finally, the total Hamiltonian reads:
\begin{equation}
H=\sum_{i=1}^N\left(-\lambda\nabla^2_i+u_{\rm ext}({\bf r}_i)\right)+\sum_{i<j}u_{\rm int}(|{\bf r}_i-{\bf r}_j|)\,,
\label{eq8}
\end{equation}
where $\lambda=\hbar^2/(2m)$ ($m$ being the particle mass) and
\begin{equation}
u_{\rm int}(|{\bf r}-{\bf r}'|)=\frac{1}{|{\bf r}-{\bf r}'|^{12}}\,,\,\,\,\,\,\,u_{\rm ext}({\bf r})=u_{\rm opt}({\bf r})+u_{\rm box}({\bf r})\,.
\label{eq9}
\end{equation}
In this system, the parameter $\lambda$ plays a role analogous to $t$ in the BH model:
The larger $\lambda$, the stronger quantum fluctuations (i.e., the more dispersed are particle worldlines).

For a few selected $N$, and fixed $\lambda$ and $V_0$, we run our PIMC code for long, measuring the equilibrium averages of the various terms in the Hamiltonian, the pressure $\Pi$, the superfluid fraction $f_s$, and the distribution $P(L)$ of the length of permutation cycles (with $L$ from 1 to $N$).
In particular, calling $A$ the area of the box and ${\cal M}\gg 1$ the number of time slices (beads per particle), $\Pi$ and $f_s$ are computed as~\cite{sindzingre1989,marienhagen2025}
\begin{eqnarray}
\Pi&=&\frac{N{\cal M}}{\beta A}-\left\langle\frac{{\cal M}}{4A\lambda\beta^2}\sum_{i=1}^N\sum_{k=1}^{\cal M}\left(\textbf{r}^k_i-\textbf{r}^{k+1}_i\right)^2\right.
\nonumber \\
&+&\left.\frac{1}{2A{\cal M}}\sum_{i<j}^N\sum_{k=1}^{\cal M}r^k_{ij}u'_{\rm int}(r^k_{ij})\right.
\nonumber \\
&+&\left.\frac{1}{2A{\cal M}}\sum_{i=1}^N\sum_{k=1}^{{\cal M}}{\bf r}_i^k\cdot\nabla u_{\rm ext}({\bf r}_i)\right\rangle
\end{eqnarray}
and
\begin{equation}
f_{s}=\frac{2m}{\lambda\beta}\frac{\langle A^2_z\rangle-\langle A_z\rangle^2}{I_{\rm cl}}\,.
\end{equation}
In the above formulae, ${\bf r}_i^k$ is the vector radius of the $k$-th bead in worldline no. $i$, $r^k_{ij}=|{\bf r}_i^k-{\bf r}_j^k|$, $I_{\rm cl}$ is the classical moment of inertia,
\begin{equation}
I_{\rm cl}=m\sum_{i=1}^N\sum_{k=1}^{\cal M} \left(x^k_ix^{k+1}_i+y^k_iy^{k+1}_i\right)\,,
\end{equation}
and $A_z$ is the total area enclosed by particle paths,
\begin{equation}
A_z=\frac{1}{2}\sum_{i=1}^N\sum_{k=1}^{\cal M}  \left(x^k_iy^{k+1}_i-y^k_ix^{k+1}_i\right)\,.
\end{equation}
For systems with PBC the formula for $f_s$ is different and uses the winding-number estimator~\cite{ceperley1995}:
\begin{equation}
f_s=\frac{\langle W^2 \rangle}{2\lambda\beta N}\,.
\end{equation}
Finally, $P(L)$ is precisely defined as the probability that a particle chosen at random belongs to a cycle of length $L$:
Calling ${\cal N}_L(C)$ the number of $L$-cycles in configuration $C$, with $\sum_LL{\cal N}_L(C)=N$, it follows $P(L)=L\langle{\cal N}_L(C)\rangle/N$, where $\langle\cdots\rangle$ denotes the equilibrium average and $\sum_LP(L)=1$.
A flattened $P(L)$ is suggestive of superfluid-like behavior, see e.g. Refs.~\cite{jain2011a,ciardi2024,ciardi2025}. 
In all our runs, the temperature stays fixed at $T=0.5$, a low value.
\begin{figure}[t]
    \centering
    \includegraphics[width=1.0\linewidth]{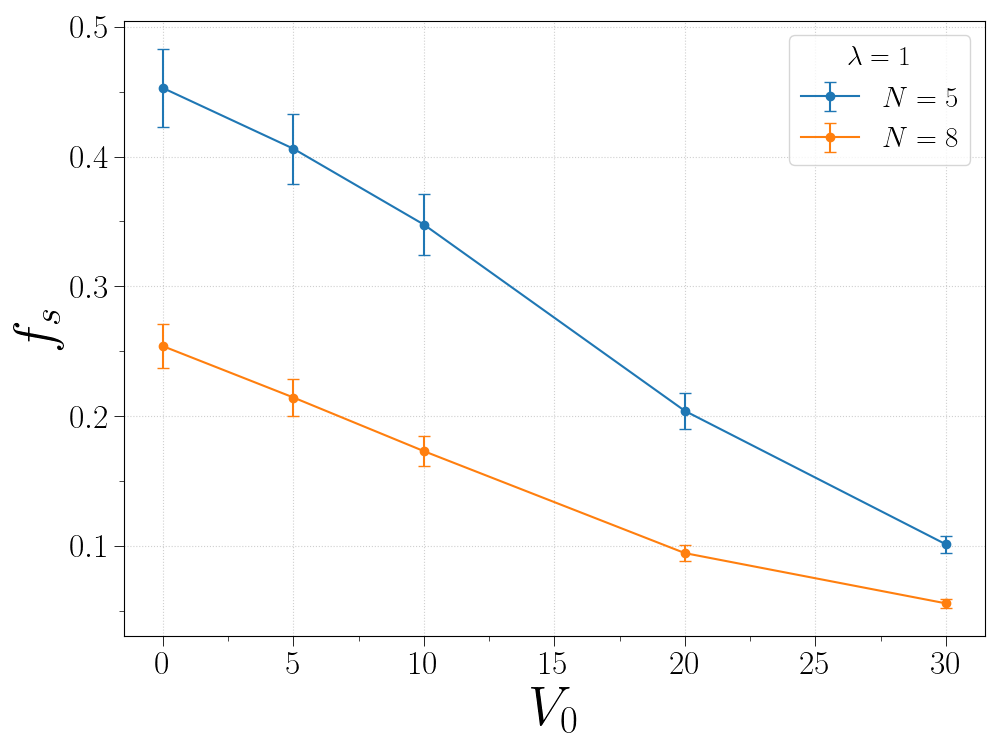}
    \caption{\label{fig11}
    Superfluid fraction for $N=5$ and $N=8$, for $\lambda=1$ and several values of $V_0$.}
\end{figure}

\begin{figure}[t]
    \centering
    \includegraphics[width=0.9\linewidth]{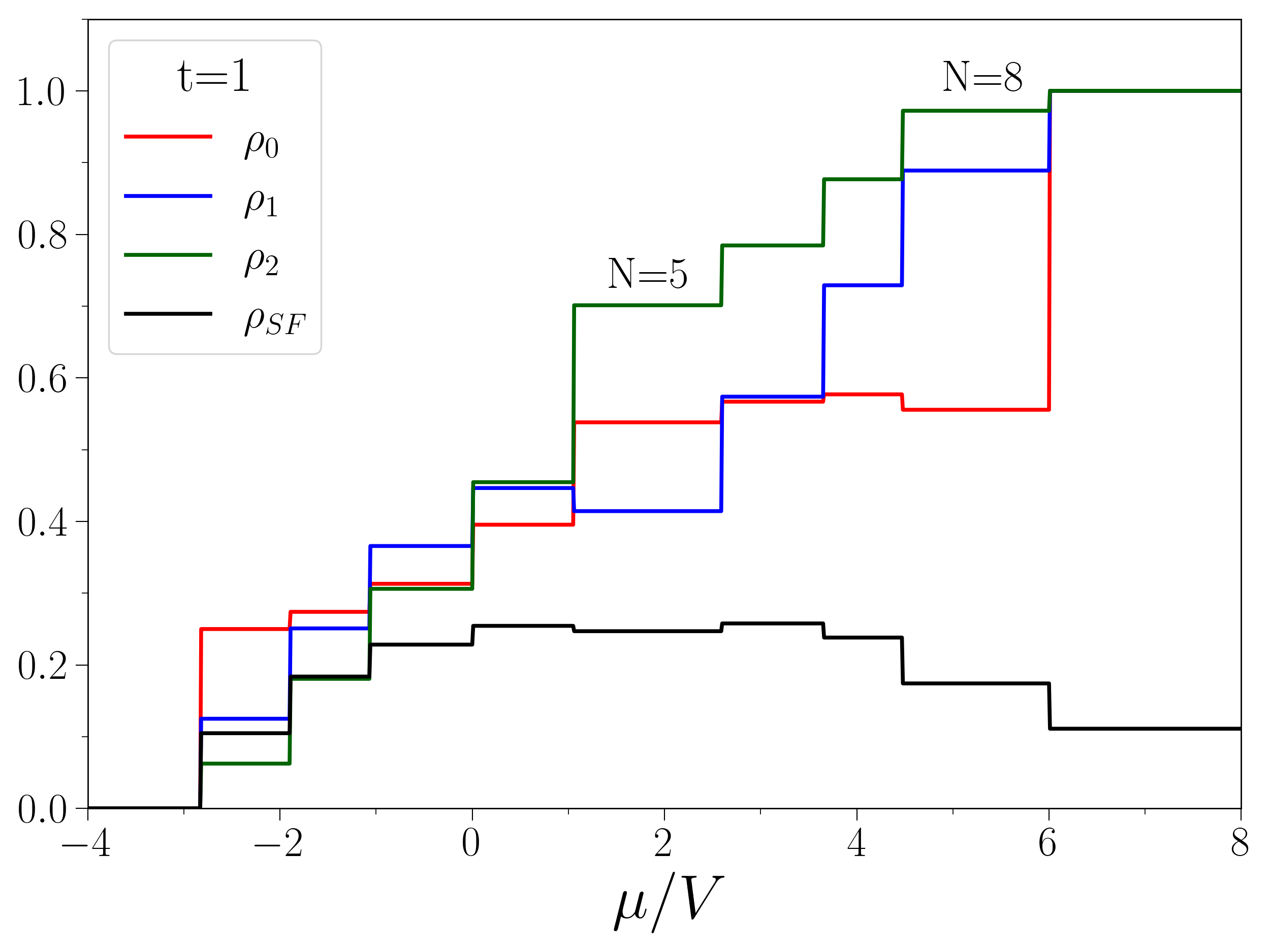}\\
    \includegraphics[width=0.9\linewidth]{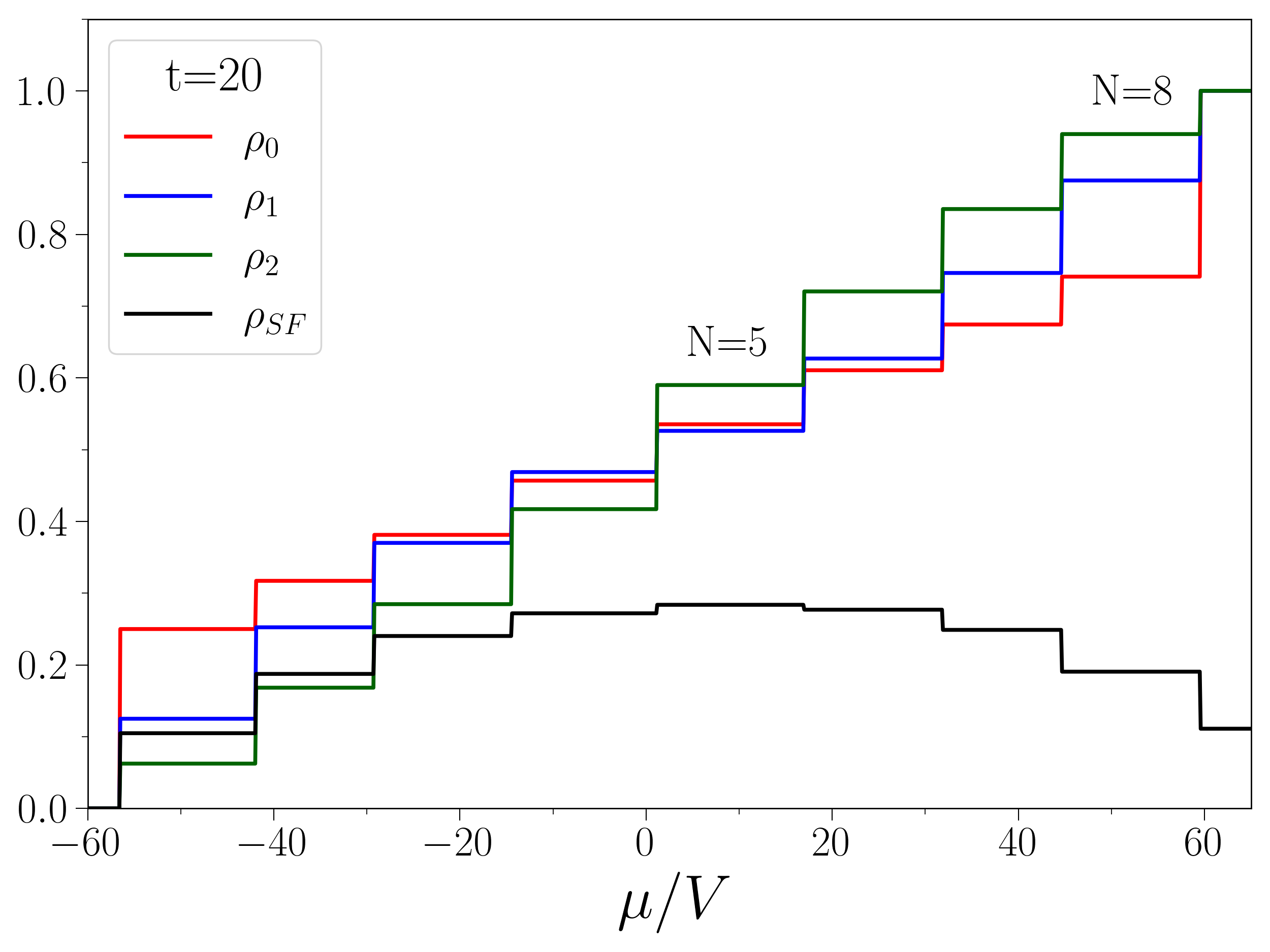}
    \caption{\label{fig12}
    Hard-core extended BH model on Square. Average site occupancies $\rho_n$ ($n=0$: central site; $n=1$: first-neighbor sites; $n=2$: second-neighbor sites) plotted as functions of $\mu$, for $t=1$ (top) and $t=20$ (bottom). Also shown is the condensate fraction $\rho_{\rm c}$ (black line). Notice the difference in $\mu$ scale between the two panels.}
\end{figure}

\begin{figure}[t]
    \centering
    \includegraphics[width=1.0\linewidth]{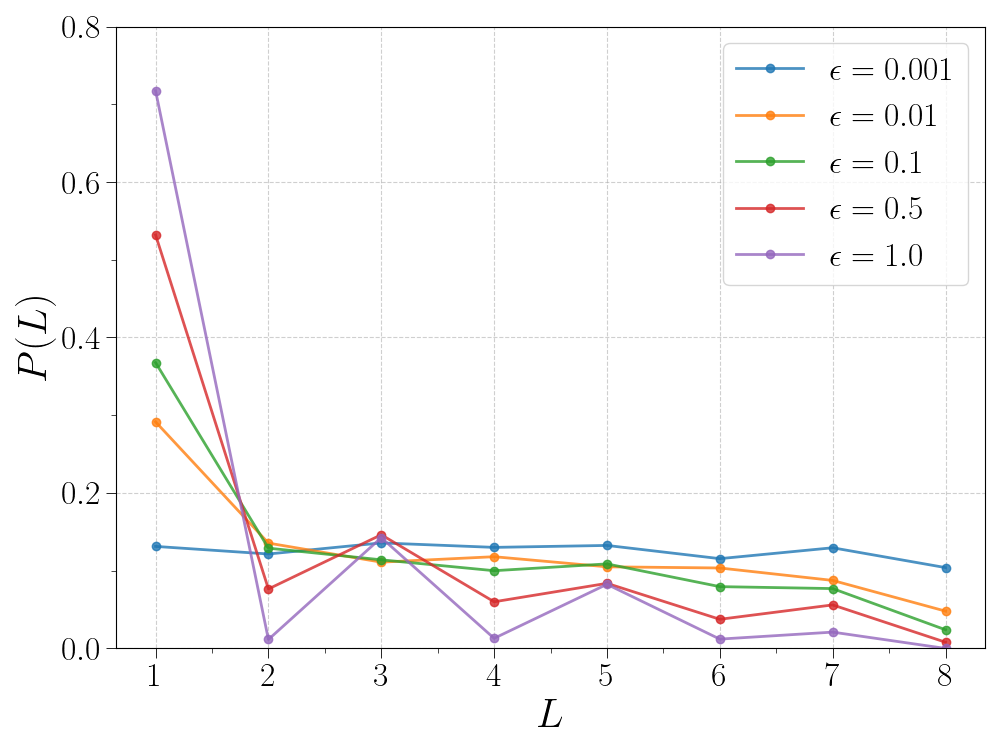}
    \caption{\label{fig13}
    Distribution $P(L)$ of exchange-cycle lengths for $\lambda=1,N=8$ and $V_0=0$, plotted as a function of $\epsilon$.}
\end{figure}

\subsection{Bosons in a $3\times 3$ box}

As a first example, we work in the same setup of Greiner's experiment~\cite{greiner2002}:
Keeping $N$ fixed and assuming a sufficiently large value of $\lambda$ (that is, 1), we increase $V_0$ progressively to see the system transitioning from superfluid-like to insulating-like conditions.
We point out that this is not the manner to look at this transition in the hard-core extended BH model;
however, if the initial state is superfluid, enhancing the strength of the confining potential more and more will increasingly suppress the mobility of particles and thus reduce the superfluid fraction.
The most interesting cases to probe are $N=5$ and $N=8$, corresponding to the states of Fig.~\ref{fig2}a.
As is clear from Fig.~\ref{fig2}b, for either $N$ we can make the entire journey from superfluid to insulator while keeping canonical conditions.
We produce several hundred million MC moves, using a number of beads per particle up to 1024;
equilibrium averages are computed over the second half of the MC trajectory.
In Fig.~\ref{fig10} we show the position-dependent density of beads as a function of $V_0$.
For both values of $N$, the expected insulating-phase structure eventually emerges, though shadows of it are already apparent for $V_0=0$, as an effect of pair repulsion only.
However, with the lattice potential switched off the density of beads is measurably non-zero in the whole box, pointing to a small superfluid response.
This is confirmed by the evolution of $f_s$ with $V_0$:
Starting from a non-zero value at $V_0=0$ ($\simeq 0.25$), $f_s$ monotonically decreases with increasing $V_0$ (see Fig.~\ref{fig11}).
It must be considered that, for $\lambda=1$ and $V_0=0$, a system of a few interacting bosons in a small box cannot be full-fledged superfluid (i.e., phase-coherent {\em and} uniform) and a much larger $\lambda$ is needed to make the density more uniform in the box.
After all, also in the EBH model on Square the average density at $T=0$ and, say, $t=1$ is all but uniform (see Fig.~\ref{fig12}), showing inhomogeneities analogous to those exhibited by the small-$t$ system.
As for the various energy terms in the Hamiltonian, we find that the averages of kinetic, confinement, and interaction contributions are roughly independent of $V_0$, while the optical contribution grows almost linearly with $V_0$ (figure not shown).
Finally, for $N=8$ the pressure increases almost linearly with $V_0$, as expected for a system becoming increasingly more incompressible.

\begin{figure*}[t]
    \centering
    \begin{minipage}[t]{0.2\linewidth} 
        \centering
        \textbf{$\lambda=0.1$}\par\medskip
        \includegraphics[width=0.99\linewidth]{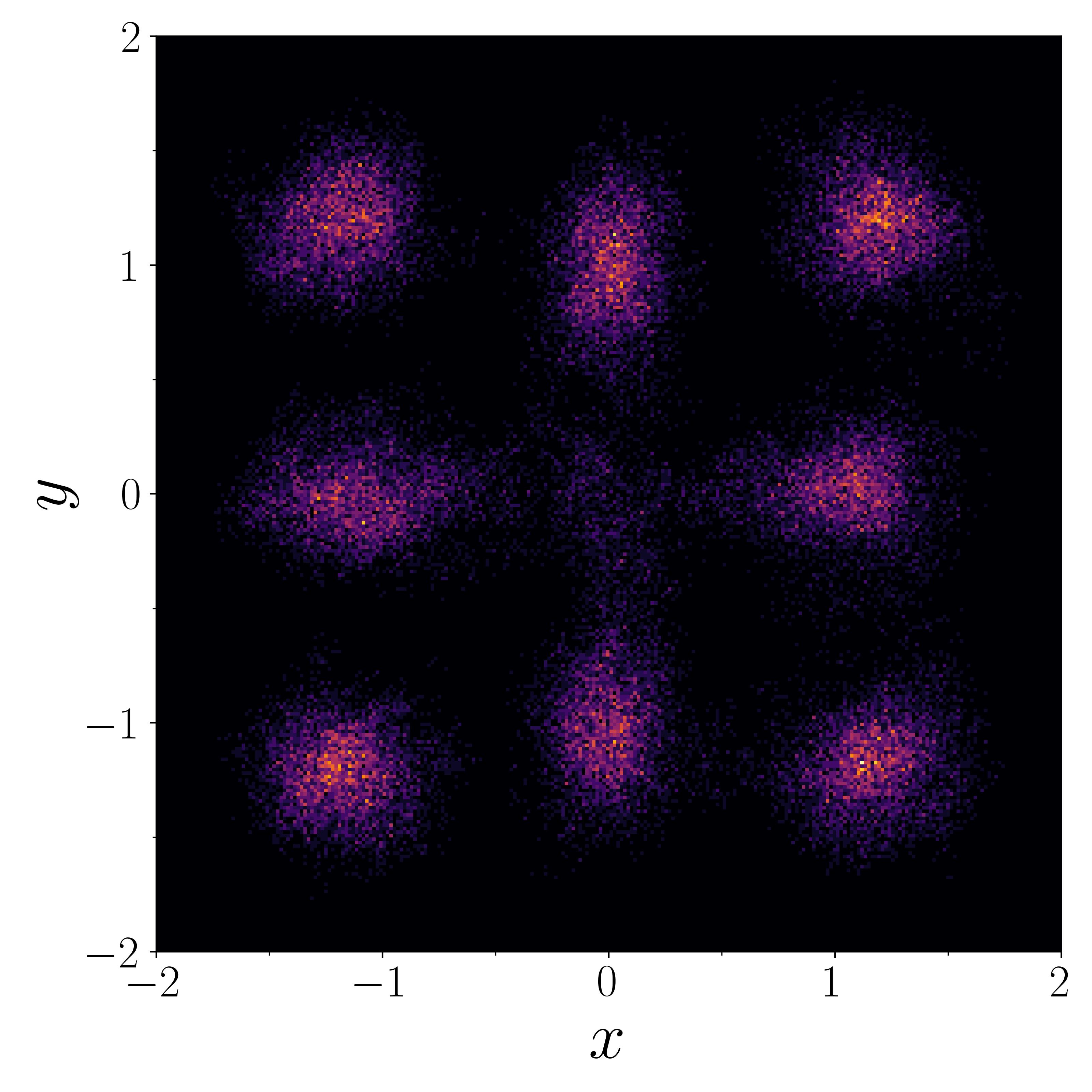}
    \end{minipage}\hfill
    \begin{minipage}[t]{0.2\linewidth} 
        \centering
        \textbf{$\lambda=0.5$}\par\medskip
        \includegraphics[width=0.99\linewidth]{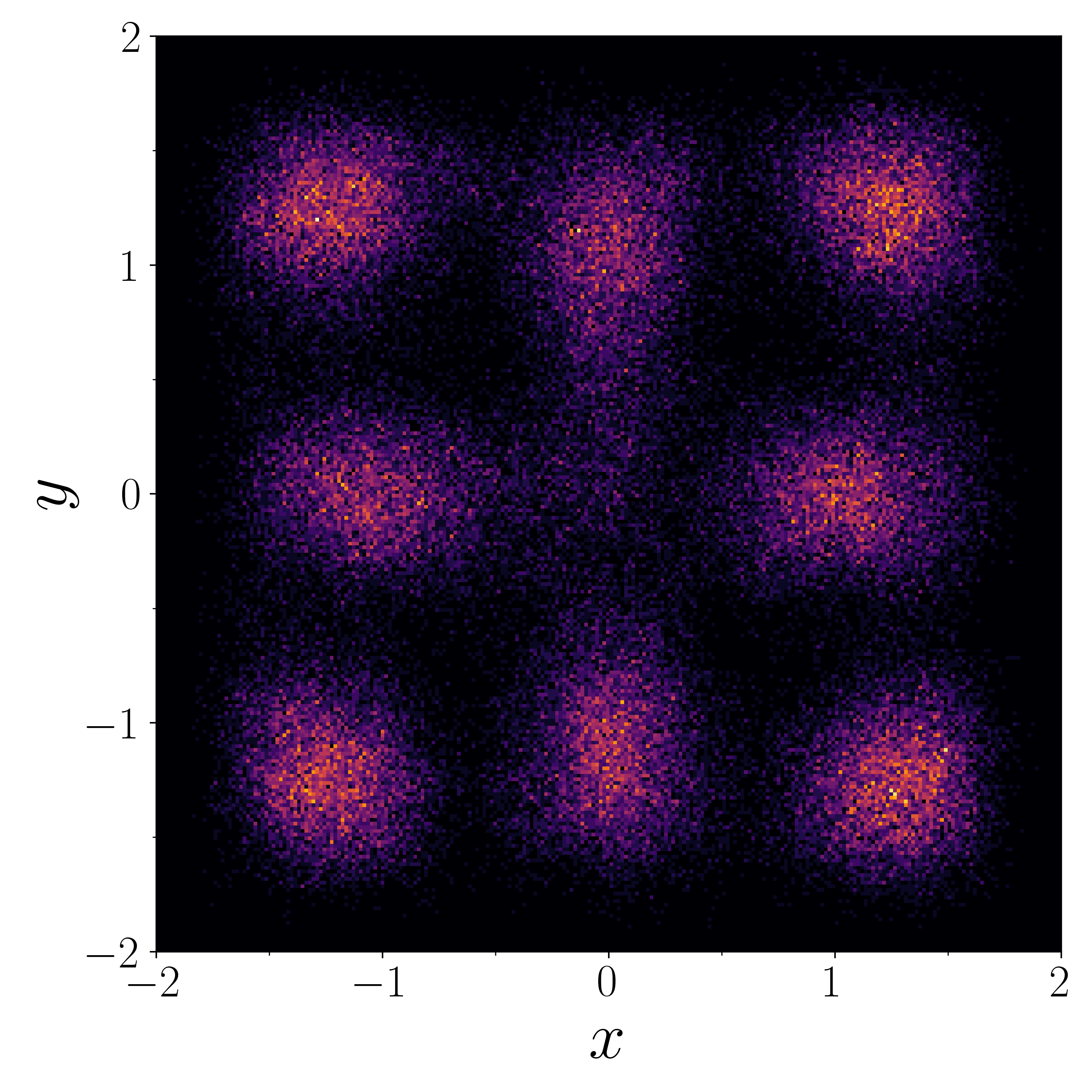}
    \end{minipage}\hfill
    \begin{minipage}[t]{0.2\linewidth} 
        \centering
        \textbf{$\lambda=1$}\par\medskip
        \includegraphics[width=0.99\linewidth]{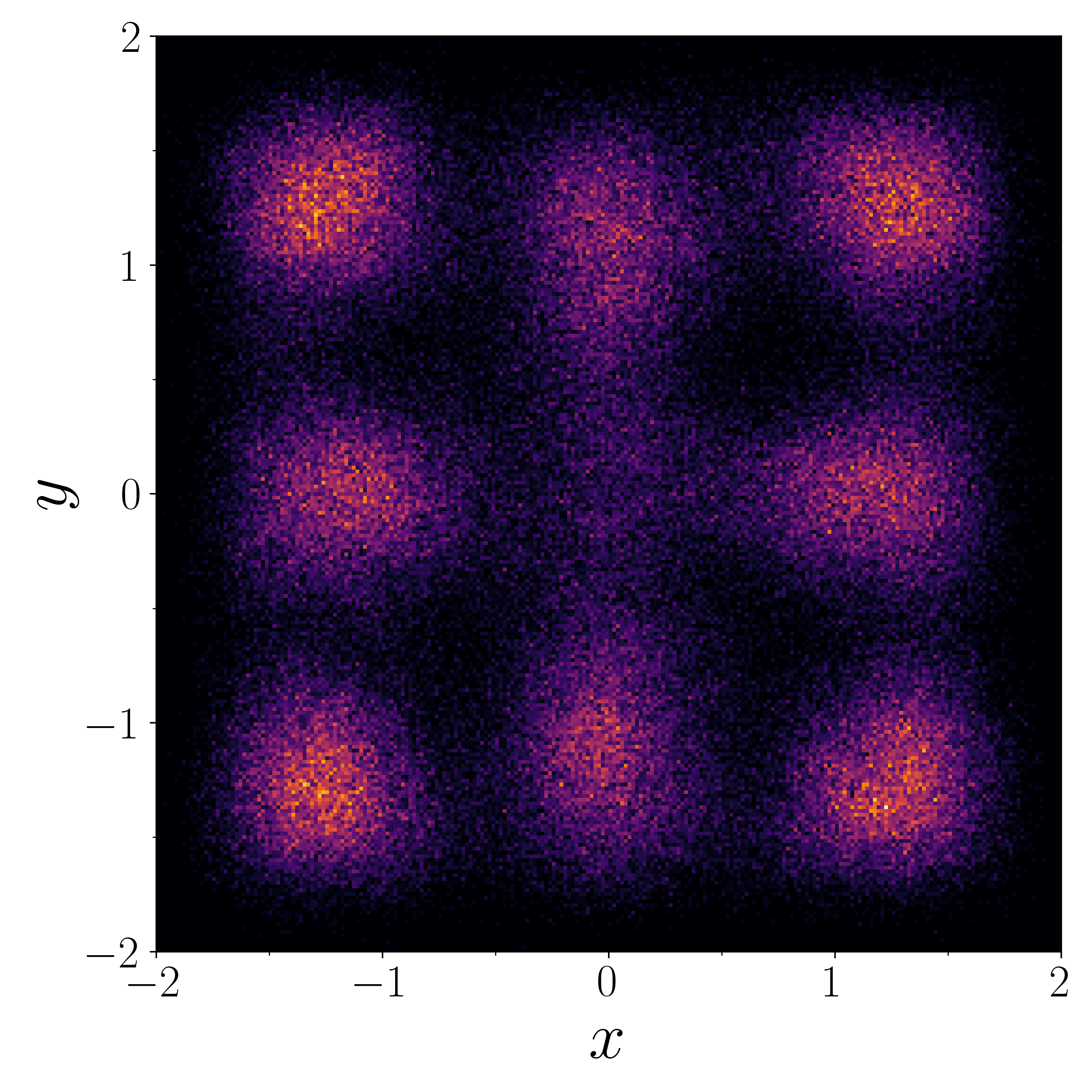}
    \end{minipage}\hfill
        \begin{minipage}[t]{0.2\linewidth}
        \centering
        \textbf{$\lambda=1.5$}\par\medskip
        \includegraphics[width=0.99\linewidth]{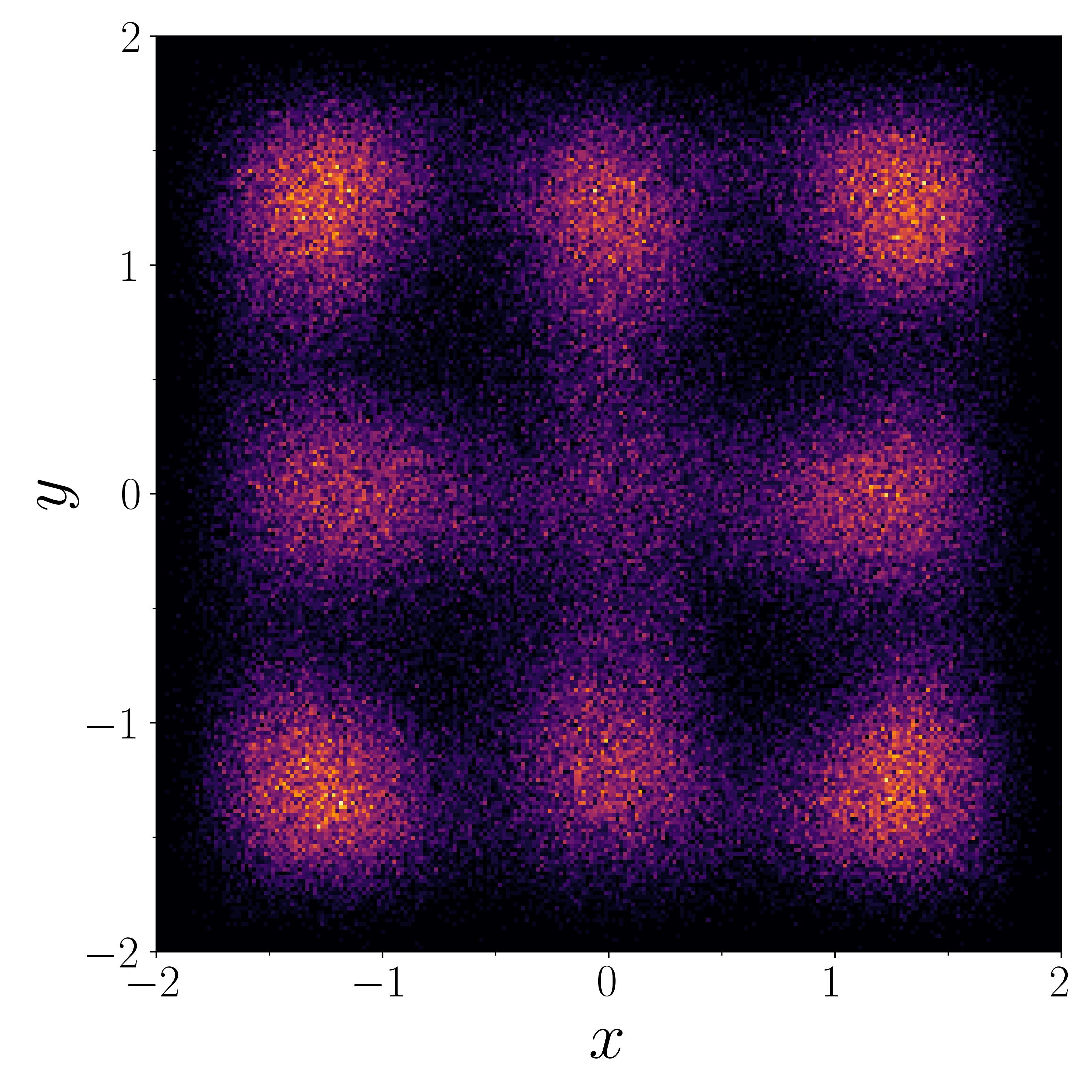}
    \end{minipage}\hfill
        \begin{minipage}[t]{0.2\linewidth} 
        \centering
        \textbf{$\lambda=2$}\par\medskip
        \includegraphics[width=0.99\linewidth]{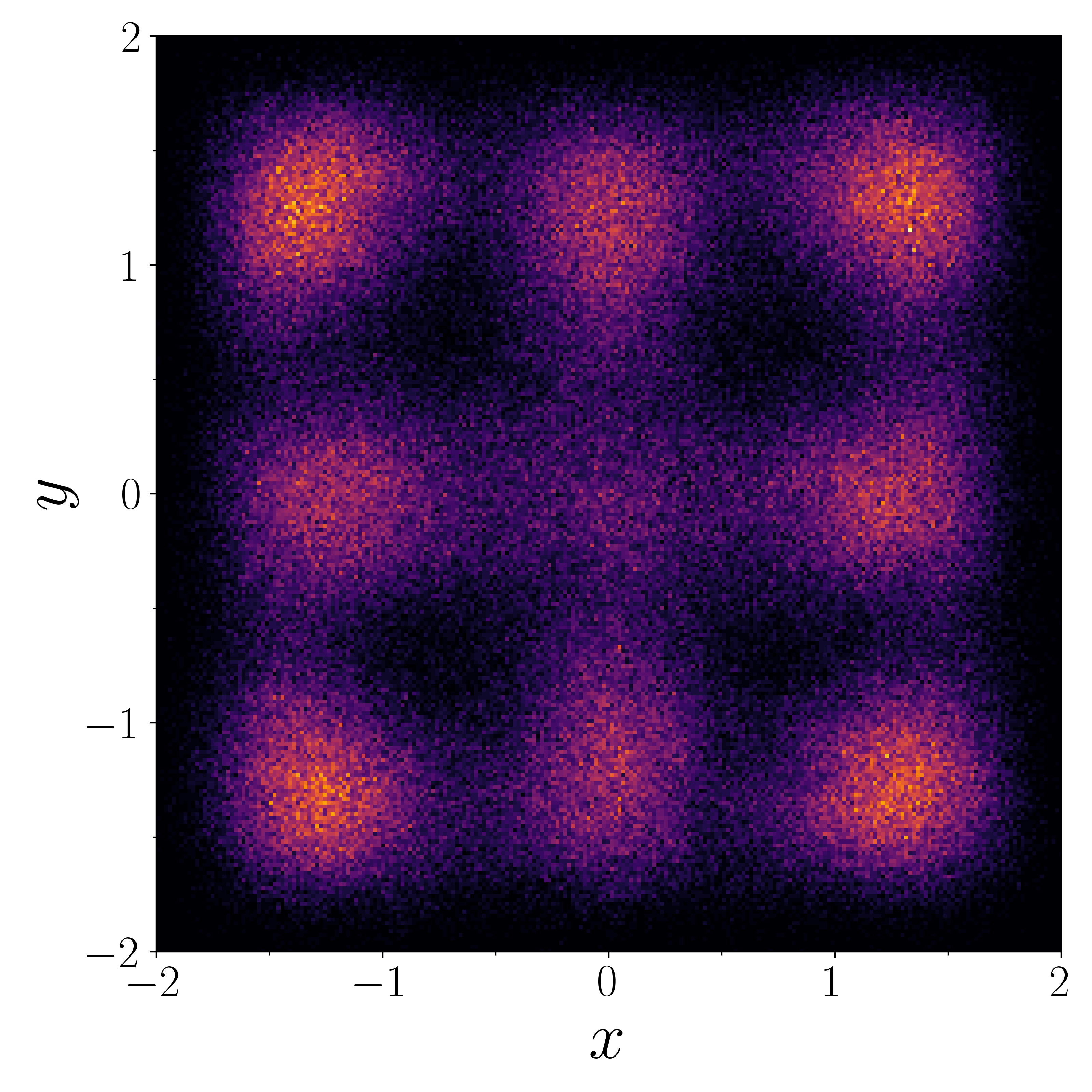}
    \end{minipage}
    \caption{\label{fig14}
    Density of beads for $N=8$ and $V_0=5$, for several values of $\lambda$.}
\end{figure*}

\begin{figure*}[t]
    \centering
    \begin{minipage}[t]{0.25\linewidth} 
        \centering
        \textbf{$V_0=0$}\par\medskip
        \includegraphics[width=1.0\linewidth]{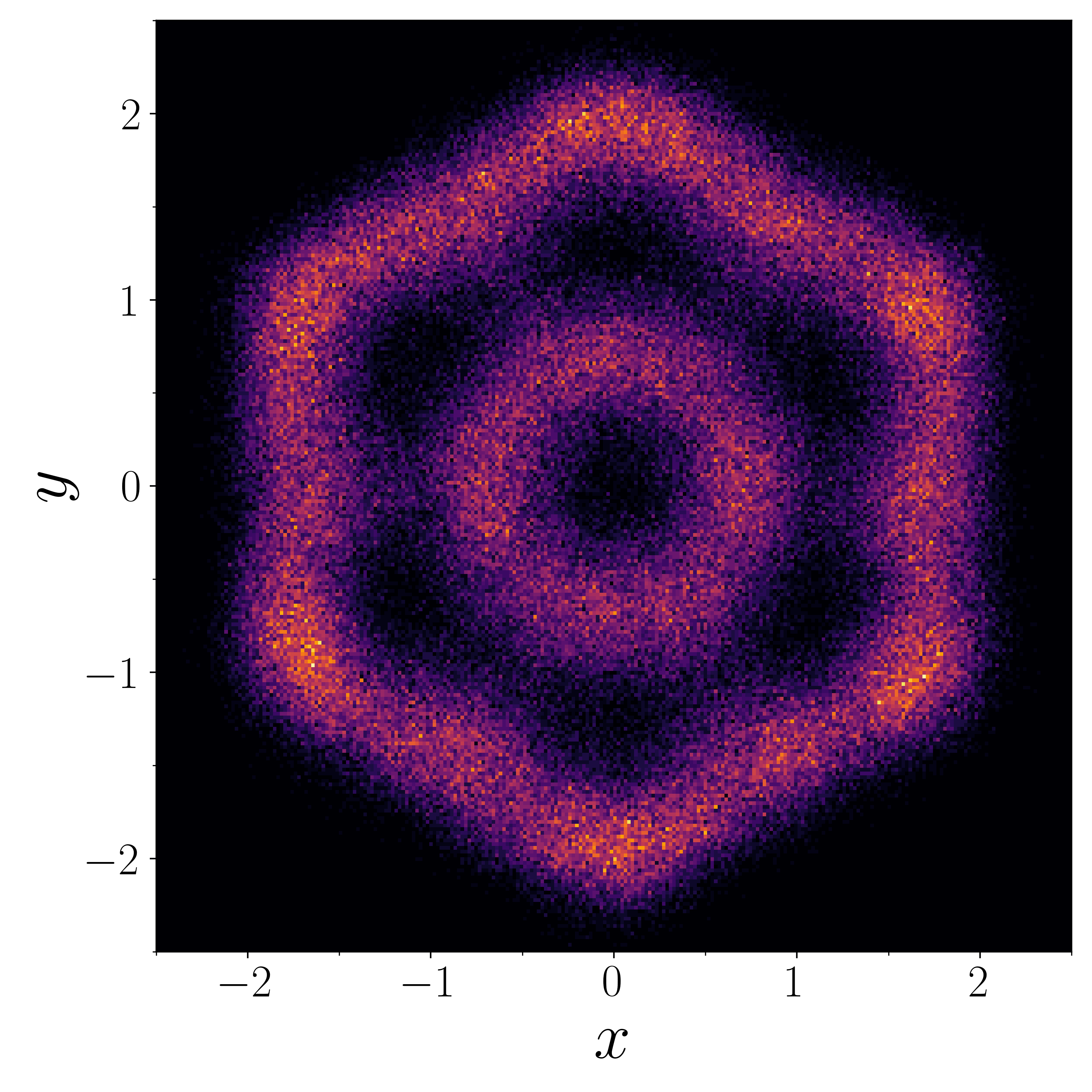}
    \end{minipage}\hfill
    \begin{minipage}[t]{0.25\linewidth} 
        \centering
        \textbf{$V_0=1$}\par\medskip
        \includegraphics[width=1.0\linewidth]{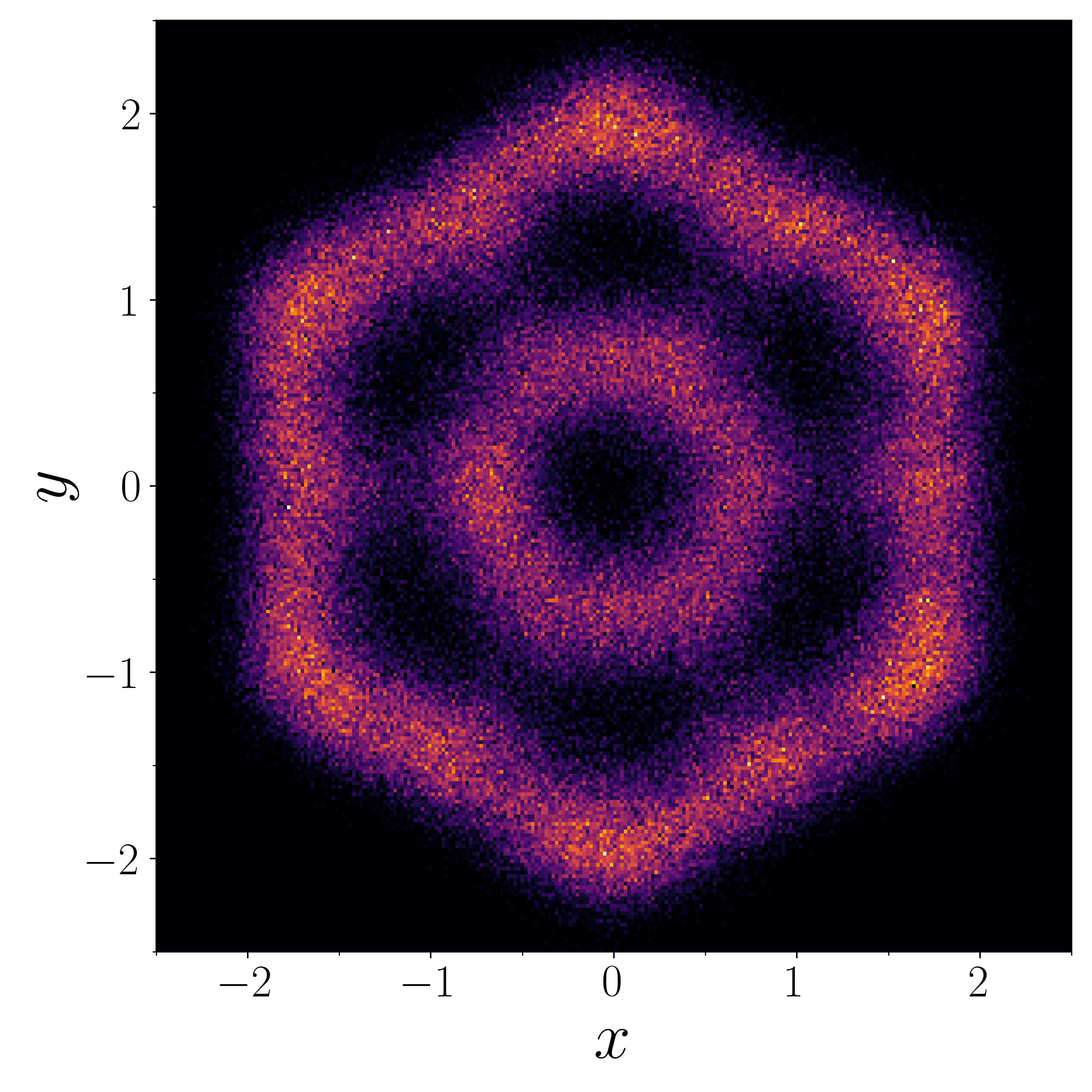}
    \end{minipage}\hfill
    \begin{minipage}[t]{0.25\linewidth} 
        \centering
        \textbf{$V_0=10$}\par\medskip
        \includegraphics[width=1.0\linewidth]{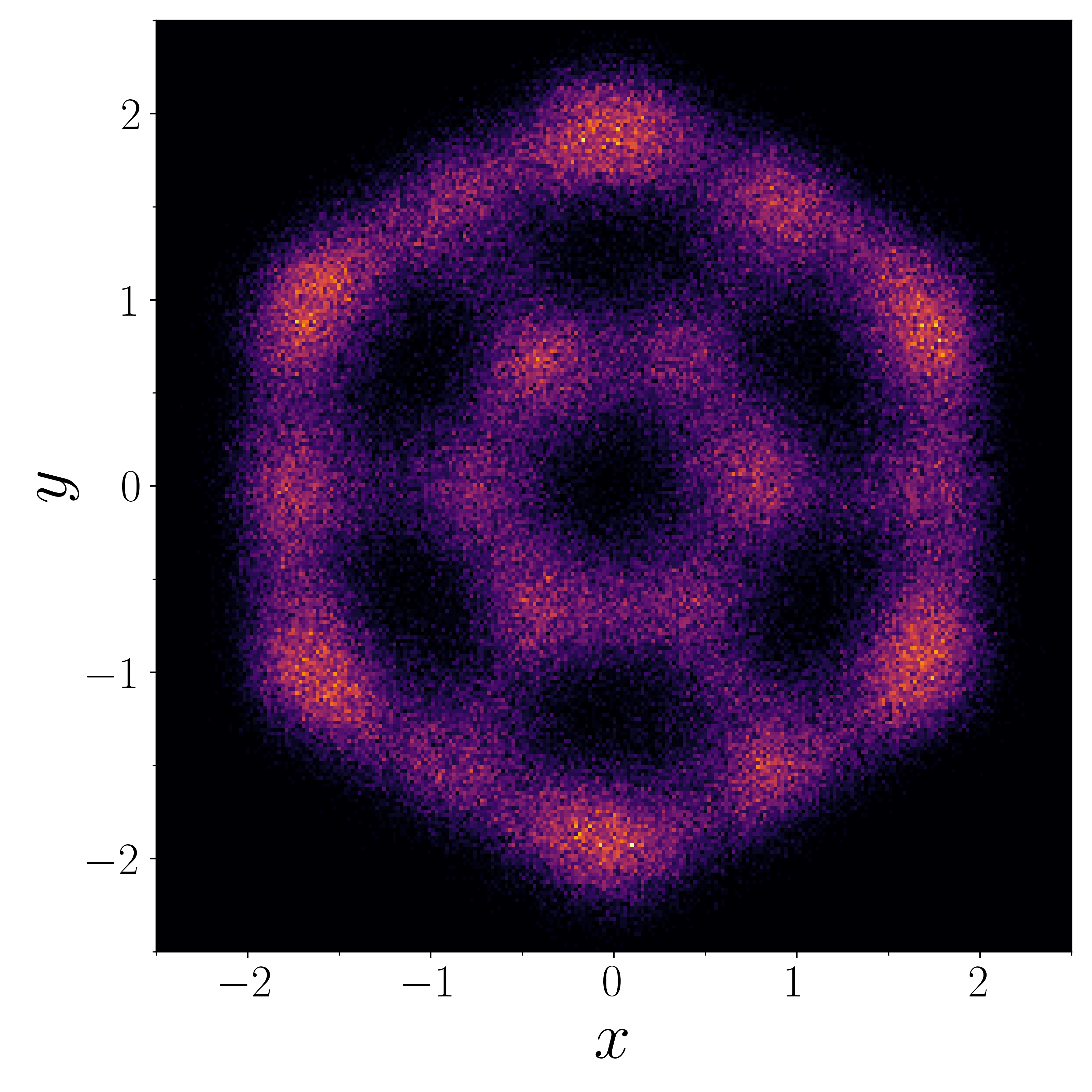}
    \end{minipage}\hfill
        \begin{minipage}[t]{0.25\linewidth}
        \centering
        \textbf{$V_0=30$}\par\medskip
        \includegraphics[width=1.0\linewidth]{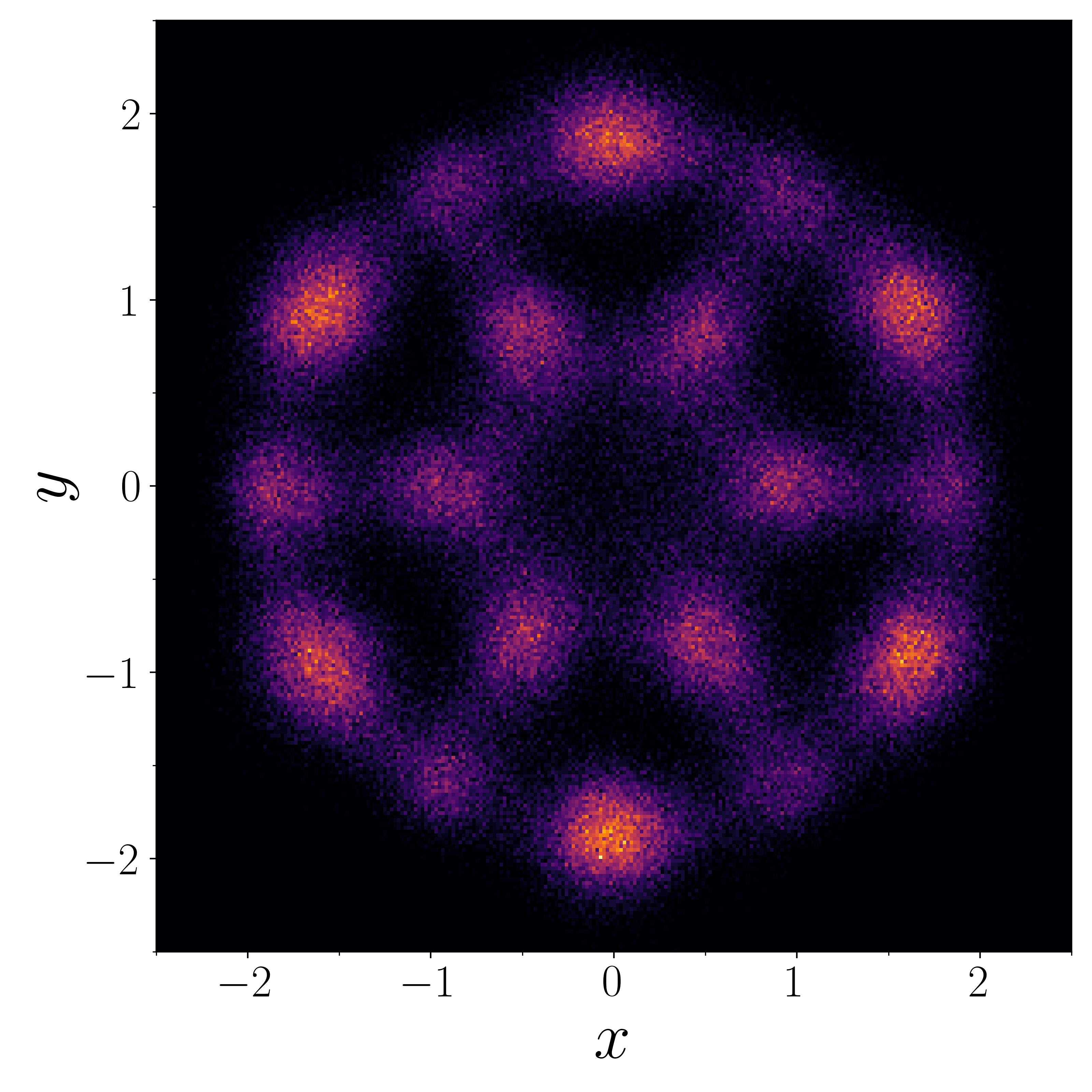}
    \end{minipage}\\
    \begin{minipage}[t]{0.25\linewidth} 
        \centering
        \includegraphics[width=1.0\linewidth]{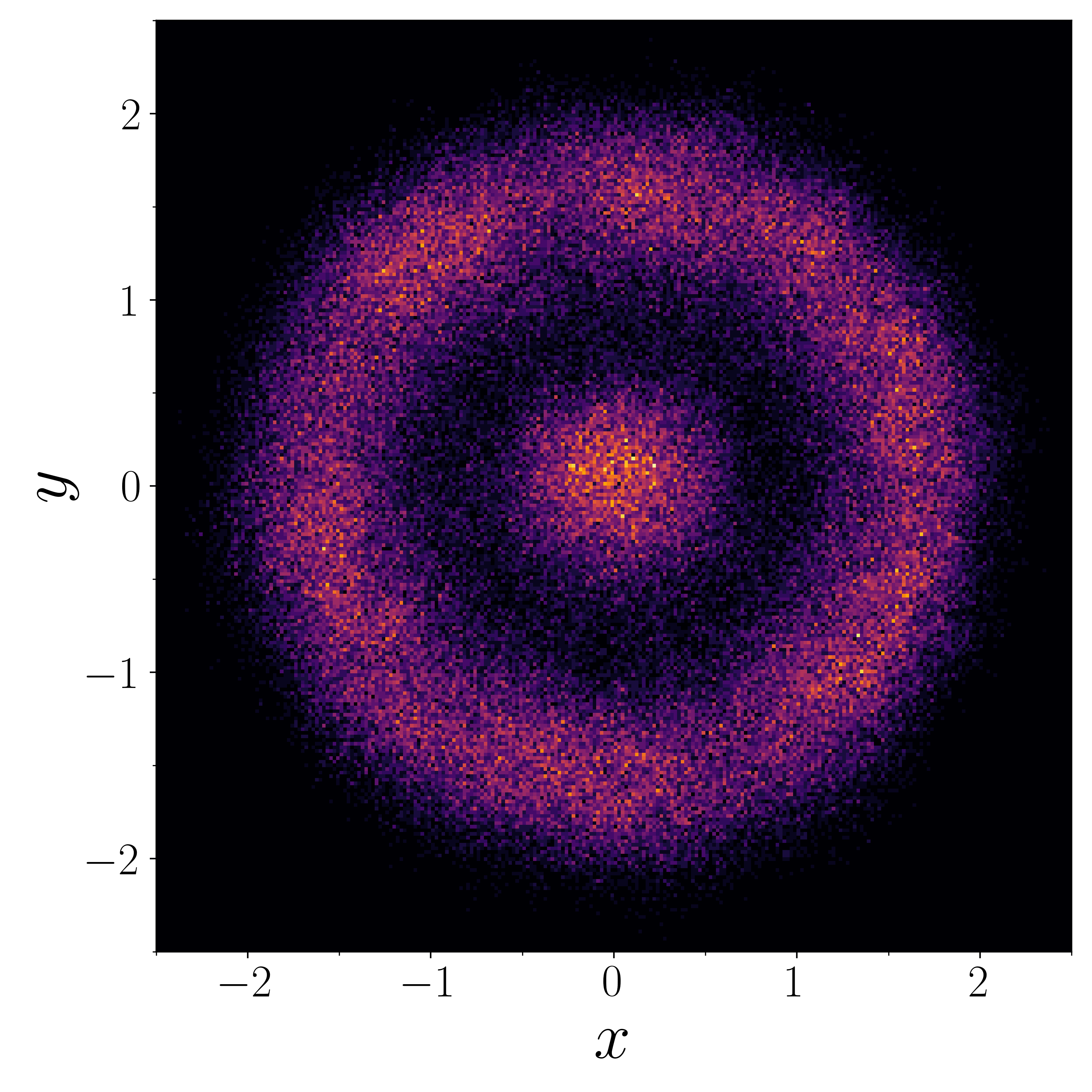}
    \end{minipage}\hfill
    \begin{minipage}[t]{0.25\linewidth} 
        \centering
        \includegraphics[width=1.0\linewidth]{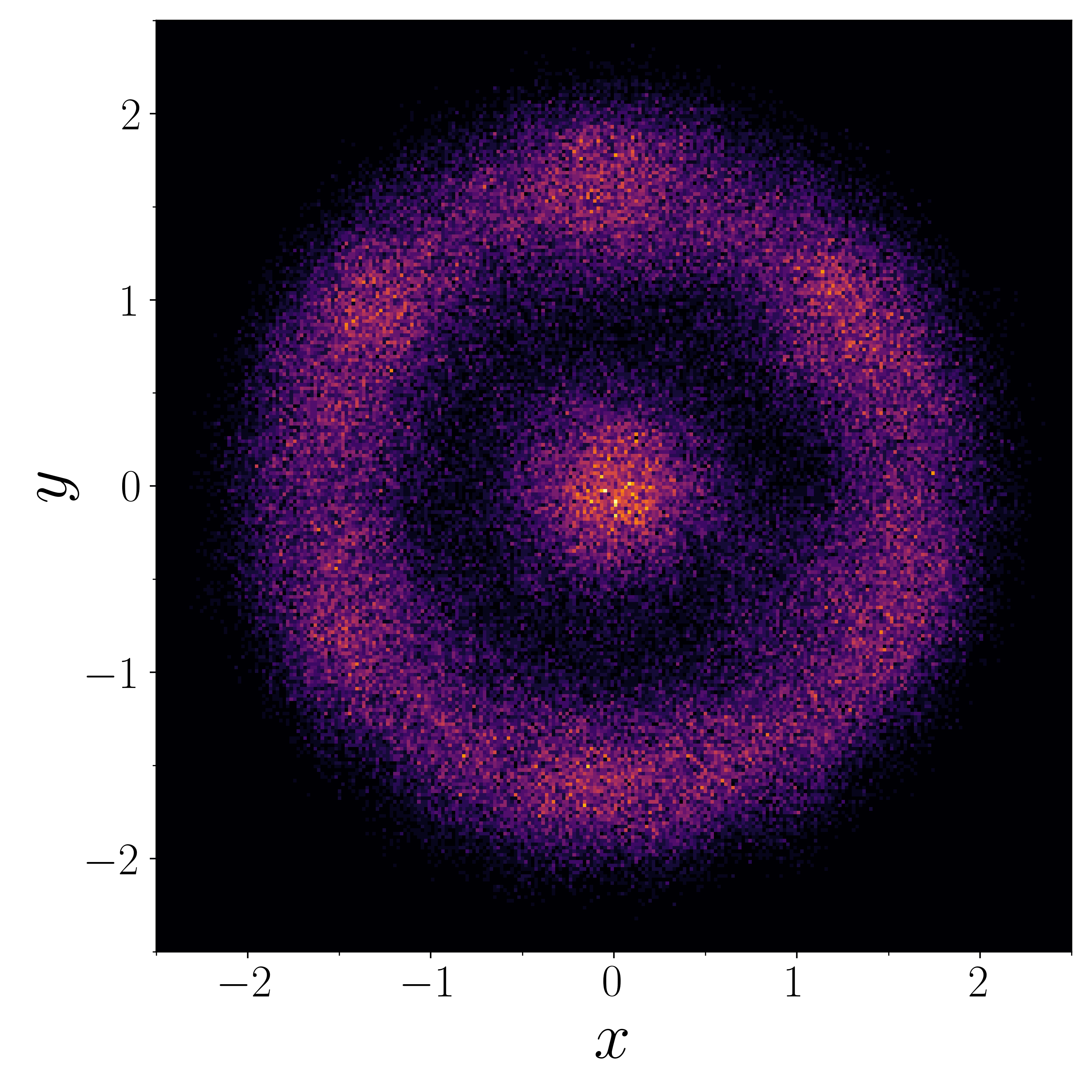}
    \end{minipage}\hfill
    \begin{minipage}[t]{0.25\linewidth} 
        \centering
        \includegraphics[width=1.0\linewidth]{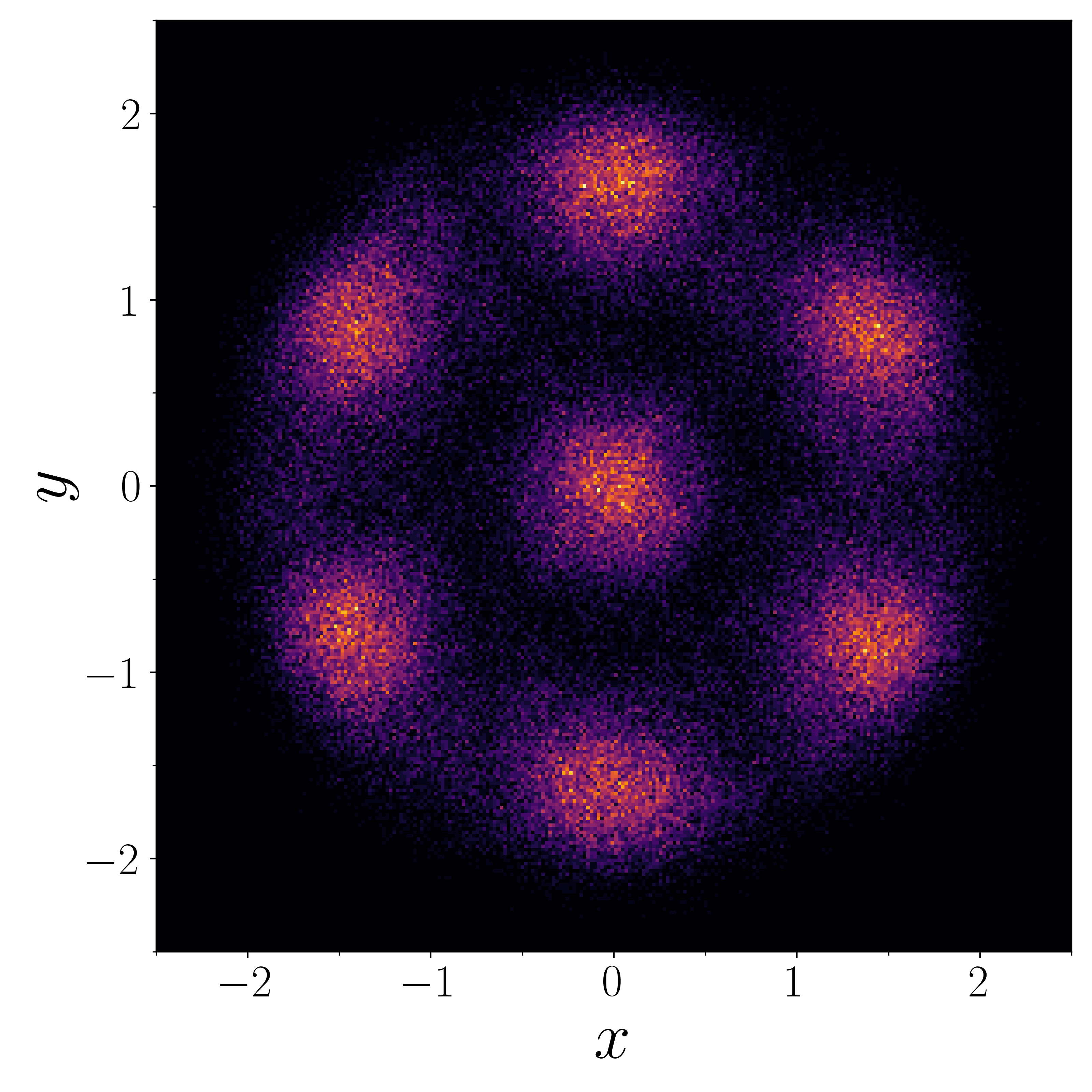}
    \end{minipage}\hfill
        \begin{minipage}[t]{0.25\linewidth}
        \centering
        \includegraphics[width=1.0\linewidth]{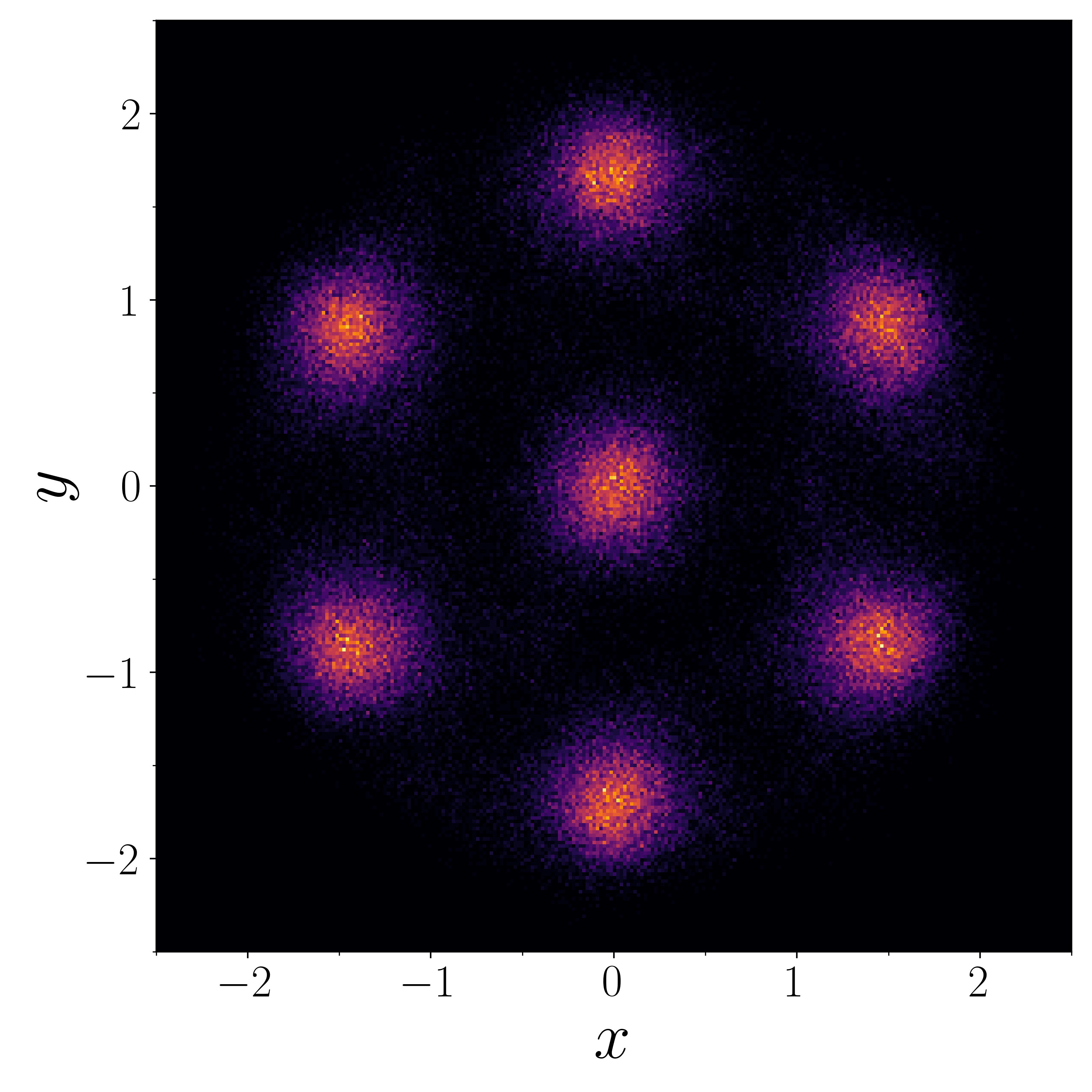}
    \end{minipage}
    \caption{\label{fig15}
    Simulation results for box shapes different from square. Density of beads for $N=12$ (top row) and $N=7$ (bottom row), for $\lambda=1$ and several values of $V_0$.}
\end{figure*}

We observe a curious behavior for $N=8$ and $V_0=0$.
In Fig.~\ref{fig13} we plot the cycle-length distribution $P(L)$ for varying strength $\epsilon$ of the interparticle repulsion (the figure is for $\lambda=1$, but the evidence is similar for $\lambda=0.5$ or 0.1).
We see that $P(L)$ oscillates, alternating maxima at odd values of $L$ with minima at even values.
When $\epsilon$ is decreased, the oscillations of $P(L)$ persist but they dampen progressively.
An oscillatory $P(L)$ also exists for $V_0>0$, but the modulation is less evident.
Instead, nothing similar is observed for $N=5$ or other values of $N$.
We have also verified that $P(L)$ flattens out when we remove the box potential and apply PBC.
The non-monotonous decay of $P(L)$ is hard to interpret:
Admittedly, it is an exclusive property of the $N=8$ system ascribable to both interparticle forces and confinement.

Next, we consider the other setup where $V_0$ is kept fixed at, say, 5, while $\lambda$ is progressively increased, starting from $\lambda=0.1$.
We expect to see a crossover from insulating-like to superfluid-like behavior, akin to the transition occurring in the BH model for increasing $t$ at fixed $N$.
As $\lambda$ grows, we make the number of particle beads accordingly larger, in order to keep sampling efficiency high as the acceptance rate of exchange MC moves decreases.
In Fig.~\ref{fig14} we show density maps for $N=8$.
Again, we see a smooth transition from a structure reminiscent of the BH ground state at $t=0$ to a distribution of beads that is slightly more uniform in the box.
In turn, the superfluid fraction grows monotonically from roughly zero at $\lambda=0.1$ to $\simeq 0.3$ at $\lambda=2$.
Of the various energy contributions, only kinetic energy varies appreciably, growing substantially with increasing $\lambda$.

\subsection{The box is removed and PBC are applied}

The density of beads should become more uniform when confinement is removed and periodic conditions are applied at the boundary of the $3\times 3$ square, much like the behavior reported in Fig.~\ref{fig3}.
With $\lambda=1$, we find similar results for $N=3$ and $N=6$.
As $V_0$ grows, we move from a roughly uniform density of beads for $V_0=10$ (where $f_s\simeq 1$) to a density function which, unlike the configurations in Fig.~\ref{fig3}a, exhibits pronounced maxima in all potential wells ($f_s\simeq 0.15$ at $V_0=100$).
This happens because rotational symmetry around the center, which is broken in the $t=0$ ground state of the BH model, is fully restored when hopping (no matter how small) is allowed.

\subsection{Other box shapes}

We have carried out a few simulations also for bosons subject to a confining potential with triangular symmetry (i.e., same as in the star-shaped grid), augmented with a box potential of hexagonal shape:

\small
\begin{align}
u_{\mathrm{opt}}(x,y) &= \frac{2}{9}V_0 \Bigg[ 3 - \cos\left(2\pi x+\frac{2 \pi y}{\sqrt{3}}\right) \notag \\
&\quad - \cos\left(\frac{4 \pi y}{\sqrt{3}}\right) - \cos\left(2\pi x-\frac{2 \pi y}{\sqrt{3}}\right) \Bigg]\,,\notag \\
u_{\mathrm{box}}(x,y) &= \left(\frac{x}{1.75}\right)^{20} + \left(\frac{x+\sqrt{3}y}{3.5}\right)^{20} + \left(\frac{-x+\sqrt{3}y}{3.5}\right)^{20}\,.
\label{eq15}
\end{align}
\normalsize

In Fig.~\ref{fig15} top we plot density maps for $N=12$ and $\lambda=1$, as functions of $V_0$.
Similarly to Fig.~\ref{fig10} we see a smooth transition from a superfluid-like to an insulating-like structure.
Contrary to expectation, $V_0=30$ is not large enough to fully recover the structure seen in Fig.~\ref{fig7}a.
This happens because the hexagonal box is unfit for totally excluding particles from approaching the six minima of the optical-lattice potential at distance two from the center. 

We have also tried to confine bosons in a circular box of radius 1.85, i.e., intermediate between the second-neighbor and the third-neighbor distance on the triangular grid.
In Fig.~\ref{fig15} bottom we report results for $N=7$ and $\lambda=1$.
For $V_0=0$, the structure is superfluid ($f_s=0.22$).
Upon increasing $V_0$, six spots at second-neighbor positions eventually appear, and the superfluid fraction eventually drops to zero ($f_s\simeq 0.04$ at $V_0=30$).
In this case, no spurious density maxima appear at distance two from the center because, with seven particles only, the interparticle repulsion is fully effective in excluding particles from these sites.

\section{Conclusions}

Perfect ordering can only take place in an infinitely large system.
In practice, already fairly small systems can anticipate some features of the ordered phase in an elementary form if the temperature is low enough.
It is worth studying how this structuring builds up in a finite quantum system.
We have chosen the BH model for a demonstration.

The BH model is a popular model of bosonic particles on a lattice, with hopping ($\propto t$) and on-site repulsion ($\propto U$) terms, describing the competition between tunneling (large $t/U$) and localization (large $U/t$) at low temperature.
Including in the model a longer-range repulsion (``extended BH model'') is a common strategy to make more varied the crystalline structure of the insulator;
in this case, also phases with mixed superfluid-crystalline nature (supersolids) may become stable.
Further peculiarities arise from the application of a confining potential (e.g., from putting particles in a trap), which makes the system inhomogeneous and strongly affects the structure of all phases (see, e.g., Refs.~\cite{prestipino2019a,ciardi2024}).
A truncated lattice is a limiting case of confinement, corresponding to a trapping potential which blows up upon crossing the edge of the grid.

We have investigated the extended BH model on small two-dimensional grids, using exact diagonalization to reconstruct the zero-temperature phase diagram in the hard-core, $U\rightarrow\infty$ limit.
Imposing single site occupancy leads to a drastic simplification, but nevertheless it is a valuable approximation when the chemical potential is not too large.
For three different grids, we have traced the boundaries between regions ($N$-sectors) comprising ground states with the same number $N$ of bosons.
Some of these lines are in a clear correspondence with the iso-density lines (and phase boundaries) of the infinite-size system, others are unexpected and point to the existence of novel finite-size ground states (``phases'') where particles gather on the boundary of the grid.
By computing the condensate fraction and the energy gap we can clearly distinguish between superfluid-like and insulating-like phases.
The same exercise has been repeated under PBC, finding that the superfluid region is now more extended.

Next, we have tried to reproduce some of the aforementioned unusual features in a few-boson system in continuous space, with the aim of exploring the connections between the low-temperature properties of confined bosonic particles and the physics of BH models.
To this purpose, we consider bosons interacting through an inverse-power repulsion and expose them to suitable optical-lattice and box potentials, tailored to the geometry of the grid.
Using PIMC simulations, we have recovered the same smooth crossover from superfluid to insulating behavior observed in the extended BH model on a small grid, including the structure of the insulating phase.

Broadening the perspective, our results provide a theoretical benchmark for several emerging experimental platforms. Specifically, the geometries analyzed here, such as the $3 \times 3$ tile and the 13-site star, can be faithfully realized using programmable optical tweezers arrays~\cite{browaeys2020,kaufman2021}. In these setups, the single-site resolution offered by quantum gas microscopy allows for a direct observation of the density profiles and insulating-like patterns we investigated~\cite{sherson2010single,bakr2009quantum}. Furthermore, the hard-core limit explored in our exact diagonalization study is naturally suited for implementation in superconducting qubit processors~\cite{roushan2017,yan2019}, where the interplay between hopping and interactions mimics the extended BH Hamiltonian.

Finally, our PIMC results in the continuum suggest that signatures of lattice-like behavior and superfluidity are robust even in small traps, a finding that could be tested in experiments with few-body systems of dipolar atoms~\cite{boettcher2021} or excitons in semiconductor nanostructures \cite{high2012, lagoin2022}.
In particular, two examples of systems where the extended BH paradigm is at play, and where our constrained continuous-space simulations can fruitfully complement the experiments, are magnetic atoms confined in an optical lattice~\cite{su2023} and semiconductor excitons on a lattice imprinted by an array of electrodes~\cite{lagoin2022a}.
In the former case, anisotropic long-range interactions can be distinctly tuned to originate a wealth of stripe-ordered states, both stable and metastable.
Other insulating phases, with integer or fractional filling, can be realized with dipolar excitons confined in customizable lattice grids.
In both these experiments, the grid contains 100-200 sites.
These numbers of sites are still too large to observe deviations of the BH behavior from the thermodynamic limit.
It would be intriguing if the same experiments could be replicated with even smaller grids, as this might reveal finite-size effects similar to those analyzed in the present paper.

By bridging the gap between discrete lattice models and continuous-space simulations~\cite{zwerger2003,pilati2012}, our study confirms that the fundamental competition between kinetic energy and repulsion, typical of Bose-Hubbard physics, emerges robustly even in systems composed of just a few particles~\cite{blume2012, wenz2013}. This offers a clear path for the design and calibration of small-scale quantum simulators~\cite{ebadi2021}.

\begin{acknowledgments}
This work was supported by the European Union through the Next Generation EU funds through the Italian MUR National Recovery and Resilience Plan, Mission 4 Component 2 - Investment 1.4 - National Center for HPC, Big Data and Quantum Computing (CUP B83C22002830001). 
F. C. acknowledges financial support from PNRR MUR Project No. PE0000023-NQSTI. 
\end{acknowledgments}

\bibliography{mybose.bib}

@article{baier2016,
author = {Baier, S. and Mark, M. J. and Petter, D. and Aikawa, K. and Chomaz, L. and Cai, Z. and Baranov, M. and Zoller, P. and Ferlaino, F.},
title = {Extended {B}ose-{H}ubbard models with ultracold magnetic atoms},
journal = {Science},
volume = {352},
number = {6282},
pages = {201-205},
year = {2016},
doi = {10.1126/science.aac9812},
key = {169},
}

@article{bloch2008b,
author = {Bloch, I. and Dalibard, J. and Zwerger, W.},
issue = {3},
journal = {Rev. Mod. Phys.},
month = {07},
pages = {885--964},
publisher = {American Physical Society},
title = {Many-body physics with ultracold gases},
volume = {80},
year = {2008},
doi = {http://link.aps.org/doi/10.1103/RevModPhys.80.885},
key = {217}
}

@article{boninsegni2006b,
author = {Boninsegni, M. and Prokof'ev, N. and Svistunov, B.},
doi = {10.1103/PhysRevLett.96.070601},
issue = {7},
journal = {Phys. Rev. Lett.},
month = {02},
numpages = {4},
pages = {070601},
publisher = {American Physical Society},
title = {Worm Algorithm for Continuous-Space Path Integral {M}onte {C}arlo Simulations},
url = {https://link.aps.org/doi/10.1103/PhysRevLett.96.070601},
volume = {96},
year = {2006},
key = {225},
}

@article{browaeys2020,
author = {Browaeys, A. and Lahaye, T.},
da = {2020/02/01},
doi = {10.1038/s41567-019-0733-z},
id = {Browaeys2020},
journal = {Nature Physics},
number = {2},
pages = {132--142},
title = {Many-body physics with individually controlled Rydberg atoms},
ty = {JOUR},
url = {https://doi.org/10.1038/s41567-019-0733-z},
volume = {16},
year = {2020},
key = {243},
}

@article{capogrosso-sansone2007,
title = {Phase diagram and thermodynamics of the three-dimensional {B}ose-{H}ubbard model},
author = {Capogrosso-Sansone, B. and Prokof'ev, N. V. and Svistunov, B. V.},
journal = {Phys. Rev. B},
volume = {75},
issue = {13},
pages = {134302},
numpages = {10},
year = {2007},
month = {04},
publisher = {American Physical Society},
doi = {10.1103/PhysRevB.75.134302},
url = {https://link.aps.org/doi/10.1103/PhysRevB.75.134302},
key = {106},
}

@article{capogrosso-sansone2010,
author = {Capogrosso-Sansone, B. and Trefzger, C. and Lewenstein, M. and Zoller, P. and Pupillo, G.},
journal = {Phys. Rev. Lett.},
month = {03},
number = {12},
numpages = {4},
pages = {125301},
publisher = {American Physical Society},
title = {Quantum Phases of Cold Polar Molecules in 2D Optical Lattices},
volume = {104},
year = {2010},
doi = {10.1103/PhysRevLett.104.125301}
}

@article{ceperley1995,
author = {Ceperley, D. M.},
doi = {10.1103/RevModPhys.67.279},
issue = {2},
journal = {Rev. Mod. Phys.},
month = {04},
numpages = {0},
pages = {279--355},
publisher = {American Physical Society},
title = {Path integrals in the theory of condensed helium},
url = {https://link.aps.org/doi/10.1103/RevModPhys.67.279},
volume = {67},
year = {1995},
key = {273},
}

@article{ciardi2024,
  title = {Supersolid Phases of Bosonic Particles in a Bubble Trap},
  author = {Ciardi, Matteo and Cinti, Fabio and Pellicane, Giuseppe and Prestipino, Santi},
  journal = {Phys. Rev. Lett.},
  volume = {132},
  issue = {2},
  pages = {026001},
  numpages = {8},
  year = {2024},
  month = {Jan},
  publisher = {American Physical Society},
  doi = {10.1103/PhysRevLett.132.026001}
}

@article{degregorio2021,
article-number = {10053},
author = {De Gregorio, D. and Prestipino, S.},
doi = {10.3390/app112110053},
journal = {Applied Sciences},
pages={10053},
number = {21},
title = {Classical and Quantum Gases on a Semiregular Mesh},
volume = {11},
year = {2021},
key = {338},
}

@article{fisher1989,
author = {Fisher, M. P. A. and Weichman, P. B. and Grinstein, G. and Fisher, D. S.},
issue = {1},
journal = {Phys. Rev. B},
month = {07},
pages = {546--570},
publisher = {American Physical Society},
title = {Boson localization and the superfluid-insulator transition},
volume = {40},
year = {1989},
doi = {doi/10.1103/PhysRevB.40.546}
}

@article{greiner2002,
author = {Greiner, M. and Mandel, O. and Esslinger, T. and H\"{a}nsch, T. W. and Bloch, I.},
date = {2002-01-03},
day = {03},
journal = {Nature},
m3 = {10.1038/415039a},
month = {01},
number = {6867},
pages = {39--44},
title = {Quantum phase transition from a superfluid to a {M}ott insulator in a gas of ultracold atoms},
ty = {JOUR},
volume = {415},
year = {2002},
doi = {10.1038/415039a},
key = {444},
}

@article{jain2011a,
author = {Jain, P. and Cinti, F. and Boninsegni, M.},
doi = {10.1103/PhysRevB.84.014534},
issue = {1},
journal = {Phys. Rev. B},
month = {07},
numpages = {9},
pages = {014534},
publisher = {American Physical Society},
title = {Structure, {B}ose-{E}instein condensation, and superfluidity of two-dimensional confined dipolar assemblies},
url = {https://link.aps.org/doi/10.1103/PhysRevB.84.014534},
volume = {84},
year = {2011},
key = {508},
}

@article{jaksch1998,
author = {Jaksch, D. and Bruder, C. and Cirac, J. I. and Gardiner, C. W. and Zoller, P.},
issue = {15},
journal = {Phys. Rev. Lett.},
month = {10},
pages = {3108--3111},
publisher = {American Physical Society},
title = {Cold Bosonic Atoms in Optical Lattices},
volume = {81},
year = {1998},
doi = {10.1103/PhysRevLett.81.3108},
key = {511},
}

@article{pollet2010,
author = {Pollet, L. and Picon, J. D. and B\"uchler, H. P. and Troyer, M.},
issue = {12},
journal = {Phys. Rev. Lett.},
month = {03},
numpages = {4},
pages = {125302},
publisher = {American Physical Society},
title = {Supersolid Phase with Cold Polar Molecules on a Triangular Lattice},
volume = {104},
year = {2010},
doi = {10.1103/PhysRevLett.104.125302}
}

@article{prestipino2019a,
author = {Prestipino, S. and Giaquinta, P. V.},
doi = {10.1103/PhysRevA.99.063619},
issue = {6},
journal = {Phys. Rev. A},
month = {06},
numpages = {17},
pages = {063619},
publisher = {American Physical Society},
title = {Ground state of weakly repulsive soft-core bosons on a sphere},
url = {https://link.aps.org/doi/10.1103/PhysRevA.99.063619},
volume = {99},
year = {2019},
key = {782},
}

@article{prestipino2021,
author = {Prestipino, S.},
doi = {10.1103/PhysRevA.103.033313},
issue = {3},
journal = {Phys. Rev. A},
month = {03},
numpages = {18},
pages = {033313},
publisher = {American Physical Society},
title = {Bose-{H}ubbard model on polyhedral graphs},
url = {https://link.aps.org/doi/10.1103/PhysRevA.103.033313},
volume = {103},
year = {2021},
key = {784},
}

@article{sindzingre1989,
author = {Sindzingre, P. and Klein, M. L. and Ceperley, D. M.},
doi = {10.1103/PhysRevLett.63.1601},
issue = {15},
journal = {Phys. Rev. Lett.},
month = {10},
numpages = {0},
pages = {1601--1604},
publisher = {American Physical Society},
title = {Path-integral {M}onte {C}arlo study of low-temperature $^{4}\mathrm{He}$ clusters},
url = {https://link.aps.org/doi/10.1103/PhysRevLett.63.1601},
volume = {63},
year = {1989},
key = {884},
}

@inproceedings{gheeraert2015,
  title={Mean-Field Theory for Extended {B}ose-{H}ubbard Model with Hard-Core Bosons},
  author={Gheeraert, Nicolas and Chester, Shai and May, Mathias and Eggert, Sebastian and Pelster, Axel},
  booktitle={Self-organization in Complex Systems: The Past, Present, and Future of Synergetics: Proceedings of the International Symposium, Hanse Institute of Advanced Studies, Delmenhorst, Germany, November 13-16, 2012},
  doi={10.1007/978-3-319-27635-9_18},
  pages={289--296},
  year={2015},
  organization={Springer}
}

@article{sheshadri1993,
  title={Superfluid and insulating phases in an interacting-boson model: Mean-field theory and the {RPA}},
  author={Sheshadri, K and Krishnamurthy, HR and Pandit, Rahul and Ramakrishnan, TV},
  journal={EPL (Europhysics Letters)},
  volume={22},
  number={4},
  pages={257--263},
  year={1993},
  doi={10.1209/0295-5075/22/4/004}
}

@article{capogrosso-sansone2008,
  title={Monte {C}arlo study of the two-dimensional {B}ose-{H}ubbard model},
  author={Capogrosso-Sansone, Barbara and S{\"o}yler, {\c{S}}ebnem G{\"u}ne{\c{s}} and Prokof’ev, Nikolay and Svistunov, Boris},
  journal={Phys. Rev. A},
  volume={77},
  number={1},
  pages={015602},
  year={2008}, doi={10.1103/PhysRevA.77.015602}
}

@article{wessel2005,
author = {Wessel, S. and Troyer, M.},
issue = {12},
journal = {Phys. Rev. Lett.},
month = {09},
numpages = {4},
pages = {127205},
publisher = {American Physical Society},
title = {Supersolid Hard-Core Bosons on the Triangular Lattice},
volume = {95},
year = {2005},
doi = {10.1103/PhysRevLett.95.127205}
}

@ARTICLE{ciardi2025,
  title     = {Effects of gravity on supersolid order in bubble-trapped bosons},
  author = {Ciardi, Matteo and Cinti, Fabio and Pellicane, Giuseppe and Prestipino, Santi},
  journal   = {Phys. Rev. B.},
  publisher = {American Physical Society (APS)},
  volume    = {111},
  number    = {2},
  pages     = {024512},
  year      = {2025},
  doi = {10.1103/PhysRevB.111.024512}
}

@article{chanda2025,
  title={Recent progress on quantum simulations of non-standard {B}ose-{H}ubbard models},
  author={Chanda, Titas and Barbiero, Luca and Lewenstein, Maciej and Mark, Manfred J and Zakrzewski, Jakub},
  journal={Reports on Progress in Physics},
  volume={88},
  number={4},
  pages={044501},
  year={2025},
  publisher={IOP Publishing},
  doi={10.1088/1361-6633/adc3a7}
}

@article{su2023,
  title={Dipolar quantum solids emerging in a {H}ubbard quantum simulator},
  author={Su, Lin and Douglas, Alexander and Szurek, Michal and Groth, Robin and Ozturk, S Furkan and Krahn, Aaron and H{\'e}bert, Anne H and Phelps, Gregory A and Ebadi, Sepehr and Dickerson, Susannah and others},
  journal={Nature},
  volume={622},
  number={7984},
  pages={724--729},
  year={2023},
  publisher={Nature Publishing Group UK London},
  doi={10.1038/s41586-023-06614-3}
}

@article{rokhsar1991,
  title={Gutzwiller projection for bosons},
  author={Rokhsar, Daniel S and Kotliar, BG},
  journal={Physical Review B},
  volume={44},
  number={18},
  pages={10328},
  year={1991},
  publisher={APS}, doi={10.1103/PhysRevB.44.10328}
}

@article{krauth1992,
  title={Gutzwiller wave function for a model of strongly interacting bosons},
  author={Krauth, Werner and Caffarel, Michel and Bouchaud, Jean-Philippe},
  journal={Physical Review B},
  volume={45},
  number={6},
  pages={3137},
  year={1992},
  publisher={APS},
  doi={doi.org/10.1103/PhysRevB.45.3137}
}

@article{zhang2022,
  title={Supersolid phases of lattice dipoles tilted in three dimensions},
  author={Zhang, Jin and Zhang, Chao and Yang, Jin and Capogrosso-Sansone, Barbara},
  journal={Physical Review A},
  volume={105},
  number={6},
  pages={063302},
  year={2022},
  publisher={APS},
  doi={10.1103/PhysRevA.105.063302}
}

@article{iskin2009,
  title={Strong-coupling perturbation theory for the extended {B}ose-{H}ubbard model},
  author={Iskin, M and Freericks, JK},
  journal={Physical Review A},
  volume={79},
  number={5},
  pages={053634},
  year={2009},
  publisher={APS},
  doi={10.1103/PhysRevA.79.053634}
}

@article{lagoin2022a,
  title={Extended {B}ose-{H}ubbard model with dipolar excitons},
  author={Lagoin, Camille and Bhattacharya, Utso and Grass, Tobias and Chhajlany, RW and Salamon, Tymoteusz and Baldwin, Kirk and Pfeiffer, Loren and Lewenstein, Maciej and Holzmann, Markus and Dubin, Fran{\c{c}}ois},
  journal={Nature},
  volume={609},
  number={7927},
  pages={485--489},
  year={2022},
  publisher={Nature Publishing Group UK London},
  doi={10.1038/s41586-022-05123-z}
}

@article{suthar2020,
  title={Staggered superfluid phases of dipolar bosons in two-dimensional square lattices},
  author={Suthar, Kuldeep and Kraus, Rebecca and Sable, Hrushikesh and Angom, Dilip and Morigi, Giovanna and Zakrzewski, Jakub},
  journal={Physical Review B},
  volume={102},
  number={21},
  pages={214503},
  year={2020},
  publisher={APS},
  doi={10.1103/PhysRevB.102.214503}
}

@article{trefzger2011,
  title={Ultracold dipolar gases in optical lattices},
  author={Trefzger, Christian and Menotti, C and Capogrosso-Sansone, B and Lewenstein, M},
  journal={Journal of Physics B: Atomic, Molecular and Optical Physics},
  volume={44},
  number={19},
  pages={193001},
  year={2011}, doi={10.1088/0953-4075/44/19/193001}
}

@article{dutta2015,
  title={Non-standard {H}ubbard models in optical lattices: a review},
  author={Dutta, Omjyoti and Gajda, Mariusz and Hauke, Philipp and Lewenstein, Maciej and L{\"u}hmann, Dirk-S{\"o}ren and Malomed, Boris A and Sowi{\'n}ski, Tomasz and Zakrzewski, Jakub},
  journal={Reports on Progress in Physics},
  volume={78},
  number={6},
  pages={066001},
  year={2015},
  publisher={IOP Publishing}, doi={10.1088/0034-4885/78/6/066001}
}

@article{vojta2003,
  title={Quantum phase transitions},
  author={Vojta, Matthias},
  journal={Reports on Progress in Physics},
  volume={66},
  number={12},
  pages={2069--2110},
  year={2003},
  doi={10.1088/0034-4885/66/12/R01}
}

@article{sachdev1999,
  title={Quantum phase transitions},
  author={Sachdev, Subir},
  journal={Phys. World},
  volume={12},
  number={4},
  pages={33--38},
  year={1999},
  doi={10.1088/2058-7058/12/4/23}
}

@article{kuebler2019,
  title = {Improving mean-field theory for bosons in optical lattices via degenerate perturbation theory},
  volume = {99},
  ISSN = {2469-9934},
  doi = {10.1103/physreva.99.063603},
  pages={063603},
  number = {6},
  journal = {Physical Review A},
  publisher = {American Physical Society (APS)},
  author = {K\"{u}bler,  M. and Sant’Ana,  F. T. and dos Santos,  F. E. A. and Pelster,  A.},
  year = {2019},
  month = jun 
}

@article{prestipino2020,
  title = {Ultracold Bosons on a Regular Spherical Mesh},
  volume = {22},
  ISSN = {1099-4300},
  DOI = {10.3390/e22111289},
  number = {11},
  journal = {Entropy},
  publisher = {MDPI AG},
  author = {Prestipino,  Santi},
  year = {2020},
  month = nov,
  pages = {1289}
}

@article{vanoosten2001,
  title = {Quantum phases in an optical lattice},
  volume = {63},
  ISSN = {1094-1622},
  DOI = {10.1103/physreva.63.053601},
  number = {5},
  pages={053601},
  journal = {Physical Review A},
  publisher = {American Physical Society (APS)},
  author = {van Oosten,  D. and van der Straten,  P. and Stoof,  H. T. C.},
  year = {2001},
  month = apr 
}

@article{ng2010,
  title = {Thermal phase transitions of supersolids in the extended {B}ose-{H}ubbard model},
  volume = {82},
  ISSN = {1550-235X},
  DOI = {10.1103/physrevb.82.184505},
  number = {18},
  pages={184505},
  journal = {Physical Review B},
  publisher = {American Physical Society (APS)},
  author = {Ng,  Kwai-Kong},
  year = {2010},
  month = nov 
}

@article{iskin2011,
  title = {Route to supersolidity for the extended {B}ose-{H}ubbard model},
  volume = {83},
  ISSN = {1094-1622},
  url = {http://dx.doi.org/10.1103/PhysRevA.83.051606},
  DOI = {10.1103/physreva.83.051606},
  number = {5},
  pages={051606},
  journal = {Physical Review A},
  publisher = {American Physical Society (APS)},
  author = {Iskin,  M.},
  year = {2011},
  month = may 
}

@article{kimura2012,
  title = {Gutzwiller study of phase diagrams of extended {B}ose-{H}ubbard models},
  volume = {400},
  ISSN = {1742-6596},  doi={10.1088/1742-6596/400/1/012032},
  number = {1},
  pages = {012032},
  journal = {Journal of Physics: Conference Series},
  publisher = {IOP Publishing},
  author = {Kimura, Takashi},
  year = {2012},
  month = dec
}

@article{ohgoe2012,
  title = {Commensurate Supersolid of Three-Dimensional Lattice Bosons},
  volume = {108},
  ISSN = {1079-7114},
  url = {http://dx.doi.org/10.1103/PhysRevLett.108.185302},
  DOI = {10.1103/physrevlett.108.185302},
  number = {18},
  pages={185302},
  journal = {Physical Review Letters},
  publisher = {American Physical Society (APS)},
  author = {Ohgoe,  Takahiro and Suzuki,  Takafumi and Kawashima,  Naoki},
  year = {2012},
  month = may 
}

@article{kurdestany2012,
  title = {The inhomogeneous extended {B}ose‐{H}ubbard model: A mean‐field theory},
  volume = {524},
  ISSN = {1521-3889},
  DOI = {10.1002/andp.201100274},
  number = {3–4},
  journal = {Annalen der Physik},
  publisher = {Wiley},
  author = {Kurdestany,  J.M. and Pai,  R.V. and Pandit,  R.},
  year = {2012},
  month = feb,
  pages = {234–244}
}

@article{yamamoto2012,
  title = {Quantum phases of hardcore bosons with long-range interactions on a square lattice},
  volume = {86},
  ISSN = {1550-235X},
  url = {http://dx.doi.org/10.1103/PhysRevB.86.054516},
  DOI = {10.1103/physrevb.86.054516},
  number = {5},
  pages={054516},
  journal = {Physical Review B},
  publisher = {American Physical Society (APS)},
  author = {Yamamoto,  Daisuke and Masaki,  Akiko and Danshita,  Ippei},
  year = {2012},
  month = aug 
}

@article{zhang2011,
  title = {Supersolid phase transitions for hard-core bosons on a triangular lattice},
  volume = {84},
  ISSN = {1550-235X},
  DOI = {10.1103/physrevb.84.174515},
  number = {17},
  pages={174515},
  journal = {Physical Review B},
  publisher = {American Physical Society (APS)},
  author = {Zhang,  Xue-Feng and Dillenschneider,  Raoul and Yu,  Yue and Eggert,  Sebastian},
  year = {2011},
  month = nov 
}

@article{kapfer2015,
  title = {Two-Dimensional Melting: From Liquid-Hexatic Coexistence to Continuous Transitions},
  volume = {114},
  ISSN = {1079-7114},
  DOI = {10.1103/physrevlett.114.035702},
  number = {3},
  pages={035702},
  journal = {Physical Review Letters},
  publisher = {American Physical Society (APS)},
  author = {Kapfer,  Sebastian C. and Krauth,  Werner},
  year = {2015},
  month = jan 
}

@article{marienhagen2025,
  title = {Calculation of thermodynamic properties using path integral {M}onte {C}arlo simulations in the canonical ensemble},
  volume = {163},
  ISSN = {1089-7690},
  DOI = {10.1063/5.0282863},
  number = {7},
  pages={074116},
  journal = {The Journal of Chemical Physics},
  publisher = {AIP Publishing},
  author = {Marienhagen,  Philipp and Meier,  Karsten},
  year = {2025},
  month = aug 
}

@article{zhang2010,
  title = {Exact diagonalization: the {B}ose–{H}ubbard model as an example},
  volume = {31},
  ISSN = {1361-6404},
  url = {http://dx.doi.org/10.1088/0143-0807/31/3/016},
  DOI = {10.1088/0143-0807/31/3/016},
  number = {3},
  journal = {European Journal of Physics},
  publisher = {IOP Publishing},
  author = {Zhang,  J M and Dong,  R X},
  year = {2010},
  month = apr,
  pages = {591–602}
}

@article{sowinski2012,
  title = {Exact diagonalization of the one-dimensional {B}ose-{H}ubbard model with local three-body interactions},
  volume = {85},
  pages={065601},
  ISSN = {1094-1622},
  DOI = {10.1103/physreva.85.065601},
  number = {6},
  journal = {Physical Review A},
  publisher = {American Physical Society (APS)},
  author = {Sowiński,  Tomasz},
  year = {2012},
  month = jun 
}

@ARTICLE{szabados2012,
  title     = "Efficient iterative diagonalization of the {Bose--Hubbard} model for ultracold bosons in a periodic optical trap",
  author    = "Szabados, {\'A}gnes and Jeszenszki, P{\'e}ter and Surj{\'a}n, P{\'e}ter R",
  doi={10.1016/j.chemphys.2011.10.003},
  journal   = "Chem. Phys.",
  publisher = "Elsevier BV",
  volume    =  401,
  pages     = {208--216},
  month     =  jun,
  year      =  2012
}

@ARTICLE{raventos2017,
  title     = "Cold bosons in optical lattices: a tutorial for exact diagonalization",
  author    = "Ravent{\'o}s, David and Gra{\ss}, Tobias and Lewenstein, Maciej and Juli{\'a}-D{\'\i}az, Bruno",
  doi={10.1088/1361-6455/aa68b1},
  journal   = "J. Phys. B At. Mol. Opt. Phys.",
  publisher = "IOP Publishing",
  volume    =  50,
  number    =  11,
  pages     = {113001},
  month     =  jun,
  year      =  2017
}

@article{Kaufman2021,
  title = {Quantum science with optical tweezer arrays of ultracold atoms and molecules},
  volume = {17},
  ISSN = {1745-2481},
  DOI = {10.1038/s41567-021-01357-2},
  number = {12},
  journal = {Nature Physics},
  publisher = {Springer Science and Business Media LLC},
  author = {Kaufman,  Adam M. and Ni,  Kang-Kuen},
  year = {2021},
  month = nov,
  pages = {1324–1333}
}

@article{roushan2017,
  title = {Spectroscopic signatures of localization with interacting photons in superconducting qubits},
  author = {Roushan, P. and others},
  journal = {Science},
  volume = {358},
  number = {6367},
  pages = {1175--1179},
  year = {2017},
  doi = {10.1126/science.aao1401},
  publisher = {American Association for the Advancement of Science}
}

@article{yan2019,
  title = {Strongly correlated quantum walks with three interacting photons},
  author = {Yan, Zhiguang and others},
  journal = {Science},
  volume = {364},
  number = {6442},
  pages = {753--756},
  year = {2019},
  doi = {10.1126/science.aaw1611},
  publisher = {American Association for the Advancement of Science}
}

@article{boettcher2021,
  title = {New states of matter with fine-tuned interactions: quantum droplets and supersolids},
  author = {B{\"o}ttcher, F. and others},
  journal = {Reports on Progress in Physics},
  volume = {84},
  number = {1},
  pages = {012401},
  year = {2021},
  doi = {10.1088/1361-6633/abb9d3},
  publisher = {IOP Publishing}
}

@article{zwerger2003,
  title={{M}ott--{H}ubbard transition of ultracold atoms in optical lattices},
  author={Zwerger, Wilhelm},
  journal={Journal of Optics B: Quantum and Semiclassical Optics},
  volume={5},
  number={2},
  pages={S9},
  year={2003},
  doi = {10.1088/1464-4266/5/2/352},
  publisher = {IOP Publishing}
}

@article{pilati2012,
  title={{B}ose-{H}ubbard Model for Cold Atoms in Optical Lattices: Beyond the Proximity Approximation},
  author={Pilati, S. and Troyer, M.},
  journal={Physical Review Letters},
  volume={108},
  number={15},
  pages={155301},
  year={2012},
  doi = {10.1103/PhysRevLett.108.155301},
  publisher = {APS}
}

@article{blume2012,
  title={Few-body physics with ultracold atoms},
  author={Blume, Doerte},
  journal={Reports on Progress in Physics},
  volume={75},
  number={4},
  pages={046401},
  year={2012},
  doi = {10.1088/0034-4885/75/4/046401},
  publisher = {IOP Publishing}
}

@article{wenz2013,
  title={From Few to Many: Observe the Emergence of a {F}ermi Sea},
  author={Wenz, A. N. and others},
  journal={Science},
  volume={342},
  number={6157},
  pages={457--460},
  year={2013},
  doi = {10.1126/science.1240516},
  publisher = {American Association for the Advancement of Science}
}

@article{high2012,
  title={Condensation of excitons in a trap},
  author={High, A. A. and others},
  journal={Nano Letters},
  volume={12},
  number={5},
  pages={2605--2630},
  year={2012},
  doi = {10.1021/nl300989n},
  publisher = {ACS Publications}
}

@article{lagoin2022,
  title={Key role of the condensate fraction in the superfluid-to-insulator transition of dipolar excitons},
  author={Lagoin, C. and others},
  journal={Physical Review B},
  volume={105},
  pages={L121303},
  year={2022},
  doi = {10.1103/PhysRevB.105.L121303},
  publisher = {APS}
}

@article{ebadi2021,
  title={Quantum phases of matter on a 256-atom programmable quantum simulator},
  author={Ebadi, Sepehr and others},
  journal={Nature},
  volume={595},
  number={7866},
  pages={227--232},
  year={2021},
  doi = {10.1038/s41586-021-03582-4},
  publisher = {Nature Publishing Group}
}

@article{sherson2010single,
  title={Single-atom-resolved fluorescence imaging of an atomic {M}ott insulator},
  author={Sherson, Jacob F and Weitenberg, Christof and Endres, Manuel and Cheneau, Marc and Bloch, Immanuel and Kuhr, Stefan},
  doi={10.1038/nature09378},
  journal={Nature},
  volume={467},
  number={7311},
  pages={68--72},
  year={2010},
  publisher={Nature Publishing Group UK London}
}

@article{bakr2009quantum,
  title={A quantum gas microscope for detecting single atoms in a {H}ubbard-regime optical lattice},
  author={Bakr, Waseem S and Gillen, Jonathon I and Peng, Amy and F{\"o}lling, Simon and Greiner, Markus},
  doi={10.1038/nature08482},
  journal={Nature},
  volume={462},
  number={7269},
  pages={74--77},
  year={2009},
  publisher={Nature Publishing Group UK London}
}

@article{yamamoto2009,
  title = {Successive phase transitions at finite temperatures toward the supersolid state in a three-dimensional extended {B}ose-{H}ubbard model},
  volume = {79},
  pages={094503},
  ISSN = {1550-235X},
  DOI = {10.1103/physrevb.79.094503},
  number = {9},
  journal = {Physical Review B},
  publisher = {American Physical Society (APS)},
  author = {Yamamoto,  Keisuke and Todo,  Synge and Miyashita,  Seiji},
  year = {2009},
  month = mar 
}

@article{alet2004,
  title = {Generic incommensurate transition in the two-dimensional boson {H}ubbard model},
  volume = {70},
  pages={024513},
  ISSN = {1550-235X},
  doi = {10.1103/physrevb.70.024513},
  number = {2},
  journal = {Physical Review B},
  publisher = {American Physical Society (APS)},
  author = {Alet,  Fabien and Sørensen,  Erik S.},
  year = {2004},
  month = jul 
}
\newpage

\appendix
\renewcommand{\thefigure}{A\arabic{figure}}
\setcounter{figure}{0}
\section{Tiling the lattice with grids}

In this appendix we report some figures which (to avoid text overloading) could not find place in Section III.

\begin{figure}[H]
\centering
        \begin{minipage}[t]{0.49\linewidth} 
        \centering
        (a)\par\medskip
        \includegraphics[width=1.0\linewidth]{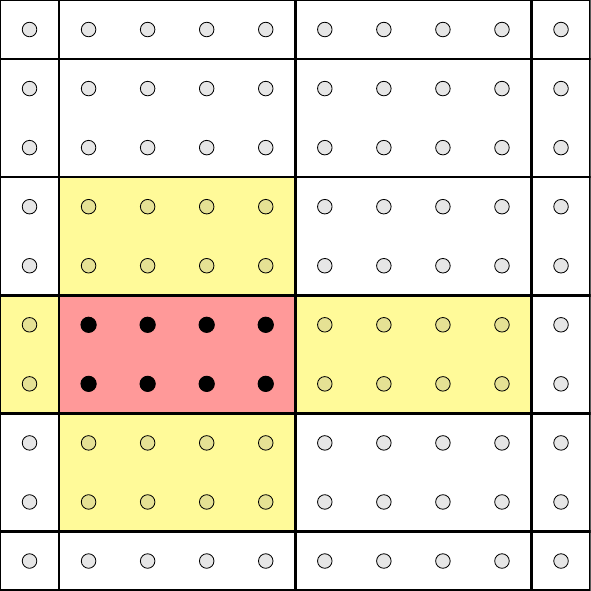}
    \end{minipage}\hfill
    \begin{minipage}[t]{0.49\linewidth} 
        \centering
        (b)\par\medskip
        \includegraphics[width=1.0\linewidth]{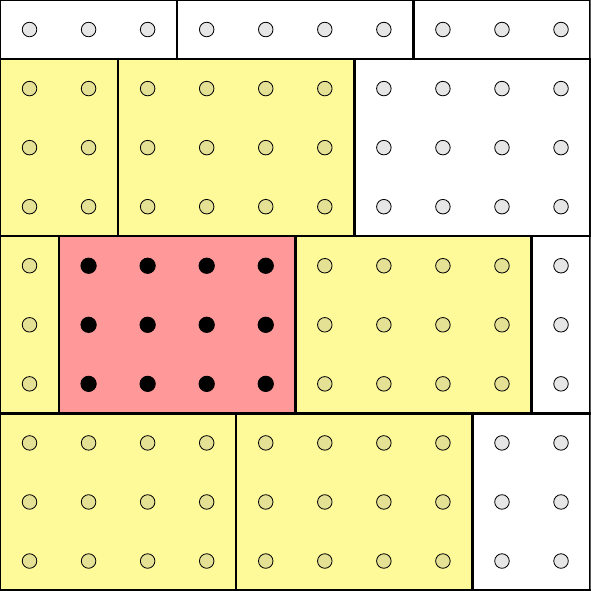}
    \end{minipage}
\caption{\label{figA1}
The rectangular grids analyzed in this paper, drawn together with their periodic images.}
\end{figure}

\begin{figure}[H]
    \centering
        \begin{minipage}[t]{0.49\linewidth} 
        \centering
        (a)\par\medskip
        \includegraphics[width=1.0\linewidth]{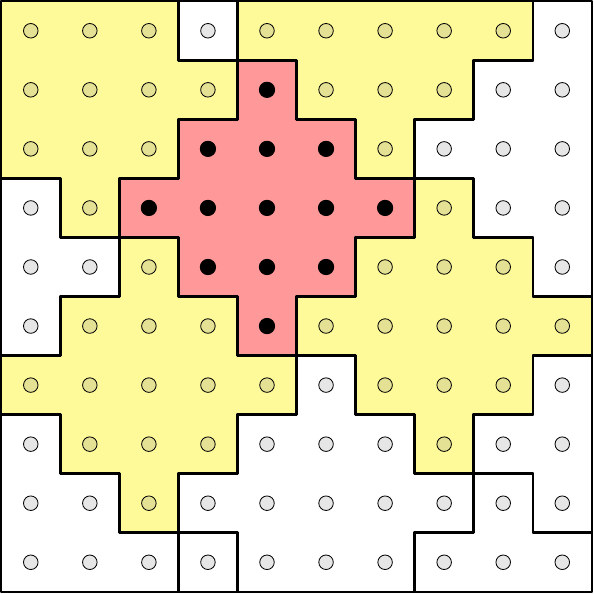}
    \end{minipage}\hfill
    \begin{minipage}[t]{0.49\linewidth} 
        \centering
        (b)\par\medskip
        \includegraphics[width=1.0\linewidth]{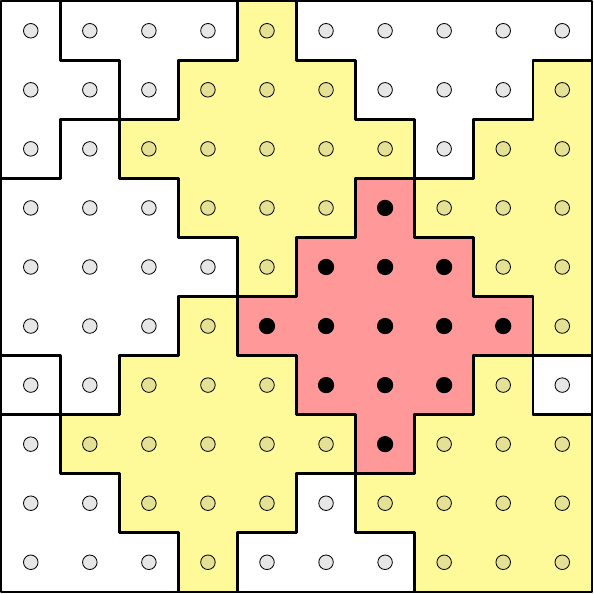}
    \end{minipage}
    \caption{\label{figA2}
    Diamond and its periodic images, chosen in two different ways. Both choices exclude realizability of checkerboard order.}
\end{figure}

\begin{figure}[H]
    \centering
        \begin{minipage}[t]{0.49\linewidth} 
        \centering
        (a)\par\medskip
        \includegraphics[width=1.0\linewidth]{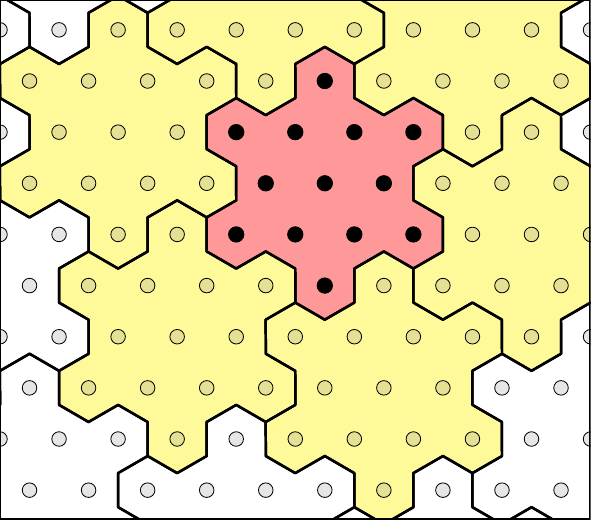}
    \end{minipage}\hfill
    \begin{minipage}[t]{0.49\linewidth} 
        \centering
        (b)\par\medskip
        \includegraphics[width=1.0\linewidth]{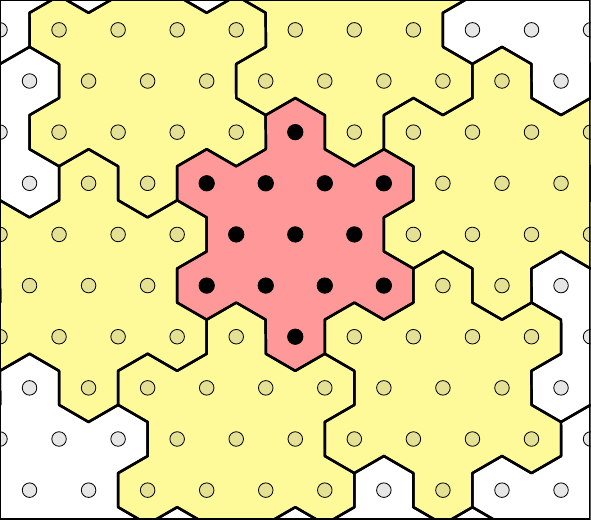}
    \end{minipage}
    \caption{\label{figA3}
    Star and its periodic images, chosen in two different ways.}
\end{figure}

In the two panels of Fig.~\ref{figA1} we show how the $4\times 2$ and $4\times 3$ rectangular grids mentioned in Section III.A are repeated periodically.
In either case, there is no a priori exclusion of checkerboard order in the model ground state.
We note that on the $4\times 2$ grid each vertical interaction has to be counted twice.

Figures \ref{figA2} and \ref{figA3} show how the infinite lattice can be covered with Diamond or Star tiles without overlaps or gaps. 
This can be done in two distinct ways, illustrated in the panels of each figure.
Though different as for the association of nearest neighbors to the boundary sites, the two choices are equivalent in terms of energy eigenvalues and eigenstates.

\section{Mean-field phase boundary on the triangular lattice}

In this appendix, the classic mean-field treatment of the ordinary BH model at $T=0$~\cite{vanoosten2001} is adapted to the extended BH model on the triangular lattice, taking into account the tripartite nature of the lattice.
The purpose of this calculation is to obtain approximate boundary lines separating the insulating phases from the superfluid phase in the thermodynamic limit.

Let A, B, and C be the three sublattices.
The Hamiltonian of the model reads:
\begin{equation}
H=-t\sum_{\langle i,j\rangle}\left(a_i^\dagger a_j+a_j^\dagger a_i\right)+H_0
\label{eqsm1}
\end{equation}
with
\footnotesize
\begin{eqnarray}
H_0&=&\frac{U}{2}\sum_{i\in{\rm A}}n_i(n_i-1)+\frac{U}{2}\sum_{i\in{\rm B}}n_i(n_i-1)+\frac{U}{2}\sum_{i\in{\rm C}}n_i(n_i-1)
\nonumber \\
&+&\frac{V}{2}\sum_{i\in{\rm A}}n_i\sum_{j\in{\rm NN}_i}n_j+\frac{V}{2}\sum_{i\in{\rm B}}n_i\sum_{j\in{\rm NN}_i}n_j+\frac{V}{2}\sum_{i\in{\rm C}}n_i\sum_{j\in{\rm NN}_i}n_j
\nonumber \\
&-&\mu\sum_{i\in{\rm A}}n_i-\mu\sum_{i\in{\rm B}}n_i-\mu\sum_{i\in{\rm C}}n_i\,.
\label{eqsm2}
\end{eqnarray}
\normalsize
In a mean-field type of approach, all sites belonging to the same sublattice are given the same occupancy.
This amounts to replace the original Hamiltonian with an effective three-site Hamiltonian $H'$, written in terms of sublattice field operators:
\small
\begin{eqnarray}
H'&=&-\frac{1}{2}t\left[a^\dagger_{\rm A}(3a_{\rm B}+3a_{\rm C})+a^\dagger_{\rm B}(3a_{\rm A}+3a_{\rm C})\right.
\nonumber \\
&&+\left.a^\dagger_{\rm C}(3a_{\rm A}+3a_{\rm B})+{\rm h.c.}\right]+H'_0
\nonumber \\
&=&-3t\left[a^\dagger_{\rm A}(a_{\rm B}+a_{\rm C})+a^\dagger_{\rm B}(a_{\rm A}+a_{\rm C})+a^\dagger_{\rm C}(a_{\rm A}+a_{\rm B})\right]+H'_0\,,
\nonumber \\
\label{eqsm3}
\end{eqnarray}
\normalsize
where
\small
\begin{eqnarray}
H'_0&=&\frac{U}{2}\left[n_{\rm A}(n_{\rm A}-1)+n_{\rm B}(n_{\rm B}-1)+n_{\rm C}(n_{\rm C}-1)\right]
\nonumber \\
&+&\frac{V}{2}\left[n_{\rm A}(3n_{\rm B}+3n_{\rm C})+n_{\rm B}(3n_{\rm A}+3n_{\rm C})+n_{\rm C}(3n_{\rm A}+3n_{\rm B})\right]
\nonumber \\
&-&\mu(n_{\rm A}+n_{\rm B}+n_{\rm C})\,.
\label{eqsm4}
\end{eqnarray}
\normalsize
Let $\{\vert x_{\rm A},x_{\rm B},x_{\rm C}\rangle\}$ be the set of Fock states in this representation.
These states are eigenstates of $H'_0$ with eigenvalue
\small
\begin{eqnarray}
\epsilon(x_{\rm A},x_{\rm B},x_{\rm C})&=&\frac{U}{2}\left[x_{\rm A}(x_{\rm A}-1)+x_{\rm B}(x_{\rm B}-1)+x_{\rm C}(x_{\rm C}-1)\right]
\nonumber \\
&+&3V(x_{\rm A}x_{\rm B}+x_{\rm A}x_{\rm C}+x_{\rm B}x_{\rm C})
\nonumber \\
&-&\mu(x_{\rm A}+x_{\rm B}+x_{\rm C})\,.
\label{eqsm5}
\end{eqnarray}
\normalsize
Using the decoupling approximation with $\langle a_\alpha\rangle=\phi_\alpha$ ($\alpha={\rm A},{\rm B},{\rm C}$), the hopping term in (\ref{eqsm3}) is linearized as
\small
\begin{eqnarray}
H'_1&\equiv&3t\left(\phi^*_{\rm A}\phi_{\rm B}+\phi^*_{\rm B}\phi_{\rm A}+\phi^*_{\rm A}\phi_{\rm C}+\phi^*_{\rm C}\phi_{\rm A}+\phi^*_{\rm B}\phi_{\rm C}+\phi^*_{\rm C}\phi_{\rm B}\right)
\nonumber \\
&&-3t\left[(\phi_{\rm B}+\phi_{\rm C})a_{\rm A}^\dagger+(\phi^*_{\rm B}+\phi^*_{\rm C})a_{\rm A}\right]
\nonumber \\
&&-3t\left[(\phi_{\rm A}+\phi_{\rm C})a_{\rm B}^\dagger+(\phi^*_{\rm A}+\phi^*_{\rm C})a_{\rm B}\right]
\nonumber \\
&&-3t\left[(\phi_{\rm A}+\phi_{\rm B})a_{\rm C}^\dagger+(\phi^*_{\rm A}+\phi^*_{\rm B})a_{\rm C}\right]\,.
\label{eqsm6}
\end{eqnarray}
\normalsize

Now, let $\vert x_{0\rm A},x_{0\rm B},x_{0\rm C}\rangle$ be the minimum-energy eigenstate of $H'_0$, while $H'_1$ is treated as a small perturbation.
Then, at first order in $t$ the effect of $H'_1$ is
\footnotesize
\begin{eqnarray}
&&\langle x_{0\rm A},x_{0\rm B},x_{0\rm C}\vert H'_1\vert x_{0\rm A},x_{0\rm B},x_{0\rm C}\rangle
\nonumber \\
&=&3t\left(\phi^*_{\rm A}\phi_{\rm B}+\phi^*_{\rm B}\phi_{\rm A}+\phi^*_{\rm A}\phi_{\rm C}+\phi^*_{\rm C}\phi_{\rm A}+\phi^*_{\rm B}\phi_{\rm C}+\phi^*_{\rm C}\phi_{\rm B}\right)\,.
\nonumber \\
\label{eqsm7}
\end{eqnarray}
\normalsize
At second order in $t$, the energy correction due to $H'_1$ is instead:
\begin{equation}
\sum_{(x_{\rm A},x_{\rm B},x_{\rm C})\ne(x_{0{\rm A}},x_{0{\rm B}},x_{0{\rm C}})}\frac{\left|\langle x_{0{\rm A}},x_{0{\rm B}},x_{0{\rm C}}\vert H'_1\vert x_{\rm A},x_{\rm B},x_{\rm C}\rangle\right|^2}{\epsilon(x_{0{\rm A}},x_{0{\rm B}},x_{0{\rm C}})-\epsilon(x_{\rm A},x_{\rm B},x_{\rm C})}\,.
\label{eqsm8}
\end{equation}
In this sum, the only non-zero terms correspond to
\small
\begin{eqnarray}
(x_{\rm A},x_{\rm B},x_{\rm C})&=&(x_{0{\rm A}}\pm 1,x_{0{\rm B}},x_{0{\rm C}})\,,(x_{0{\rm A}},x_{0{\rm B}}\pm 1,x_{0{\rm C}})\,,
\nonumber \\
&&(x_{0{\rm A}},x_{0{\rm B}},x_{0{\rm C}}\pm 1)\,,
\label{eqsm9}
\end{eqnarray}
\normalsize
being, e.g.,
\begin{eqnarray}
&&\langle x_{0\rm A},x_{0\rm B},x_{0\rm C}\vert H'_1\vert x_{0\rm A}\pm 1,x_{0\rm B},x_{0\rm C}\rangle
\nonumber \\
&=&\left\{
\begin{array}{l}
-3t(\phi^*_{\rm B}+\phi^*_{\rm C})\sqrt{x_{0\rm A}+1}\\
-3t(\phi_{\rm B}+\phi_{\rm C})\sqrt{x_{0\rm A}}
\end{array}
\right.
\label{eqsm10}
\end{eqnarray}
Moreover,
\begin{eqnarray}
&&\epsilon(x_{0{\rm A}},x_{0{\rm B}},x_{0{\rm C}})-\epsilon(x_{0{\rm A}}+1,x_{0{\rm B}},x_{0{\rm C}})
\nonumber \\
&&=\mu-Ux_{0{\rm A}}-3V(x_{0{\rm B}}+x_{0{\rm C}})\,;
\nonumber \\
&&\epsilon(x_{0{\rm A}},x_{0{\rm B}},x_{0{\rm C}})-\epsilon(x_{0{\rm A}}-1,x_{0{\rm B}},x_{0{\rm C}})
\nonumber \\
&&=-\mu+U(x_{0{\rm A}}-1)+3V(x_{0{\rm B}}+x_{0{\rm C}})\,,
\label{eqsm11}
\end{eqnarray}
etc.
The perturbed energy is then
\small
\begin{eqnarray}
&&E=\epsilon(x_{0{\rm A}},x_{0{\rm B}},x_{0{\rm C}})
\nonumber \\
&&+3t\left(\phi^*_{\rm A}\phi_{\rm B}+\phi^*_{\rm B}\phi_{\rm A}+\phi^*_{\rm A}\phi_{\rm C}+\phi^*_{\rm C}\phi_{\rm A}+\phi^*_{\rm B}\phi_{\rm C}+\phi^*_{\rm C}\phi_{\rm B}\right)
\nonumber \\
&&+9t^2\left|\phi_{\rm B}+\phi_{\rm C}\right|^2\left[{\rm A}\right]+9t^2\left|\phi_{\rm A}+\phi_{\rm C}\right|^2\left[{\rm B}\right]
\nonumber \\
&&+9t^2\left|\phi_{\rm A}+\phi_{\rm B}\right|^2\left[{\rm C}\right]
\nonumber \\
&\equiv&\epsilon(x_{0{\rm A}},x_{0{\rm B}},x_{0{\rm C}})+\sum_{\alpha,\beta}\phi^*_\alpha A_{\alpha,\beta}\phi_\beta\,,
\label{eqsm12}
\end{eqnarray}
\normalsize
where $A_{\alpha,\beta}$ is a real symmetric matrix and
\begin{eqnarray}
[{\rm A}]&=&-\frac{x_{0{\rm A}}+1}{Ux_{0{\rm A}}+3V(x_{0{\rm B}}+x_{0{\rm C}})-\mu}
\nonumber \\
&+&\frac{x_{0{\rm A}}}{U(x_{0{\rm A}}-1)+3V(x_{0{\rm B}}+x_{0{\rm C}})-\mu}\,,
\label{eqsm13}
\end{eqnarray}

\begin{eqnarray}
[{\rm B}]&=&-\frac{x_{0{\rm B}}+1}{Ux_{0{\rm B}}+3V(x_{0{\rm A}}+x_{0{\rm C}})-\mu}
\nonumber \\
&+&\frac{x_{0{\rm B}}}{U(x_{0{\rm B}}-1)+3V(x_{0{\rm A}}+x_{0{\rm C}})-\mu}\,,
\label{eqsm14}
\end{eqnarray}
and
\begin{eqnarray}
[{\rm C}]&=&-\frac{x_{0{\rm C}}+1}{Ux_{0{\rm C}}+3V(x_{0{\rm A}}+x_{0{\rm B}})-\mu}
\nonumber \\
&+&\frac{x_{0{\rm C}}}{U(x_{0{\rm C}}-1)+3V(x_{0{\rm A}}+x_{0{\rm B}})-\mu}\,.
\label{eqsm15}
\end{eqnarray}
If the Hermitian form in (\ref{eqsm12}) is of definite sign, then we are either in an insulating phase ($\phi^*A\phi>0$ for any $\phi\ne 0$) or in the superfluid phase ($\phi^*A\phi<0$, signaling an instability towards non-zero $\phi$ values).
Hence, the phase boundary is given by the condition that the form is semidefinite:
\begin{equation}
\begin{vmatrix}
3t([{\rm B}]+[{\rm C}]) & 1+3t[{\rm C}] & 1+3t[{\rm B}]\\
1+3t[{\rm C}] & 3t([{\rm A}]+[{\rm C}]) & 1+3t[{\rm A}]\\
1+3t[{\rm B}] & 1+3t[{\rm A}] & 3t([{\rm A}]+[{\rm B}])
\end{vmatrix}
=0\,.
\label{eqsm16}
\end{equation}
Equation (\ref{eqsm16}) defines the locus of $(t,\mu)$ points providing the upper limit of stability of the insulating phase for the given $x_{0{\rm A}}$, etc.
To treat the hard-core case, we took $U=1000V$.
\end{document}